\documentclass[acmtog, nonacm]{acmart}

\setcopyright{none}
\makeatletter
\@ACM@acmcpfalse
\makeatother
\newcommand{\final}{1}

\PassOptionsToPackage{dvipsnames,svgnames,table}{xcolor}
\usepackage{xcolor}

\usepackage{booktabs}

\usepackage{graphicx} 
\usepackage[export]{adjustbox} 
\usepackage{subfig} 
\usepackage{wrapfig} 

\usepackage{amsmath}
\usepackage{amsthm}
\usepackage{mathtools} 

\usepackage{pifont} 
\newcommand{\cmark}{\ding{51}}%
\newcommand{\xmark}{\ding{55}}%

\usepackage[most]{tcolorbox}
\usepackage[export]{adjustbox}
\usepackage{algorithm}
\usepackage{algpseudocode}

\usepackage{colortbl}
\usepackage{array}
\makeatletter
\providecommand\insert@pcolumn{\insert@column}
\makeatother
\usepackage{ragged2e}
\usepackage{tabularx} 
\usepackage{makecell} 
\usepackage{multirow} 

\usepackage{cleveref} 

\usepackage{alltt}
\usepackage{newlfont} 
\usepackage{floatflt} 
\usepackage{float} 

\usepackage{CJKutf8} 
\usepackage{multirow}
\usepackage{xspace}
\usepackage{fontawesome5}
\usepackage[table]{xcolor}
\definecolor{NavyBlue}{RGB}{0,0,128}
\newcommand{\greencheck}{{\color{green!40!black}\cmark}\xspace}
\newcommand{\green}{\cellcolor{green!12.5}\greencheck}
\newcommand{\yellowcheck}{{\color{orange}(\cmark)}\xspace}
\newcommand{\yellow}{\cellcolor{orange!12.5}\yellowcheck}
\newcommand{\redcheck}{{\color{red}\xmark}\xspace}
\newcommand{\red}{\cellcolor{red!12.5}\redcheck}

\usepackage{needspace}

\usepackage{tikz}
\usetikzlibrary{calc,arrows.meta,positioning,decorations.pathreplacing}
\usepackage{tikz-3dplot}

\setcitestyle{square}
\def\textrightarrow{$\rightarrow$}

\let\oldcaption\caption
\renewcommand{\caption}[2][]{\oldcaption[#1]{{\em #1} #2}}

\definecolor{SithColor}{rgb}{0.7,0,0} 
\newcommand{\praneeth}[1]{{\color{SithColor} Praneeth: #1 $\qed$}}
\newcommand{\zihao}[1]{{\color{SithColor} Zihao: #1 $\qed$}}
\newcommand{\todo}[1]{}

\definecolor{ConsularColor}{rgb}{0,0.4,0} 
\definecolor{GuardianColor}{rgb}{0,0,0.8} 
\newcommand{\yoda}[1]{{\color{ConsularColor} Yoda: #1 $\qed$}}
\newcommand{\warning}[1]{}
\newcommand{\note}[1]{}
\newcommand{\nothing}[1]{}

\definecolor{AudioColor}{rgb}{0.56,0.34,0.62}
\newcommand{\audio}[1]{}

\definecolor{figred}{rgb}{1,0,0}
\definecolor{figgreen}{rgb}{0,0.6,0}
\definecolor{figblue}{rgb}{0,0,1}
\definecolor{figpink}{rgb}{1,0.63,0.63}
\definecolor{HengyuColor}{rgb}{0.5, 0, 0.5}

\definecolor{ConsularColor}{rgb}{0,0.4,0} 
\newcommand{\pc}[1]{}
\newcommand{\anzhou}[1]{}
\newcommand{\hy}[1]{}

\ifthenelse{\equal{\final}{1}}
{
\renewcommand{\praneeth}[1]{}
\renewcommand{\yoda}[1]{}
\renewcommand{\warning}[1]{}
\renewcommand{\note}[1]{}
\renewcommand{\zihao}[1]{}
\renewcommand{\pc}[1]{}
}
{}

\newcommand{\pseudocode}{Algorithm}
\floatstyle{plain}
\newfloat{algorithm}{tbhp}{lop}
\floatname{algorithm}{\pseudocode}

\newcommand{\filename}[1]{\url{#1}}
\newcommand{\foldername}[1]{\url{#1}}

\ifdefined\email
\else
\newcommand{\email}[1]{\url{#1}}
\fi

\newcommand{\ie}{\emph{i.e., }}
\newcommand{\eg}{\emph{e.g., }}

\graphicspath{
{figs/handdrawn/}
{figs/raster/}
}

\begin{document}
\title{Ellipsography: Single-Shot Speckle-Free Holography via Vectorial Interference Shaping}

\author{Anzhou Wen}
\email{awen@unc.edu}
\affiliation{
  \institution{The University of North Carolina at Chapel Hill}
  \department{Dept. of Computer Science}
  \city{Chapel Hill}
  \state{NC}
  \country{USA}
}

\author{Praneeth Chakravarthula}
\email{cpk@cs.unc.edu}
\affiliation{
  \institution{The University of North Carolina at Chapel Hill}
  \department{Dept. of Computer Science}
  \city{Chapel Hill}
  \state{NC}
  \country{USA}
}
\renewcommand\shortauthors{Wen and Chakravarthula}
\authorsaddresses{}

\begin{teaserfigure}
\centering
  \includegraphics[width=0.98\textwidth]{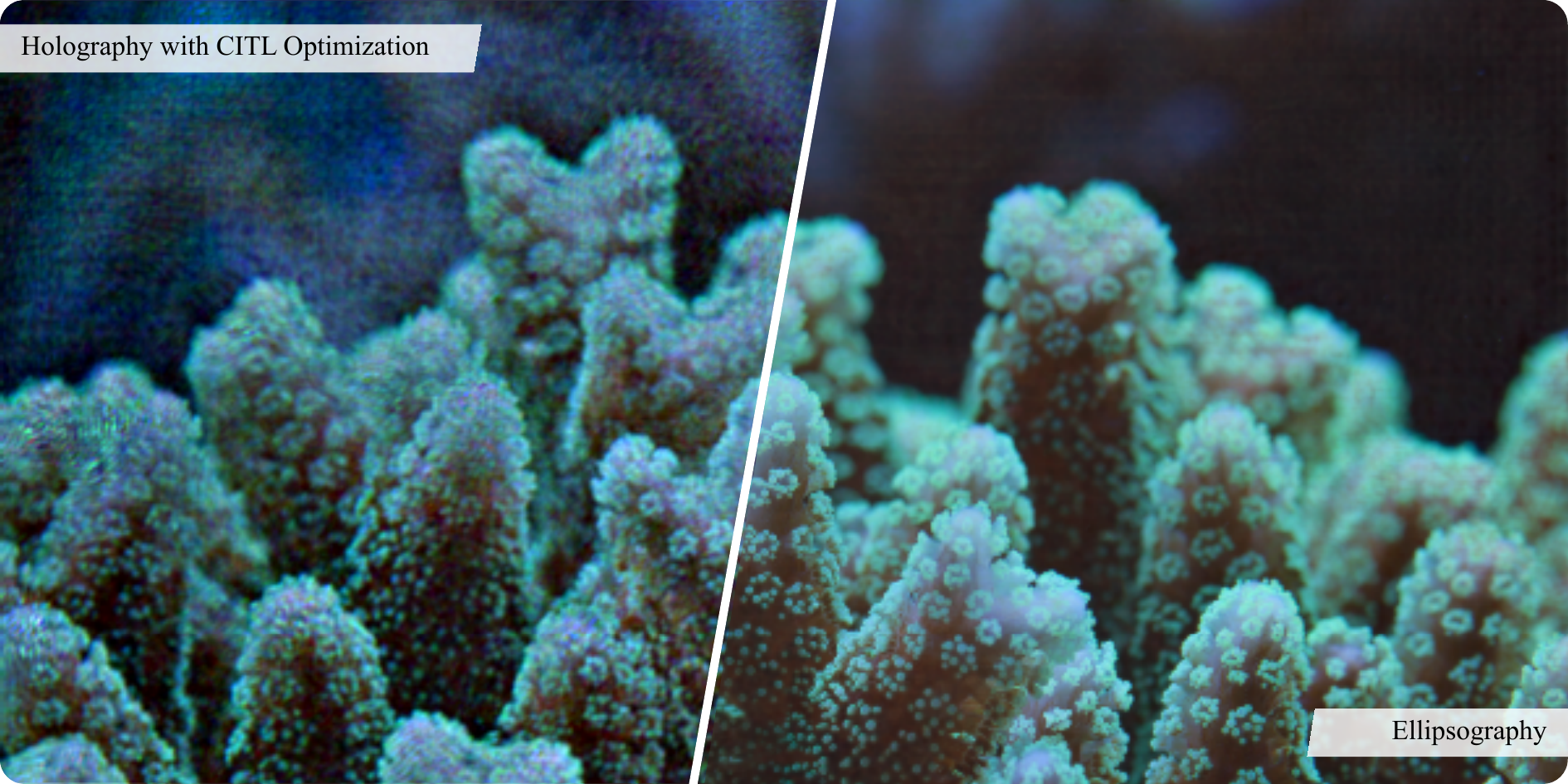}
  \vspace{-2mm}
  \caption{
  \textit{Random Phase Holograms Captured on a Prototype Holographic Display}.
  Camera-in-the-loop (CITL) optimization, which relies on active camera feedback-based iterative refinement to mitigate noise, has long defined the performance limits of experimental holographic displays.
  We introduce \textit{Ellipsography}, a single-shot speckle-free holography technique that jointly modulates phase and polarization to structurally control interference at the optical level. 
  Without frame averaging or active camera-feedback loops, Ellipsography achieves noise suppression in a single frame, while preserving high spatial resolution and realistic depth cues. This represents a pivotal step toward scalable, high-fidelity, noise-free holographic displays using standard SLM hardware.
  }
  \label{fig:teaser}
\end{teaserfigure}

\begin{abstract}

Holographic displays are widely regarded as the `ultimate' display technology, promising immersive 3D visuals with natural depth cues, continuous parallax, and perceptual realism. 
Realizing this potential, however, has remained elusive due to persistent image quality limitations---most notably speckle noise, a byproduct of the random interference inherent to coherent light. This is typically further exacerbated by the hologram's phase randomness required for maintaining uniform energy distribution across the eyebox.
While speckle suppression techniques like temporal multiplexing or smooth-phase heuristics exist, they often necessitate high-speed hardware and introduce visual artifacts, hindering their practical adoption.

We introduce \textit{Ellipsography}, a single-shot holography technique that achieves \textit{near-limit speckle suppression}, reaching the image fidelity equivalent to averaging a million conventional scalar holograms---in a single frame in simulation. 
By jointly modulating the phase and polarization of light, we structure optical interference and suppress speckle at its source. 
We present a full pipeline including a vectorial wave model, an end-to-end hologram synthesis algorithm, and a functional prototype display. 
Our experiments demonstrate substantial improvements in visual clarity, depth continuity, and focus cues over current state-of-the-art methods, achieving high-quality reconstructions approaching 30dB PSNR on a real holographic display for the first time---a 10dB improvement over the best existing techniques. 
By pushing holographic reconstruction closer to the perceptual quality expected of modern displays, Ellipsography sets a new benchmark for
practical, high-fidelity, speckle-free holography.

\end{abstract}

\maketitle

\section{Introduction}

For decades, we have envisioned a display so immersive, so true to life, that it would be indistinguishable from reality. Holography promises exactly that. By reconstructing the full light field, holography has the potential to render images that 
are perceptually indistinguishable from the real world.
Unlike conventional LCD and OLED displays that simulate depth through stereoscopic disparity and lens-based tricks, holography encodes and reconstructs the actual wavefront of light, enabling natural human accommodation responses 
\cite{kim2022accommodative,kim2024holographic}, and software-driven optical and eye aberration correction \cite{maimone2017holographic,shi2021towards}.
These capabilities make holography particularly well-suited for next-generation near-eye displays in compact, eyeglass-style form factors \cite{maimone2017holographic,gopakumar2024full}. 

Despite its potential and decades of progress, holography continues to face significant image quality issues, including limited resolution, shallow depth of field, and visual noise that prevent it from competing with mature display technologies like LCD and OLED, see \Cref{tab:display_comparison}.
These shortcomings remain major barriers to widespread adoption, let alone holography replacing existing display technologies. 
To close this gap, it is essential to develop noise-free, high-resolution holographic displays that can rival or surpass the visual quality of today's mobile screens.
Doing so would not only overcome longstanding limitations, but also establish holography as a viable mass-market display technology.

Phase randomness plays a key role in holographic image synthesis. It is widely used to diffuse light evenly across the display, avoiding concentrated energy hot spots in the eyebox, minimizing unwanted diffraction artifacts, and improving brightness uniformity \cite{kuo2023multisource,chakravarthula2022pupil}. 
Random phase modulation also helps suppress aliasing and ghosting, leading to more stable 3D reconstructions \cite{chakravarthula2022hogel,choi2021neural}. 
However, this necessary phase randomness introduces a major problem: \emph{speckle noise}. 
Due to destructive spatial interference among overlapping image-plane pixels, speckle appears as high-frequency grain across the image \cite{makowski2013minimized}.

A range of techniques have been proposed to reduce speckle, typically by averaging out multiple independently generated noise patterns over time or space \cite{park2020hologram,bianco2018strategies,goodman2013speckle,nam2023depolarized,choi2022time}.
While effective in principle, these methods come with trade-offs. 
Decreasing coherence of light source can suppress speckle but softens depth cues and reduces spatial detail \cite{deng2017coherence,peng2021speckle}.
Smooth phase holograms remove speckle by eliminating phase randomness altogether, but leads to uneven eyebox energy \cite{chakravarthula2022pupil,kuo2023multisource} and unnatural focal cues, undermining depth realism critical for perceptual immersion \cite{yoo2021optimization, kim2022accommodative,chakravarthula2022hogel}.
Temporal multiplexing offers a compromise by displaying multiple phase-randomized holograms in rapid succession to visually average out speckle, but requires kilohertz-rate spatial light modulators (SLMs) \cite{choi2022time, Chu2025RealTime} and still suffers from diminishing returns as speckle reduction improves only \textit{sub-linearly} with frame count ($\sqrt{N}$ improvement from averaging $N$ frames).

In practice, this creates a fundamental tension: \textit{random phase is needed for uniformity and realism, but it inevitably produces speckle}.
Overcoming this trade-off and achieving random-phase, speckle-free holograms in a single frame remains one of the key unsolved challenges in holographic display technology.

In this work, we introduce \textit{Ellipsography}, a new class of holography that, for the first time, achieves \textit{speckle-free random-phase holographic reconstructions from a single frame}. Ellipsography fundamentally breaks the longstanding trade-off between phase diversity, essential for uniform light distribution, and the speckle noise it inevitably introduces. 
Our approach jointly modulates the phase and polarization of a vectorial optical field, and projects it into a scalar field with a pixel-wise analyzer. 
This vector-to-scalar projection enables deterministic control of coherent interference, yielding high-fidelity image formation while preserving phase randomness.
Leveraging joint modulation of phase and polarization, ellipsography alters the image formation process itself rather than relying on frame averaging, coherence reduction, or additional spatial or temporal multiplexing for speckle suppression.

We present a complete ellipsographic display framework consisting of a vectorial wave propagation model, an end-to-end hologram optimization pipeline, a display system architecture, and a calibration procedure. 
We validate our framework through extensive simulation and physical experiments, and demonstrate substantial improvements in image clarity, speckle suppression, depth fidelity, and focus cues over current state-of-the-art approaches. 
Overall, ellipsography establishes a new foundation for high-fidelity, single-shot holographic displays, bringing them measurably closer to the visual quality and usability required for next-generation near-eye and consumer-facing applications.

\begin{table}[t]
    \setlength{\tabcolsep}{0em}
    \centering
    \footnotesize
    \caption{
        Comparison of display technologies, where the per-frame performance of each criterion is indicated as high 
        \green, moderate \yellow, or low \red.
        }
    \vspace{-2mm}
    \begin{tabularx}{\linewidth}{m{0.24\linewidth}XXXXX}
    \toprule
    &
    
    {\footnotesize LCD/OLED Displays} &
    {\footnotesize Smooth Phase Holography} &
    {\footnotesize Random Phase Holography} &
    {\footnotesize \textbf{Ours}} &
    {\footnotesize \textbf{Ultimate Display}} \\
    \midrule
    \multicolumn{6}{l}{\textbf{Display Characteristics}}\\ 
    \midrule
    Image Quality & 
    
    \green & 
    \green &
    \red &
    \green& 
    \green \\
    Eyebox & 
    \green & 
    \red & 
    \green & 
    \green & 
    \green \\
    Focus Cues & 
    \red & 
    \yellow & 
    \green & 
    \green & 
    \green \\
    Depth Realism & 
    \red & 
    \red & 
    \yellow & 
    \green & 
    \green \\
    Noise Suppression & 
    \green & 
    \green & 
    \red & 
    \green &

    \green\\
    Compactness &
    \green & 
    \red & 
    \red & 
    \red &
    \green  \\
    \bottomrule
    \end{tabularx}
    \label{tab:display_comparison}
    \vspace{-5mm}
\end{table}

We summarize our contributions as follows:

\begin{itemize}
    \item We \textit{introduce} Ellipsography, a single-shot holography framework that achieves speckle-free reconstructions while preserving phase randomness, overcoming fundamental limitations in phase-only holographic display methods.
    \item We \textit{integrate} Jones formalism into holography framework for jointly modulating phase and polarization using phase-only SLMs and for vector-to-scalar field projection. 
    \item We \textit{analyze} the performance of our approach across key dimensions, including eyebox energy uniformity, focal cue accuracy, speckle contrast, and overall image quality.
    \item We \emph{design and build} a prototype display system that implements ellipsography using off-the-shelf optics and phase-only SLMs, and demonstrate substantial improvements in image quality over state-of-the-art approaches, without relying on heuristic phase encoding or relaxation schemes.
\end{itemize}

\paragraph{Overview of Limitations}
While our proposed method achieves unparalleled image quality, the current prototype relies on 4f relay system and additional optical components for polarization modulation, resulting in a non-compact setup. 
Nevertheless, we outline a path toward a more compact architecture in \Cref{sec:discussion}, recognizing that reducing form factor is essential for practical holographic displays, especially for near-eye augmented and virtual reality applications.
In our current implementation, joint phase and polarization modulation is achieved using a dual-SLM setup and a pixel-wise analyzer is emulated by optically filtering the polarization states at individual pixels.
While dual-SLM configurations have been explored previously to improve scalar holography \cite{kuo2023multisource,choi2021optimizing}, our approach is the first to demonstrate near-total speckle suppression from a single random-phase hologram frame, without requiring temporal or spatial averaging.
Looking ahead, we envision this work to lay the foundation for a new class of SLM and analyzer designs that support simultaneous, independent phase and polarization control at each pixel.
Such hardware advances would unlock the full potential of ellipsography, enabling compact, high-performance, perceptually compelling holographic displays.

\section{Related Work}

Maintaining phase randomness and image quality simultaneously is challenging but crucial for the success of holographic displays.
While the proposed ellipsography approach fundamentally differs from traditional scalar holography, we discuss relevant works on holography and speckle reduction in the following.

\paragraph{Random Phase Holograms}
Random phase at the image plane, which mimics scattering from a diffuse object, enables natural defocus blur \cite{yoo2021optimization} and uniform energy distribution within the eyebox \cite{chakravarthula2022pupil}.
However, phase randomness also introduces speckle noise due to interference from randomly varying optical path lengths. 
For 2D images, allowing the image plane phase to vary freely for iterative hologram optimization algorithms is shown to minimize speckle for specific viewing angles \cite{fienup1982phase, gerchberg1972practical}.
Gaze-contingent hologram optimization \cite{chakravarthula2021gaze} or introducing spatial ``don't care'' regions \cite{georgiou2008aspects,georgiou2023visual} can further improve image quality in the targeted regions of interest.
In contrast, for 3D imagery, the limited space-bandwidth on the SLM make it impossible to uniformly suppress speckle across the entire viewing volume \cite{lohmann1996space}. 
While optimizing holograms with out-of-focus regions left unconstrained improves the clarity of in-focus regions, it introduces additional speckle artifacts in the defocused areas \cite{choi2021neural,kuo2020high,chakravarthula2020learned}.
Recent studies demonstrate 3D holographic imagery with reduced artifacts via complex optimization schemes \cite{chakravarthula2022hogel,choi2022time,kavakli2023realistic,schiffers2023stochastic}.
However, achieving natural defocus blur with low speckle throughout the entire 3D volume cannot be accomplished through algorithmic optimization alone, and demands additional despeckling techniques \cite{kuo2023multisource,schiffers2025holochrome}.
In contrast, \textit{ellipsography overcomes the above limitations and achieves speckle-free images from random phase holograms in a single frame}. 

\paragraph{Smooth Phase Holograms}
Smooth phase holograms minimize speckle by enforcing a nearly constant phase at the image plane, thereby avoiding random interference between neighboring points \cite{maimone2017holographic,shi2021towards}.
Achieving a specific phase pattern at the image plane typically requires complex modulation at the SLM (hologram plane). 
However, prior research has introduced solutions like the double phase amplitude coding method, which encodes complex amplitude into two phase-only values to effectively suppress speckle without adding system complexity \cite{hsueh1978computer}.
Additionally, studies have shown that a relatively lower phase variation at the image plane can be achieved through gradient descent optimization \cite{chakravarthula2019wirtinger,peng2020neural}.
While smooth-phase holograms can generate high-quality, speckle-free 2D images, they are limited by unnatural and restricted defocus blur, often resulting in undesirable ringing artifacts \cite{padmanaban2019holographic,shi2021towards,chakravarthula2022hogel}.
Moreover, these holograms tend to concentrate light into a small region of the eyebox, making them highly sensitive to eye movements and optical imperfections \cite{chakravarthula2022pupil,kuo2023multisource,Chu2025RealTime}.
Recent studies also indicate that smooth-phase holograms fail to effectively trigger eye's natural accommodation responses \cite{kim2022accommodative}, significantly limiting their practicality for immersive display applications.
\paragraph{Speckle Suppression via Spatial, Temporal or Spectral Multiplexing}
Most popular approaches to mitigate speckle noise involve multiplexing many holograms, either spatially or temporally, each with a unique speckle pattern \cite{bianco2018strategies}. 
While speckle contrast depends on the temporal coherence of light, the image sharpness is guided by the spatial coherence \cite{deng2017coherence}.
Despeckling can be achieved via partially coherent light sources such as light emitting diodes (LEDs) \cite{kozacki2022led} and superluminescent LEDs (SLEDs) \cite{markley2023simultaneous,peng2021speckle,primerov2019p}, and
wavelength or angle diversity to illumination \cite{george1974space,wang2013speckle,tran2016speckle,nomura2008image,deng2017coherence,kuo2023multisource,kavakli2023multi,schiffers2025holochrome}. 
However, these approaches result in loss of spatial detail and depth of field \cite{lee2020light,lee2020speckle,lee2022high}. 
State-of-the-art temporal multiplexing methods achieve speckle suppression by averaging multiple frames using high-speed modulators such as digital micromirror devices (DMDs) \cite{curtis2021dcgh,lee2020light}, ferro-electronic liquid crystal on silicon (FLCoS) \cite{lee2022high}, or micro-electromechanical systems (MEMS) based SLMs \cite{choi2022time}. 
However, these devices often have limited bit depth, leading to quantization errors and reduced diffraction efficiency, degrading image quality \cite{ketchum2021diffraction,choi2022time,ding2024digital}.
In addition, \cite{chan2025holospeed} shows this strategy manifests motion blur and stroboscopic artifacts in the context of dynamic scenes. 
Moreover, speckle suppression scales only sub-linearly with frame count \cite{goodman2007speckle}, and current algorithms lack support for content updates between subframes, which is needed for temporal multiplexing at kilohertz refresh rates.
Reducing the number of required frames not only lowers compute cost but also frees temporal bandwidth.

\paragraph{Speckle Suppression via Polarization Multiplexing}
Recent approach by \citet{nam2023depolarized} uses polarization diversity to average two hologram frames of orthogonal polarization for speckle suppression, achieving a factor of at most ${\sqrt{2}}$ speckle reduction---far below what is possible with temporal multiplexing and still insufficient for artifact-free images. As opposed to using polarization as an extra degree of freedom, \textit{our work uses polarization to structure the interference itself, achieving speckle reduction from a single random-phase hologram, eliminating the need for spatial or temporal averaging}.
\paragraph{Camera-feedback for Noise Suppression}
To address imperfections arising from speckle and other hardware nonidealities, previous works attempted to learn a differentiable model of the hardware optical system by using experimentally captured calibration data, which can then be used as an ``offline'' calibration model while computing holograms \cite{chakravarthula2020learned,kavakli2022learned}.
Another alternative is to use an ``online'' camera-in-the-loop (CITL) calibration where each SLM pattern is fine-tuned to an individual image based on active camera feedback \cite{peng2020neural}.
While this method does not generalize to new content, CITL approach effectively demonstrates the performance limits of a given experimental setup \cite{kuo2023multisource}.
In this work, we \textit{surpass the performance limits of CITL-based approaches, without relying on active camera feedback}.

\section{Primer}
\label{sec:Primer}

\subsection{Background on Jones Calculus}
\label{subsec:jones-calculus}
Jones calculus is a widely used framework for describing the polarization state of light and its transformation through optical elements.
We use Jones calculus to model vectorial wave propagation and polarization-dependent image formation in ellipsography, and briefly review the relevant concepts below. 
For a more comprehensive discussion of polarization optics, we refer readers to \citet{collett2005field} and \citet{yakovlev2015polarization}.

In the remainder of the paper, boldface symbols (\eg $\mathbf{E}$) denote Jones vectors
representing the local polarization state of the optical field, while arrowed symbols (\eg $\Vec{E}$) refer to physical spatial field vectors contributing to interference.

The polarization state of fully polarized light at a given location can be represented by a $2\times1$ Jones vector:

\begin{equation}
\mathbf{E} = \begin{pmatrix}
    E_{x} \\
    E_{y}
\end{pmatrix},
\label{eq:jones-vector}
\end{equation}

where $E_x = A_x e^{j\phi_{x}} $ and $E_y = A_y e^{j\phi_{y}} $ are the complex amplitudes of the electric field components oscillating along the $x$- and $y$- axes, respectively. Any fully polarized optical field can be expressed as a linear combination of horizontal and vertical basis vectors:

\begin{align}
\mathbf{J}_{x}&=\begin{pmatrix}
1 \\
0
\end{pmatrix}
&
\mathbf{J}_{y}&=\begin{pmatrix}
0\\
1
\end{pmatrix}.
\end{align}

Optical elements that alter polarization, such as polarizers or wave plates, are modeled using $2 \times 2$ Jones matrices:

\begin{equation}
\mathbf{J} = \begin{pmatrix}
J_{11} & J_{12} \\
J_{21}& J_{22}
\end{pmatrix}.
\end{equation}

The resulting output polarization state is computed by multiplying the Jones matrix with the input Jones vector:

\begin{equation}
    \mathbf{E}_\text{out} = \mathbf{J} \cdot \mathbf{E}_\text{in}.
\end{equation}

In the case of spatially varying vector fields, Jones calculus is applied locally on a per-pixel basis to account for spatial variations in the polarization state across the wavefront. 
We build on this local vectorial formulation for modeling the
polarization-dependent wave interference and measurement models introduced in the
following sections.

\subsection{Speckle in Traditional Holography}
\label{subsec:speckle-in-holograms}
Speckle noise, arising from the random interference of coherent light waves, is a major source of image degradation and a fundamental limitation in holographic displays. Unlike algorithmic reconstruction errors and external noise, speckle is an inherent characteristic of coherent optical systems and cannot be eliminated without modifying the underlying interference structure itself. 

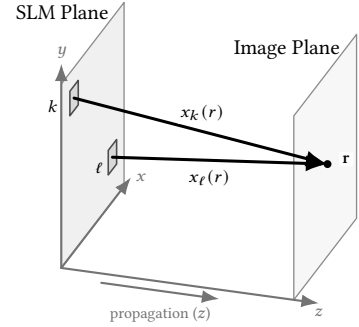
\begin{figure}[H]
\centering
\tdplotsetmaincoords{65}{18}
\begin{tikzpicture}[tdplot_main_coords, >=Latex, line cap=round, line join=round]

\tikzset{
  planeA/.style={draw=black!45, fill=black!6, line width=0.5pt},
  planeB/.style={draw=black!45, fill=black!3, line width=0.5pt},
  pix/.style={draw=black!70, fill=black!15, line width=0.6pt},
  rayMain/.style={->, line width=1.15pt},
  axis/.style={->, black!55, line width=0.8pt},
  lbl/.style={
    font=\scriptsize,
    fill=white, fill opacity=0.95, text opacity=1,
    inner sep=1.2pt, outer sep=0.8pt
  },
  plbl/.style={ 
    font=\scriptsize,
    fill=none,
    inner sep=0.5pt, outer sep=0.5pt
  },
waveCue/.style={draw=black!55, line width=0.8pt, opacity=0.65},
waveVol/.style={draw=black!55, line width=0.55pt, opacity=0.18},
}

\def\D{3.25}   
\def\L{2.70}   
\def\ps{0.18}  
\def\kshift{0.00} 
\def\xm{0.65} 

\path[planeA] (0,0,0) -- (0,\L,0) -- (0,\L,\L) -- (0,0,\L) -- cycle;
\path[planeB] (\D,0,0) -- (\D,\L,0) -- (\D,\L,\L) -- (\D,0,\L) -- cycle;

\node[lbl, font=\small, fill=none, xshift=-22pt,yshift=7pt] at (0,2.55,2.35) { SLM Plane};
\node[lbl, font=\small, fill=none, xshift=-25pt,yshift=5pt] at (\D,2.55,2.35) { Image Plane};

\coordinate (sk) at (0,{0.55+\kshift},2.25);
\coordinate (sl) at (0,2.20,0.65);

\coordinate (r)  at (\D,1.45,1.35);

\path[pix] (0,{0.55+\kshift-\ps},{2.25-\ps}) -- (0,{0.55+\kshift+\ps},{2.25-\ps})
          -- (0,{0.55+\kshift+\ps},{2.25+\ps}) -- (0,{0.55+\kshift-\ps},{2.25+\ps}) -- cycle;
\path[pix] (0,{2.20-\ps},{0.65-\ps}) -- (0,{2.20+\ps},{0.65-\ps})
          -- (0,{2.20+\ps},{0.65+\ps}) -- (0,{2.20-\ps},{0.65+\ps}) -- cycle;

\node[plbl, xshift=-8pt, yshift=-3pt] at (sk) {\(k\)};
\node[plbl, xshift=-5pt, yshift=-4pt]  at (sl) {\(\ell\)};

\fill (r) circle (1.5pt);
\node[lbl, xshift=7pt, yshift=2pt] at (r) {\(\mathbf{r}\)};

\draw[rayMain] (sk) -- (r);
\draw[rayMain] (sl) -- (r);

\path (sk) -- (r) coordinate[pos=0.34] (mk);
\path (sl) -- (r) coordinate[pos=0.28] (ml);
\node[lbl, xshift=16pt, yshift=-2pt, anchor=south] at (mk) {\(x_k(r)\)};
\node[lbl, xshift=14pt, yshift=-2pt, anchor=north] at (ml) {\(x_\ell(r)\)};

\draw[axis] (0,0,0) -- (3.6,0,0) node[lbl, fill=none, anchor=north] {$z$};

\draw[axis] (0,0,0) -- (0,3.0,0) node[lbl, fill=none, anchor=west] {$x$};

\draw[axis] (0,0,0) -- (0,0,3.0) node[lbl, fill=none, anchor=south] {$y$};

\draw[->, black!55, line width=0.8pt] (0.65,-0.28,0) -- (2.30,-0.28,0)
  node[lbl, fill=none, midway, below, yshift=-4pt] {propagation (\(z\))};

\end{tikzpicture}
\caption{
Contributions of SLM pixels toward image formation in a holographic display.
}
\label{fig:klr_geometry}
\end{figure}
In holographic displays, the finite aperture size and limited diffraction angle of SLMs cause individual pixels (or voxels) to be projected over a broad area, instead of forming a sharp point.
This overlap causes diffracted fields from multiple SLM pixels to contribute to the same image plane location, where they interfere. 
As illustrated in \Cref{fig:klr_geometry}, these contributions may arise from distinct SLM pixel locations, but their propagated fields contribute to and overlap at the same image plane location.

Consider an image-plane location $r$ that receives two coherent contributions from the SLM.
Let the corresponding complex scalar fields on the image plane at $r$ be $\Vec{E_1}(r)=A_1 e^{i\phi_1}$ and $\Vec{E_2}(r)=A_2 e^{i\phi_2}$.
Their superposition produces an intensity
\begin{equation}
    I = |\Vec{E_1} + \Vec{E_2}|^2 = A_1^2 + A_2^2 + 2A_1A_2\cos(\phi_1 - \phi_2),
    \label{eq:pixel-interference}
\end{equation}
where the third term is an interference cross-term which depends on the relative phase of the two interfering fields.
This simple two-contribution case generalizes to a dense summation over all SLM pixel contributions to the image plane in real systems. 

\label{sec:prior}

\begin{figure*}[t]

  \includegraphics[width=0.95\textwidth]{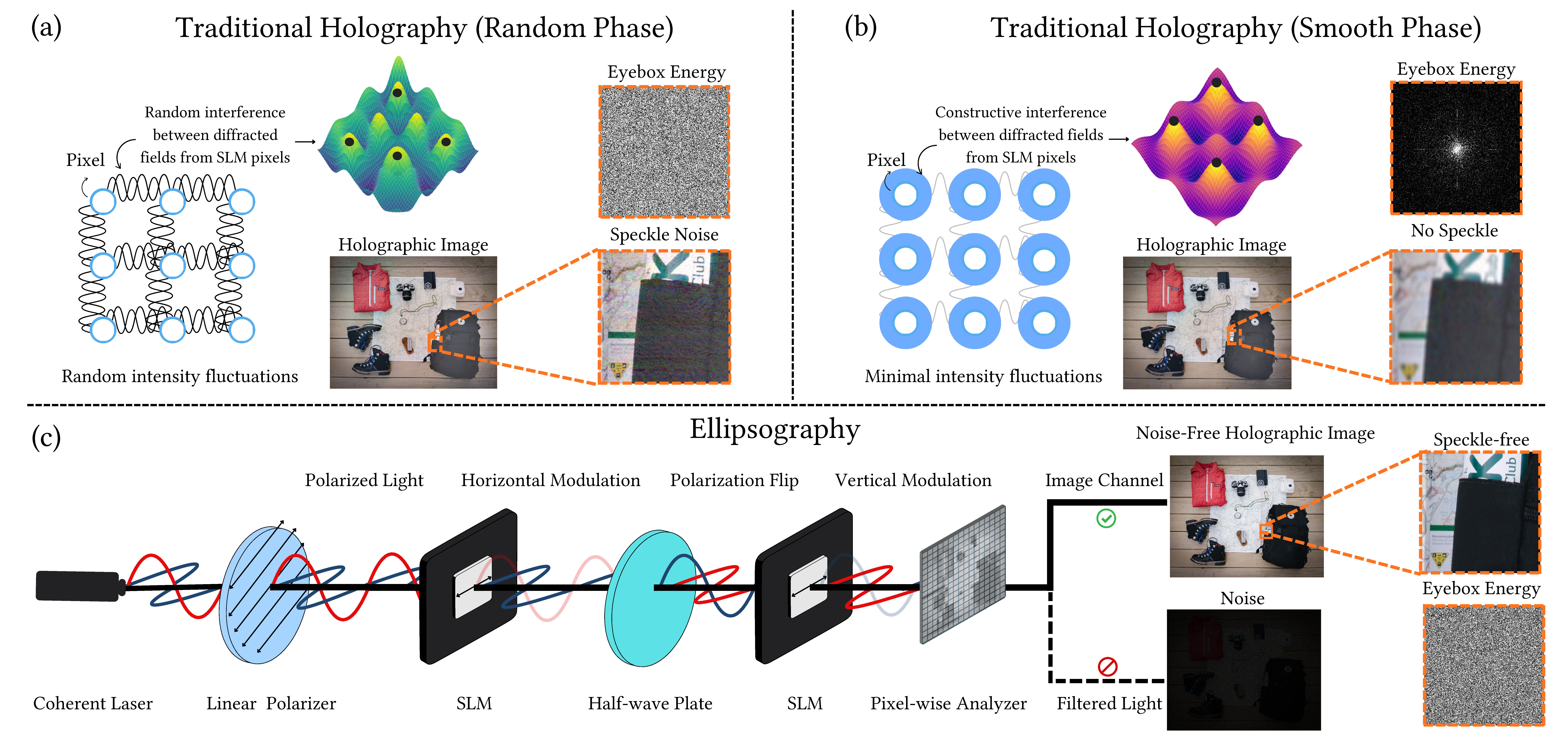}
  \vspace{-4mm}
  \caption{\textit{Comparison of Random Phase, Smooth Phase, and Ellipsography-based Holograms}.
  Random phase holograms achieve uniform energy distribution but suffer from speckle. Smooth phase holograms reduce speckle, but at the cost of spatial resolution, contrast and reduced eyebox. 
  Ellipsography achieves the best of both. By jointly modulating phase and polarization of light, it preserves phase randomness for uniformity while eliminating speckle. Any residual noise is routed into an orthogonal polarization channel and optically filtered, yielding high-fidelity speckle-free reconstructions. 
  } 
  \label{fig:motivation_figure}

\end{figure*}

The complex field $E_{\text{trad}}(r)$ at the image location $r$, as computed in traditional holographic methods \cite{maimone2017holographic}, is:
\begin{equation}
E_{\text{trad}}(r) = \sum_{k=1}^N x_k(r),
\label{eq:scalar_sum}
\end{equation}
where $\{x_k(r)\}\in\mathbb C$ denotes the field contributions from the $k-$th pixel and $N$ represents the total number of contributing SLM pixels. 
The corresponding image intensity is:
\begin{equation}
I_{\text{trad}}(r)
= \left| \sum_{k=1}^N x_k(r) \right|^2
= \sum_k |x_k(r)|^2 + \sum_{k\neq \ell} x_k(r)x_\ell(r)^{*},
\label{eq:scalar_intensity_expand}
\end{equation}
where the second summation includes all distinct pairwise interference terms, with $\{\}^* $ denoting the complex conjugate. 
In practice, the phase differences $(\phi_k(r)-\phi_\ell(r))$ vary rapidly and non-uniformly across space causing the cross-terms to fluctuate randomly, producing the characteristic granular speckle noise that significantly degrade image clarity (see \Cref{fig:motivation_figure}).

A fundamental limitation of scalar holography is that the same variables
$\{x_k(r)\}$ are responsible for both forming the desired image and producing the interference terms
in Eq.~\eqref{eq:scalar_intensity_expand} that produces speckle. 
This prevents direct control over speckle, and any attempt to suppress speckle also acts on the image-forming fields themselves, leading to a tradeoff between image fidelity and speckle suppression.
As a result, existing speckle suppression strategies primarily focus on reducing the variance of the interference term in \Cref{eq:scalar_intensity_expand}, rather than completely eliminating it. 

The most common approach to suppress this is to enforce phase smoothness across the hologram by constraining phase to vary slowly across the SLM, so that pixels with strongly overlapping  diffracted field exhibit similar phases ($\phi_k \approx \phi_\ell$). This biases the interference term towards \textit{constructive interference}, thereby suppressing speckle; see \Cref{fig:motivation_figure}. However, this also leads to reduced image contrast and effectively focuses energy into a narrow eyebox, typically less than a millimeter wide \cite{maimone2017holographic,chakravarthula2022pupil}.
Introducing random phase variations, on the other hand, can expand the eyebox by distributing energy more uniformly 
\cite{chakravarthula2022pupil}, but this increases the randomness in the cross-terms and inevitably increases speckle noise \cite{yoo2021optimization}.
Thus, \textit{traditional scalar holography inherently faces a tradeoff between speckle suppression and display usability}.

Alternative strategies aim to suppress speckle \textit{statistically} by averaging multiple uncorrelated holograms. 
These include temporal multiplexing \cite{choi2022time,schiffers2023stochastic}, and using polychromatic \cite{kuo2023multisource,schiffers2025holochrome} or broadband light sources \cite{peng2021speckle}.
These methods reduce speckle noise proportionally to $1/\sqrt{N}$, where $N$ is the number of statistically independent frames or wavelength channels being averaged \cite{goodman2007speckle}.
This reduction follows from the central limit theorem,
where as the number of uncorrelated cross terms increases, their sum averages out to zero.
The limiting case, with infinitely many cross terms, leaves only the signal terms:
$I(r) \rightarrow \sum_k |x_k(r)|^2$.
However, such averaging-based techniques suppress speckle asymptotically, requiring an impractically large number of independent frames, leading to high computational and temporal overhead. 

\textit{Despite several decades of research, speckle suppression in scalar holography remains a core bottleneck}.

\subsection{Polarimetric Vectorial Interference}
To overcome the core limitations of traditional speckle reduction methods, we propose a fundamentally different strategy: interference control via structured polarization fields. 

Note that this is different from polarization multiplexing, which passively averages orthogonal polarization channels to reduce speckle variance. Instead, we engineer polarization for suppressing speckle at the level of individual pixel contributions.

Consider two coherent contributions at the image-plane location (originating from two SLM pixels whose diffracted fields overlap at that location; see \Cref{fig:klr_geometry}):

\begin{equation}
\begin{aligned}
\Vec{E_{1}} &= A_{1} e^{j\phi_{1}} \hat{p_{1}}, \\
\Vec{E_{2}} &= A_{2} e^{j\phi_{2}} \hat{p_{2}},
\end{aligned}
\end{equation}
where $\hat{p_{1}}, \hat{p_{2}}$ are unit polarization vectors describing the respective fields.
For simplicity, consider linearly polarized fields whose unit polarization vectors can be expressed in the orthonormal basis $\hat{x}, \hat{y}$ as
\begin{equation}
\begin{aligned}
\hat{p_{1}} &= \cos(\theta_{1})\,\hat{x} + \sin(\theta_{1})\,\hat{y},\\
\hat{p_{2}} &= \cos(\theta_{2})\,\hat{x} + \sin(\theta_{2})\,\hat{y},
\end{aligned}
 \end{equation}
where $\theta_{1},\theta_{2}$ denote the polarization angles relative to $\hat{x}$. 
Extension to general elliptical states is straightforward by adding phase retardations between the basis components.

The resulting complex electric field from their superposition is:
\begin{equation}
    \Vec{E} = \Vec{E_1} + \Vec{E_2},
\end{equation}
and the corresponding intensity is:
\begin{equation}
\begin{aligned}
I &= |\Vec{E}|^2 = \Vec{E}\cdot \Vec{E}^{*}, \\
&= (A_{1} e^{j\phi_{1}} \hat{p_{1}} + A_{2} e^{j\phi_{2}} \hat{p_{2}}) \cdot (A_{1} e^{-j\phi_{1}} \hat{p_{1}} + A_{2} e^{-j\phi_{2}} \hat{p_{2}}), \\
&= |A_{1}|^{2}+|A_{2}|^2 + A_1A_2 [e^{j(\phi_{1} - \phi_{2})} + e^{-j(\phi_{1} - \phi_{2})} ] (\hat{p_{1}} \cdot \hat{p_{2}}),
\end{aligned}
\label{eq:polar-interfere}
\end{equation}
where $\{\cdot\}$ denotes the vector dot product. 
The dot product of the unit polarization vectors can be expressed as:
\begin{equation}
\begin{aligned}
    \hat{p_1}\cdot \hat{p_2} &= \cos(\theta_1) \cos(\theta_2) + \sin(\theta_1) \sin(\theta_2), \\
    &= \cos(\theta_1 - \theta_2).
\end{aligned}
\end{equation}
Simplifying the interference term in \Cref{eq:polar-interfere} using Euler's identity:
\begin{equation}
    \cos(\theta) = \text{Re}\{ e^{j\theta} \} = \frac{e^{j\theta}+e^{-j\theta}}{2},
\end{equation}
gives the intensity of the superposed field as:
\begin{equation}
   \boxed{ I = |A_{1}|^{2}+|A_{2}|^2 + 2A_1A_2 \cos(\phi_1 - \phi_2)\cos(\theta_1 - \theta_2).}
   \label{eq:polar-interfere-final}
\end{equation}

This expression reveals a \textbf{key insight}: unlike traditional scalar holography, \textit{vectorial interference is controlled by both phase difference $(\phi_1 - \phi_2)$ and polarization alignment $(\theta_1 - \theta_2)$}. Particularly, when the polarizations are orthogonal ($\theta_1-\theta_2=\pi/2$), the interference term vanishes regardless of the phase difference. This introduces a new mechanism for interference cancellation that is not accessible in traditional scalar holography. 

However, real-world holographic images are formed by the superposition of fields from many SLM pixels.
The full field at the image plane location $r$ is given by:
\begin{equation}
\mathbf{E}(r) = \sum_{k=1}^{N} \mathbf{u}_k(r), 
\quad \text{with} \quad
\mathbf{u}_k(r) =
\begin{bmatrix}
u_{k,H}(r) \\
u_{k,V}(r)
\end{bmatrix}
\in \mathbb{C}^2,
\label{eq:vector_field_sum}
\end{equation}
where $\mathbf u_k(r)$ represents the contribution from the $k$-th SLM pixel, with the subscripts $H$ and $V$ denoting the horizontal and vertical polarization components, respectively.

When this vector field is measured by a polarization-insensitive detector (\eg a conventional camera or the human eye), the measured intensity image is simply the sum of squared magnitudes of each polarization component:
\begin{equation}
\begin{aligned}
I_{\mathrm{eye}}(r)
&= |E_H(r)|^2 + |E_V(r)|^2, \\
&= \left|\sum_k u_{k,H}(r)\right|^2
  + \left|\sum_k u_{k,V}(r)\right|^2.
\end{aligned}
\label{eq:hv_intensity_expand}
\end{equation}
Expanding this reveals that each component has its own set of cross terms from distinct pixels:
\begin{equation}
\begin{aligned}
|E_H(r)|^2
&= \sum_k |u_{k,H}(r)|^2
 + \sum_{k\neq\ell} u_{k,H}(r)\,u_{\ell,H}(r)^*,\\
|E_V(r)|^2
&= \sum_k |u_{k,V}(r)|^2
 + \sum_{k\neq\ell} u_{k,V}(r)\,u_{\ell,V}(r)^*.
\end{aligned}
\label{eq:hv_cross_terms}
\end{equation}
Although orthogonal polarization components do not interfere with each other, each polarization channel still contains its own internal speckle.
As a result, polarization diversity alone (as used in polarization multiplexing methods \cite{nam2023depolarized}) offers only a statistical speckle reduction at best, similar to temporal multiplexing.

Our core contribution lies in recognizing that \textit{polarization must not merely partition coherent contributions into orthogonal channels} for statistical averaging, but must be \textit{explicitly coupled to the image formation process to deterministically shape interference}. 
We accomplish this via a pixel-wise polarization analyzer which performs a controlled projection of the vectorial interference illustrated by \Cref{eq:vector_field_sum}. This vector-to-scalar projection defines how the vector field contributes to the measured scalar image, which we formalize next. 

\subsection{Overview of Ellipsography}

Ellipsography achieves \textit{vectorial interference shaping} via controlled vector-to-scalar projection,
enabling direct control over interference structure in the final image, and, thereby, single-shot speckle suppression.

We model the polarization analyzer at position $r$ using a unit Jones vector:
\begin{equation}
\mathbf{J}_r =
\begin{bmatrix}
\cos\theta(r) \\
\sin\theta(r)
\end{bmatrix},
\label{eq:analyzer-jones}
\end{equation}
which defines the polarization axis along which the electric field is projected and measured. 
This analyzer may be fixed (global) or spatially varying (per-pixel).
The projected complex field is given by:
\begin{equation}
s(r) = \mathbf J_r^{\dagger} \mathbf E(r),
\qquad
I(r) = |s(r)|^2,
\label{eq:analyzer_measurement}
\end{equation}
where $\mathbf E(r) \in \mathbb{C}^2$ is the Jones vector representing the total image-plane field, $s(r)$ denotes the projected scalar field, and $(\cdot)^{\dagger}$ denotes the Hermitian transpose. 
Note that while $\mathbf J_r$ is real-valued and $\mathbf J_r^{\dagger}$ reduces to a transpose in practice, we retain the Hermitian form for generality.
Substituting the vector field decomposition from \Cref{eq:vector_field_sum}, each individual contributor projects as:
\begin{equation}
s(r)=\sum_k a_k(r),
\qquad \text{where} \quad
a_k(r)=\mathbf J_r^{\dagger} \mathbf u_k(r).
\end{equation}
The measured intensity then becomes:
\begin{equation}
I(r)=\sum_k |a_k(r)|^2 + \sum_{k\neq \ell} a_k(r)a_\ell(r)^* .
\label{eq:projected_intensity_expand}
\end{equation}
Note that this has the same form as traditional scalar holography, but with the key difference that each $a_k$ is a \textit{projection of a vector field}, not a scalar.

To make the role of polarization explicit, we define the \textit{pair-wise polarization coherence matrix} between overlapping pixels:
\begin{equation}
\mathbf C_{k\ell}(r)=\mathbf u_k(r)\mathbf u_\ell(r)^{\dagger}
\in \mathbb C^{2\times 2}.
\label{eq:coherence_matrix}
\end{equation}
This formulation extends the polarization coherence matrix, as proposed by \citet{wolf1959coherence, wolf2003unified}, from local polarization statistics to nonlocal polarization correlations between spatially distinct contributors.

Each projected cross term can then be written as:
\begin{equation}
\boxed{
a_k(r)a_\ell(r)^*
= \mathbf J_r^{\dagger}\,\mathbf C_{k\ell}(r)\,\mathbf J_r .
}
\label{eq:projected_cross_term}
\end{equation}

This quadratic form captures the core mechanism of ellipsography: \textit{the contribution of each interference term to the final image intensity is actively shaped by the analyzer, which interacts with the polarization coherence between any pair of pixels}.

By introducing a spatially structured polarization projection operator $\mathbf{J}_r$ that defines how vectorial coherence is projected into the scalar field, ellipsography allows the interference terms $\mathbf J_r^{\dagger}\mathbf C_{k\ell}(r)\mathbf J_r$ at each image-plane location to be directly shaped.
By jointly designing the vectorial contributions $\{\mathbf{u}_k(r)\}$ (via SLM phase modulation) and the analyzer $\mathbf{J}_r$ (via polarization modulation), ellipsography enables structured interference control and deterministic speckle suppression in a single shot.

\section{Ellipsography}
\label{sec:method}

In this section, we introduce the Ellipsography framework which jointly modulates the phase and polarization of a vectorial wave field and projects it into a scalar wave field using a pixel-wise analyzer, generating high-fidelity speckle-free holographic images. 

We begin by describing the optical setup for achieving joint phase and polarization modulation using phase-only SLMs, then derive the ellipsographic image formation model and present an optimization framework for computing holograms.

\subsection{Achieving Joint Phase and Polarization Modulation}
\label{sec:motivation}

Most commercial liquid crystal (LC) SLMs are designed for \textit{phase-only modulation} of light with a single, fixed linear polarization state, typically aligned with the long axis of the LC molecules. 
See Supplementary Material for additional discussion.
While these devices are widely adopted in display systems, they are fundamentally constrained to modulation of scalar wavefronts, limiting their ability to manipulate vectorial wave fields. 

To overcome this limitation, we employ a dual-SLM optical architecture to enable independent phase modulation of orthogonal polarization components. As illustrated in \Cref{fig:setup}, we relay two SLMs optically and insert a half-wave plate (HWP) between them to rotate the polarization by $90^\circ$. 
This ensures that each SLM modulates a different linear polarization component: horizontal on the first SLM and vertical on the second, or vice versa.
A HWP, composed of a birefringent crystal with ordinary and extraordinary axes, introduces a phase shift between these components.
This configuration enables the joint phase and polarization modulation essential to Ellipsography. 
By independently modulating orthogonal polarization components, we can synthesize arbitrary \textit{per-pixel polarization} states to precisely control vectorial interference.

Finally, a per-pixel analyzer projects the local vector field at each pixel onto a selected polarization basis.
By encoding a distinct polarization state at each pixel, the resulting interference pattern is engineered to deterministically suppress speckle.

\subsection{Vectorial Image Formation Model}

Building on our dual-SLM architecture, we now present the ellipsographic image formation model. 
An illustration of this process is provided in \Cref{fig:motivation_figure} and the hardware configuration in \Cref{fig:setup}.

As illustrated in \Cref{fig:setup}, we use a laser source linearly polarized at $45^{\circ}$ as input, whose normalized Jones vector is expressed as:
\begin{equation}
\mathbf{E}_\mathrm{in}= \frac{1}{\sqrt{2}}
\begin{pmatrix}
1 \\
1
\end{pmatrix},
\end{equation}
with the horizontal $E_x$ and vertical $E_y$ components having equal amplitudes. 
The first SLM modulates the horizontal polarization component of the input beam.
Subsequently, a half-wave plate oriented at $45^{\circ}$ can be represented as:
\begin{equation}
\mathbf{J}_{\mathrm{HWP}}=
\begin{pmatrix}
0 & 1\\
1 & 0
\end{pmatrix},
\end{equation}
which swaps the horizontal and vertical polarization components between the two SLMs, while preserving their amplitudes, ensuring optimal modulation efficiency.

Note that, after passing through the HWP, the horizontal polarization component previously modulated by the first SLM is converted to vertical polarization, while the previously unmodulated vertical component is transformed to horizontal component; refer to the illustration in \Cref{fig:motivation_figure} (c).
The second SLM then modulates this newly horizontal (previously unmodulated vertical) polarization component.

The output field after modulation by the two (relayed) SLMs can be expressed as the superposition of orthogonal components:
\begin{equation}
\mathbf{E}_\mathrm{SLM}= \mathbf{J}_\mathrm{HWP} \cdot \Biggl( \begin{pmatrix} 
E_x
\\
0
\end{pmatrix} \cdot e^{j\Phi_{1}} + \begin{pmatrix} 
0
\\
E_y
\end{pmatrix} \cdot e^{j\Phi_{2}}\Biggr),
\end{equation}
where $\Phi_{1}$, $\Phi_{2}$ denote the respective \textit{per-pixel phases} on the SLMs. 
For collimated uniform laser illumination, the complex amplitude of electric field components can be set to unity for all practical purposes, without loss of generality. 
The output field with a HWP at $45^{\circ}$ is then given by
\begin{equation}
\mathbf{E}_\mathrm{SLM}= \mathbf{J}_\mathrm{HWP} \cdot \Biggl( \begin{pmatrix} 
1
\\
0
\end{pmatrix} \cdot e^{j\Phi_{1}} + \begin{pmatrix} 
0
\\
1
\end{pmatrix} \cdot e^{j\Phi_{2}} \Biggr) =\begin{pmatrix}
e^{j\Phi_{2}}\\
e^{j\Phi_{1}}
\end{pmatrix}.
\end{equation}

We use the angular spectrum method (ASM) \cite{goodman2005introduction} to model the free space light propagation, with modifications to handle polarization effects.
Specifically, we leverage the principle that each polarized component of light, defined by its amplitude and phase, propagates independently in free space.
The free space propagation of a complex electric field $E = Ae^{j\Phi}$ over a propagation distance $z$ can be modeled using ASM as:
\begin{equation}\label{eq:fprop}
\begin{split}
f_\text{ASM}(E, z) &=\iint \mathcal{F}(E) \mathbb{H}(f_{x}, f_{y},\lambda,z) e^{j 2\pi(f_{x}x  + f_{y} y )} d{f_{x}}d{f_{y}},\\
\mathbb{H}(f_{x}, f_{y},\lambda,z) &= \begin{cases}
&e^{j{2\pi}z \sqrt{\frac{1}{\lambda^2} - f_{x}^2 - f_{y}^2}}, \qquad \text{if} \quad \sqrt{f_{x}^2 + f_{y}^2} <\frac{1}{\lambda},  \\
&0,  \qquad \qquad \qquad \qquad \;\ \text{otherwise.}
\end{cases}
\end{split}
\end{equation}
where $f_\text{ASM}(\cdot)$ is the field propagation operator, $\mathcal{F}\{ \cdot \}$ denotes the Fourier transform, $\mathbb{H}$ represents the optical transfer function, $\lambda$ denotes the wavelength of light, and $f_{x}$ and $f_{y}$ are the corresponding spatial frequencies.
To model wave propagation in ellipsography, we extend the standard ASM to handle vectorial electric fields described using the Jones formalism \cite{shao2024full}.
For a fully polarized electric field $\mathbf{E} = (E_x, E_y)^T$, the polarimetric free space propagation is applied independently to each orthogonal component:
\begin{equation}
    f_\text{polar}(\mathbf{E},z) = 
\begin{pmatrix} 
f_\text{ASM}(E_x,z)
\\
f_\text{ASM}(E_y,z)
\end{pmatrix}.
\label{eq:asm_prop}
\end{equation}
In our dual-SLM configuration, the two phase patterns $\Phi_1$ and $\Phi_2$ independently modulate the horizontal and vertical polarization components at each pixel.
Therefore, the resulting output field at the image plane is
\begin{equation}
  \mathbf{E}_\text{out} = f_\text{polar}(\mathbf{E}_\text{SLM},z) = 
\begin{pmatrix} 
f_\text{ASM}(e^{j\Phi_2},z)
\\
f_\text{ASM}(e^{j\Phi_1},z)
\end{pmatrix} = 
\begin{pmatrix}
    A_x e^{j\Psi_x}
    \\
    A_y e^{j\Psi_y}
\end{pmatrix},
\label{eq:img-field}
\end{equation}
where $(A_x,\Psi_x)$ and $(A_y,\Psi_y)$ denote the per-pixel amplitude and phase of the horizontal and vertical polarization components, respectively, at the image plane. 
The resulting local linear polarization orientation of the propagated vector field at the image plane is given by:
\begin{equation}
\vartheta =
\frac{1}{2}
\arctan\!\left(
\frac{2\cos(\Psi_x-\Psi_y)}
{\frac{A_x}{A_y}-\frac{A_y}{A_x}}
\right).
\label{eq:img-pol-angle}
\end{equation}
As seen in \Cref{eq:img-pol-angle}, the orientation angle of the local polarization ellipse at each image pixel is fully determined by the per-pixel amplitude ratio $A{x}/A{y}$ and the relative phase $\Psi_{x}-\Psi_{y}$ between the orthogonal components. 
By optimizing the phase patterns $\Phi_1$ and $\Phi_2$ on the two SLMs, we control these parameters and enable independent modulation of both the holographic phase and polarization state.

Finally, the synthesized wavefronts are passed through a \textit{per-pixel analyzer} as shown in \Cref{fig:motivation_figure}. 
Using the analyzer Jones vector $J_r$ introduced in
Equation~(\ref{eq:analyzer-jones}), the projected scalar image field is
\begin{equation}
s_{\mathrm{img}}(r)=J_r^\dagger E_{\mathrm{out}}(r), \qquad
I(r)=|s_{\mathrm{img}}(r)|^2.
\label{eq:final-img-field}
\end{equation}
This filtering step transmits the desired local projected polarization component defined by $J_r$ and suppresses any residual interference encoded into the orthogonal linear component, yielding a clean, speckle-free reconstruction. This defines the upper bound of ellipsographic interference control. In practice, a global analyzer provides a deployable approximation, while per-pixel analyzer behavior is emulated in our prototype for analysis. A detailed mathematical treatment of the polarization transformations in our display system is provided in the Supplementary Material. 

\subsection{Analyzer Placement as a Generalized Interference Operator}
\label{sec:analyzer-placement}

The projection formalism in Equations~(18)--(23) is not restricted to a
particular physical plane. Let $p$ denote any chosen analysis plane after the final modulation stage, and let the propagated vector field at that plane be
\begin{equation}
E_p(r)=\sum_k u_{k,p}(r), \qquad u_{k,p}(r)\in \mathbb{C}^2.
\end{equation}
With analyzer Jones vector $J_p(r)$ defined on plane $p$, the projected
scalar field is
\begin{equation}
s_p(r)=J_p^\dagger(r)\,E_p(r), \qquad I_p(r)=|s_p(r)|^2.
\end{equation}
Each individual contribution projects as
\begin{equation}
a_{k,p}(r)=J_p^\dagger(r)\,u_{k,p}(r),
\end{equation}
so the intensity decomposes into
\begin{equation}
I_p(r)=\sum_k |a_{k,p}(r)|^2+\sum_{k\neq \ell} a_{k,p}(r)a_{\ell,p}^*(r).
\end{equation}
Defining the polarization coherence matrix at plane $p$ as
\begin{equation}
C_{k\ell,p}(r)=u_{k,p}(r)\,u_{\ell,p}^\dagger(r),
\end{equation}
each projected cross term can be written as
\begin{equation}
a_{k,p}(r)a_{\ell,p}^*(r)=J_p^\dagger(r)\,C_{k\ell,p}(r)\,J_p(r).
\end{equation}
Thus, the analyzer acts as the operator that determines how vectorial
coherence contributes to the observed scalar intensity. The interference-shaping mechanism of ellipsography therefore holds at any chosen analysis plane $p$; what changes with plane is the propagated field $E_p(r)$, not the projection principle itself.

If the analyzer is spatially uniform, $J_p(r)\equiv J_0$,
then the projection commutes with the polarization-independent propagation
model in Equation~(\ref{eq:asm_prop}):
\begin{equation}
J_0^\dagger f_{\mathrm{polar}}(E,\Delta z)
=
f_{\mathrm{ASM}}\!\left(J_0^\dagger E,\Delta z\right).
\label{eq:ga_inv}
\end{equation}
where $\Delta z$ denotes post-analyzer propagation distance.
Hence, within a polarization-preserving relay after the final
polarization-dependent modulation stage, the axial placement of a uniform analyzer does not change the measured scalar field at the observation plane. In particular, the linear polarizer used before the camera in our prototype is equivalent to on-sensor projection under the propagation model used here.
\begin{figure}[!t]
  \centering
  \includegraphics[width=\linewidth]{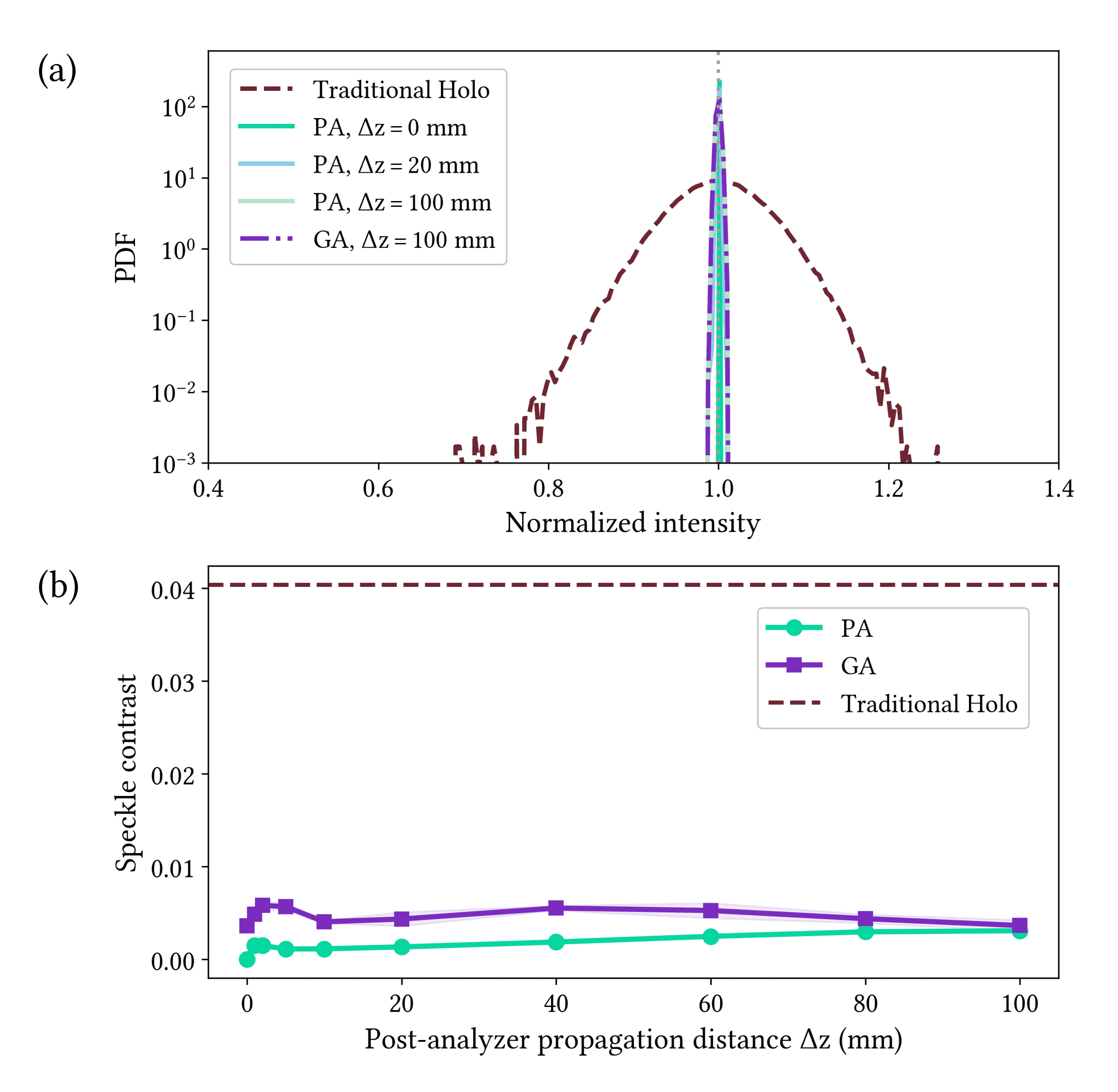}
  \caption{\textit{Speckle suppression for different analyzer placement.} We evaluate speckle suppression using a spatially uniform intensity target for different analyzer placements.
  (a) Probability density functions (PDFs) of reconstructed intensity for representative analyzer placements $\Delta z=0,20,100$~mm, comparing the exact off-plane per-pixel analyzer (PA) and global analyzer (GA) formulations against traditional random-phase holography.
(b) Speckle contrast as a function of post-analyzer propagation distance $\Delta z$. For each $\Delta z$, the analyzer is explicitly placed at its physical plane during optimization and the final reconstruction is evaluated at the image plane. PA degrades gradually with increasing $\Delta z$ but remains well below the traditional holography baseline throughout the sweep, while GA stays nearly flat, consistent with the placement invariance implied by \Cref{eq:ga_inv} under the propagation model used here. 
}

  \label{fig:analyzer_placement}
\end{figure}

If the analyzer varies spatially,
\begin{equation}
J_p(r)=
\begin{bmatrix}
\cos\theta_p(r)\\
\sin\theta_p(r)
\end{bmatrix},
\end{equation}
the projection is pointwise and generally does \emph{not} commute with propagation. More precisely, there does not in general exist a local analyzer field $\widehat{J}(r)$ at the observation plane such that
\begin{equation}
\left[f_{\mathrm{ASM}}\!\left(J_p^\dagger E_p,\Delta z\right)\right](r)
=
\widehat{J}^\dagger(r)\,\left[f_{\mathrm{polar}}(E_p,\Delta z)\right](r)
\label{eq:PAcommute}
\end{equation}
for all fields $E_p$, where $J_p^\dagger E_p$ denotes the scalar field
$\rho \mapsto J_p^\dagger(\rho)E_p(\rho)$ on plane $p$.
Therefore, a per-pixel analyzer must be realized at the plane for which $J_p(r)$ is defined, or at an optically conjugate plane through a polarization-preserving relay. Otherwise, propagation after the analyzer mixes neighboring analyzer locations and the local quadratic form in Equation~(\ref{eq:projected_cross_term}) is no longer preserved. 

To evaluate this regime, we use a spatially uniform target, which isolates speckle independently of image texture, and sweep the post-analyzer propagation distance $\Delta z$ while keeping the SLM-to-image distance fixed. See supplementary material for operator analysis.
For each $\Delta z$, we optimize the exact analyzer placement and then evaluate the reconstructed intensity at the image plane. \Cref{fig:analyzer_placement} shows that ellipsography continues to suppress speckle substantially away from the image plane: both the per-pixel analyzer (PA) and the global analyzer (GA) produce narrower normalized intensity distributions and lower speckle contrast than traditional random-phase holography. The PA curve degrades gradually with increasing $\Delta z$, consistent with increasing nonlocal mixing after projection, whereas the GA curve remains nearly invariant, consistent with the commutation relation in \Cref{eq:ga_inv}. Thus, the generalized interference-shaping mechanism remains effective when the analyzer is modeled at its physical plane.

In our prototype, the results with per-pixel analyzer are emulated at the sensor plane using the micro-polarizer array of the polarization camera, whereas the global analyzer configuration is implemented physically using an off-the-shelf linear polarizer. A detailed analysis of post-analyzer propagation, locality conditions, and nonlocal interference effects is provided in the Supplementary Material.
\subsection{Synthesizing Optimal Phase Patterns}
\label{sec:optimize}

\definecolor{lightgreen}{rgb}{0.56, 0.93, 0.56}
\definecolor{moonstoneblue}{rgb}{0.45, 0.66, 0.76}
\definecolor{lightblue}{rgb}{0.4, 0.9, 0.4}
\definecolor{darkgreen}{rgb}{0.0, 0.4, 0.0}

\begin{algorithm}[t]
\begin{tcolorbox}[
    enhanced,
    attach boxed title to top left={xshift=6mm,yshift=-3mm},
    colback=moonstoneblue!20,
    colframe=moonstoneblue,
    colbacktitle=moonstoneblue,
    title=Ellipsography,
    fonttitle=\bfseries\color{black},
    boxed title style={size=small,colframe=moonstoneblue},
    sharp corners,
    grow to right by=0mm
]
\medskip
\begin{algorithmic}
\State {$N$ : Number of iterations for hologram update}
\State {$M$ : Number of iterations for analyzer map update}
\State {$\alpha$ : Learning rate}
\State {$I_\text{target}$ : Target image}
\State {$f_\text{prop}$} : Wave propagation model
\State \textbf{$\mathcal{L}$} : Loss function defined in \Cref{eq:mseloss}
\State Initialize \textbf{$\Phi_{1}, \Phi_{2} \sim \mathcal{U}(0,2\pi)$}{} {\color{darkgreen} \Comment{\textit{SLM phase patterns}}}
\State Initialize \textbf{$\theta$}{} {\color{darkgreen} \Comment{\textit{Analyzer angle map}}}

\For{$n = 1, \dots, N$}
  \State $\Phi_{1,2} \leftarrow \Phi_{1,2}-\alpha \nabla_{\Phi_{1,2}}\mathcal{L}(I(\Phi_{1,2}, \theta, f_\text{prop}), I_\text{target})$
  \\{\color{darkgreen} \Comment{$\theta$ \textit{remains unchanged}}}
  \For{$m = 1, \dots, M$}
    \State $\theta \leftarrow \theta-\alpha \nabla_{\theta}\mathcal{L}(I(\Phi_{1,2}, \theta, f_\text{prop}), I_\text{target})$
    \\{\color{darkgreen} \Comment{$\Phi_{1,2}$ \textit{remains unchanged}}}
  \EndFor
\EndFor
\State \textbf{return} $\Phi_{1}, \Phi_{2}, \theta$
\end{algorithmic}
\end{tcolorbox}
\vspace{-3mm}
\caption{Ellipsography optimization for speckle-free interference.} 
\label{alg:code}
\end{algorithm}

Unless otherwise specified, we optimize the local-projection model of \Cref{eq:final-img-field}, corresponding to the intended observation plane or a polarization-preserving conjugate analyzer plane.
Given a target image $I_\text{target}$, our goal is to compute SLM phase patterns that jointly modulates phase and polarization, and the per-pixel analyzer to project vectorial field into a scalar field, to synthesize a speckle-free reconstruction $I$ at the image plane. 
Using the model derived in \Cref{eq:final-img-field}, the output intensity is given by:
\begin{equation}
I(\mathbf r) = |s_{\mathrm{img}}(\mathbf r)|^2 .
\end{equation}
Note that the image intensity $I(\mathbf r)$ is a function of the two SLM phase patterns $\Phi_1(\mathbf r), \Phi_2(\mathbf r)$, and the resulting per-pixel polarization state $\theta(\mathbf r)$ (\ie per-pixel analyzer angle), as illustrated by \Cref{eq:img-field,eq:final-img-field}. 

We frame the hologram synthesis task as a constrained optimization problem that minimizes the per-pixel error between target and reconstructed images:
\begin{equation}
    \Phi_1^*, \Phi^*_2, \theta^*  =\underset{\Phi_1, \Phi_2, \theta}{\textit{argmin}} \quad \mathcal{L}(I(\Phi_1, \Phi_2, \theta), I_\text{target}),
    \label{eq:opt-objective}
\end{equation}
where $\mathcal{L}$ is a loss function designed for image fidelity. In our implementation, we define:
\begin{equation}
    {\mathcal{L}} = \mathcal{L}_\text{MSE},
    \label{eq:mseloss}
\end{equation}
where $\theta$ is the analyzer angle map and $\mathcal{L}_\text{MSE}$ is the per-pixel mean-squared error.

While a pixel-wise analyzer results in the optimal performance and accurate pixel-wise interference, there are no readily available dynamic per-pixel analyzers. In contrast, a global analyzer that filters all pixels to the same polarization state can be implemented using off-the-shelf polarizers and is a practical option. In order to optimize the SLM phase patterns for a global analyzer, the optimization problem can be updated as:
\begin{equation}
    \Phi_1^*, \Phi^*_2, \theta_{GA}^*  =\underset{\Phi_1, \Phi_2, \theta_{GA}}{\textit{argmin}} \quad \mathcal{L}(I(\Phi_1, \Phi_2, \theta_{GA}), I_\text{target}),
    \label{eq:opt-objective-global}
\end{equation}
where $\theta_{GA}$ is the orientation of the global analyzer.

\section{Implementation}
\label{sec:implementation}

\subsection{Software}
We evaluated our ellipsography framework in simulation using PyTorch, running on an NVIDIA GeForce RTX 4090 GPU.
The simulated SLM configuration matches our physical prototype, with a resolution of $1080 \times 1920$ pixels and an $8 \mu$m pixel pitch.
To solve the optimization problem in \Cref{eq:opt-objective}, we use the Adam optimizer \cite{kingma2014adam} with a learning rate of $0.01$ and default parameters.
Both SLMs are initialized with random phases in the range of $[0, 2\pi]$ drawn from a uniform distribution.
The optimization converges reliably in simulation, typically synthesizing full color holograms in under 2.2 seconds.

\subsection{Hardware Prototype}
\begin{figure}[]
    \centering
    \includegraphics[width=\linewidth,trim=95 8 110 3,clip]{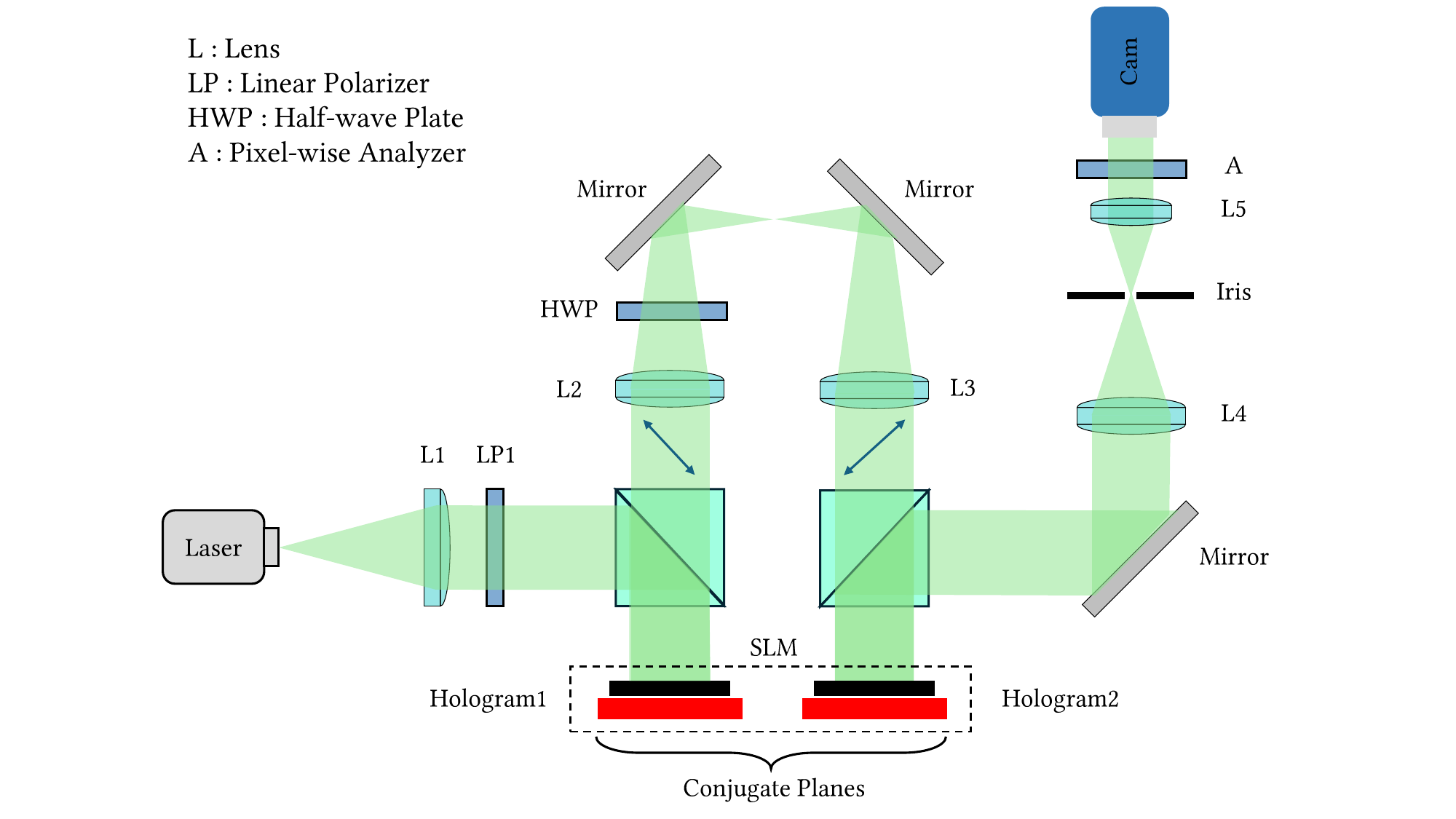}

    \vspace{-3mm}
    \caption{
    \textit{Ellipsography Hardware Setup.} Our prototype display system employs a dual-modulation configuration to achieve simultaneous control of phase and polarization. A 4f relay optics setup transfers the modulated wavefront from the first SLM to the second, with a half-wave plate (HWP) at an intermediate plane swapping the horizontal and vertical polarization channels to enable independent modulation of orthogonal components.
    The laser beam, sequentially modulated by both SLMs, is imaged onto a camera through relay and focusing optics. An iris placed at the Fourier plane filters any higher diffraction orders, ensuring clean reconstruction. The global analyzer is implemented physically with a linear polarizer before the camera, while the per-pixel analyzer is emulated at the sensor plane by the polarization camera's micro-polarizer array.
    }
    \label{fig:setup}
\end{figure}
We constructed a holographic display prototype, shown in \Cref{fig:setup}, using a Holoeye PLUTO liquid crystal on silicon (LCoS) reflective phase-only SLM.
The device uses a parallel aligned nematic liquid crystal material in its LCoS microdisplay. 
Unlike twisted nematic liquid crystals commonly used in conventional LCD display screens, parallel aligned nematic crystals enable phase-only modulation of one linear polarization component, while leaving its orthogonal component unaffected. 
Due to the limited availability of SLM hardware, we partition the SLM into two spatial regions, each displaying a distinct phase pattern, thereby enabling independent modulation of orthogonal polarization components across the two halves. 

For illumination, we use a FISBA RGBeam fiber-coupled laser module containing three optically aligned laser diodes emitting at 450 nm, 520 nm and 660 nm, each with independent per-diode power control. 
The laser module is mounted to a kinematic stage (Thorlabs KM100T) using a FC/APC fiber adapter, and collimated by achromatic doublets (Thorlabs AC254-030-AB). 
The collimated beam passes through a linear polarizer (Thorlabs LPNIRB100) mounted on a motorized rotation stage (ThorLabs PRM1/MZ8). 
The beam is then modulated by the left half of the SLM via a non-polarizing plate beamsplitter (Thorlabs BS010, $50:50$ energy split ratio), and relayed to the right half of the SLM using a second set of achromatic doublets (Thorlabs AC254-030-AB). 
An iris (Thorlabs ID25) is positioned between the lenses to block any higher diffraction orders. 

To accommodate space constraints, the optical path is extended using three dielectric mirrors (Thorlabs BB1-E02) and a right-angle prism (Thorlabs PS908L-A), which redirect the beam to a zero-order half-wave plate (Thorlabs WPH05ME) mounted on a motorized precision rotation stage (Thorlabs PRM1/MZ8). 
The modulated output beam is then relayed and focused using achromatic doublets (Thorlabs AC254-150-AB and AC254-100-AB) and imaged onto a polarization camera sensor (FLIR BFS-U3-51S5PC-C) mounted on a motorized translation stage (Thorlabs MZ9) for analysis. 
We leverage the micro-polarizer on top of each pixel in the polarization camera to act as a per-pixel analyzer. Specifically, the polarization camera records four linear polarization channels $I_{0},\; I_{45},\; I_{90},\; I_{135}$ in one exposure, from which the Stokes parameters $S_0, S_1, S_2$ are reconstructed.
Given the optimized analyzer map $\theta(\mathbf r)$, the experimental results with per-pixel analyzer are computed as:
\begin{equation}
I_{\mathrm{PA}}(\mathbf r) \;=\;\frac{1}{2}\Big(S_0(\mathbf r) + S_1(\mathbf r)\cos 2\theta(\mathbf r) + S_2(\mathbf r)\sin 2\theta(\mathbf r)\Big). 
\end{equation}
A linear polarizer (Thorlabs LPNIRB100) mounted on a motorized rotation stage (ThorLabs PRM1/MZ8) is placed before the camera for experiments with global analyzer.
For accurate spatial registration between the two halves of the SLM, 
we map the optical field from the first SLM region to the coordinate space of the second, ensuring proper alignment at the shared image plane.
Experiments for all methods are conducted using the same optical path configuration for fair comparison.
Additional hardware and calibration details are provided in the Supplemental Material.

\section{Synthetic Evaluation}
\label{sec:result}

Here, we validate the proposed ellipsography framework and compare its performance against conventional holography methods using both quantitative metrics and qualitative visualization.

\vspace{-3mm}

\subsection{Speckle Suppression}
\label{subsec:pol-diversity}

\begin{figure}[!t]
  \centering
  \includegraphics[width=\linewidth]{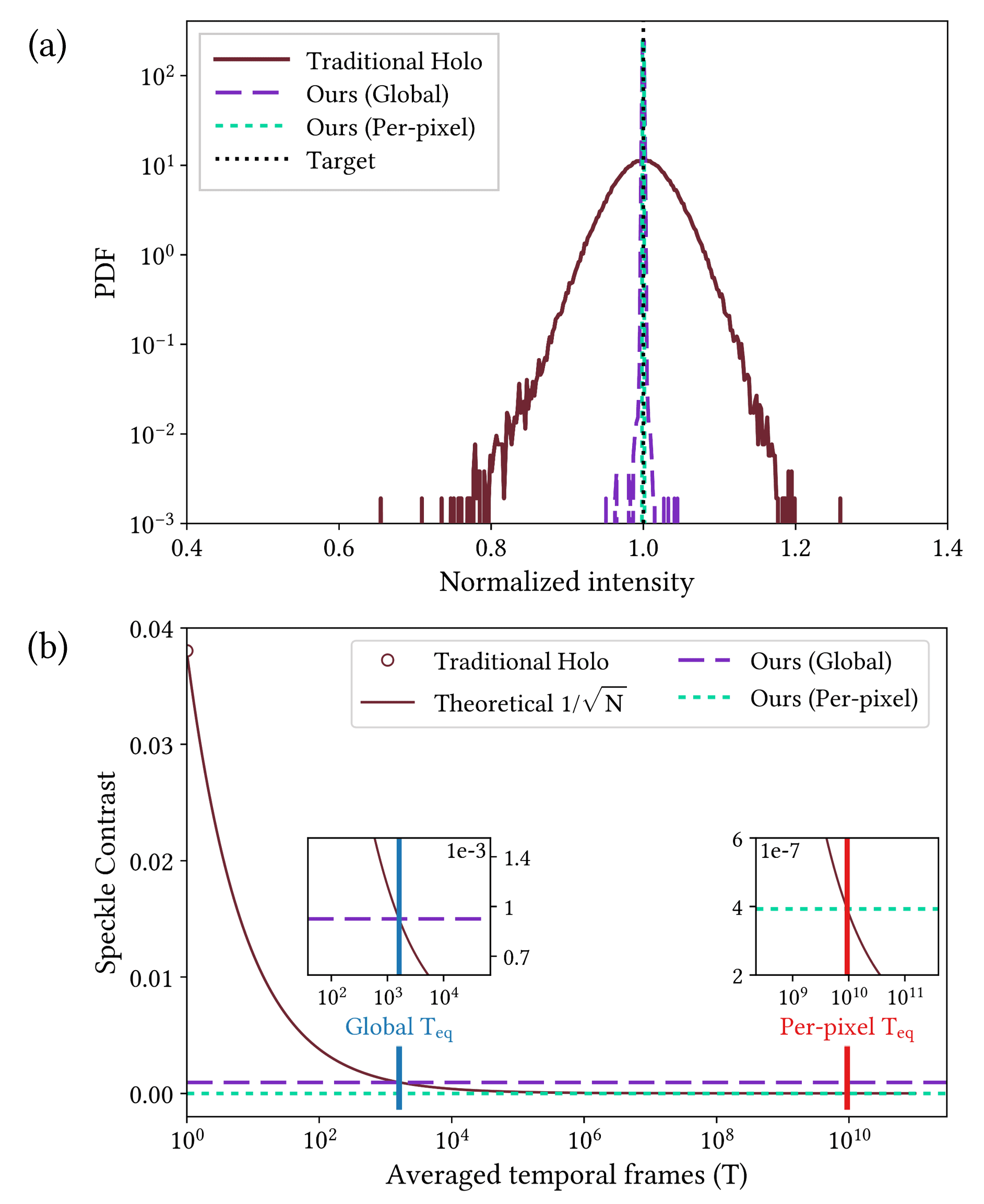}
  \caption{\textit{Synthetic Validation of Speckle Suppression.} We evaluate speckle suppression using a spatially uniform intensity target, a worst-case scenario for coherent holographic displays.
  (a) Probability density functions (PDFs) of reconstructed intensity values, normalized by their spatial mean. The red vertical line indicates the ideal delta distribution corresponding to perfectly uniform intensity. Conventional scalar holography exhibits a broad distribution with significant variance due to residual coherent interference, whereas ellipsography (using both global and per-pixel analyzers) produces sharply concentrated distributions near the target value.
(b) Speckle contrast as a function of the number of temporally averaged frames $N$ (log scale). Temporal multiplexing follows the expected $1/\sqrt{N}$ decay (dashed black), while ellipsography achieves comparable or lower speckle contrast in a single coherent frame, without temporal averaging.}

  \label{fig:poincare}
\end{figure}

Unlike polarization multiplexing, the central role of polarization in ellipsography is to \textit{structure the interference at pixel-level} rather than mere statistical averaging of intensity frames. By jointly optimizing the phase and polarization at each image pixel, ellipsography effectively eliminates the random per-pixel interference (as illustrated by \Cref{eq:polar-interfere-final}) that results in speckle noise.

To rigorously evaluate ellipsography's speckle suppression independent of image texture, we use a spatially uniform intensity target which represents one of the most challenging cases for coherent displays. 
Since the ideal reconstruction requires constant intensity while allowing arbitrary phase on the image plane, any residual coherent interference directly appears as visible speckle.
This makes uniform-field reconstruction a \textit{stringent stress test} for any speckle suppression technique.

We simulate reconstructions of smooth, speckle-free fields with uniform intensity using the traditional random-phase holography, time multiplexing of optimized holograms, and ellipsography using global and per-pixel analyzers. 
For each method, we compute the empirical probability density function (PDF) of reconstructed intensity and the speckle contrast. 
The PDF characterizes the distribution of intensity fluctuations, while speckle contrast quantifies their magnitude. 
For temporal multiplexing, we report speckle contrast as a function of the number of averaged frames $N$, with each frame optimized until convergence. Speckle suppression is achieved solely through averaging $N$ independently optimized frames. All results are reported in \Cref{fig:poincare}.

Figure~\ref{fig:poincare}(a) shows the empirical PDFs of normalized intensity reconstructions. Traditional holography exhibits a wide distribution with high intensity variance, indicating strong residual speckle. In contrast, ellipsography produces a sharply peaked distribution approaching the ideal delta function, evidencing strong suppression of interference-induced intensity fluctuations in a single frame.
Figure~\ref{fig:poincare}(b) quantifies speckle contrast across methods.
Temporal multiplexing follows the expected $1/\sqrt{N}$ decay, requiring tens of thousands of frames for significant speckle reduction.
Ellipsography, on the other hand, surpasses the speckle contrast of \textit{a million frame average} in a single shot.

Together, these results confirm that ellipsography suppresses speckle by structurally reshaping coherent interference. 
Rather than averaging out noise over time, it suppresses speckle in a single frame, marking a significant advance for high-fidelity holography.

\subsection{Comparison with Traditional Holography}
\label{sec:sim_rgb}

\setlength{\tabcolsep}{1.0pt}
\renewcommand{\arraystretch}{0.6}

\newcommand{\colw}{3.1cm}
\newcommand{\titlesize}{\fontsize{9}{11}\selectfont}
\newcommand{\coltitle}[1]{%
  \parbox[c][2.7ex][c]{\colw}{\centering\titlesize #1}%
}

\begin{figure*}[t]
\centering
\resizebox{\textwidth}{!}{%
\begin{tabular}{cccccc}

\multicolumn{1}{c}{\coltitle{Target}} &
\multicolumn{1}{c}{\coltitle{DPAC}} &
\multicolumn{1}{c}{\coltitle{Neural Holo.}} &
\multicolumn{1}{c}{\coltitle{Depolarized Holo.}} &
\multicolumn{1}{c}{\coltitle{Ellipsography (GA)}} &
\multicolumn{1}{c}{\coltitle{Ellipsography (PA)}} \\
[-0.2ex]

\includegraphics[width=\colw]{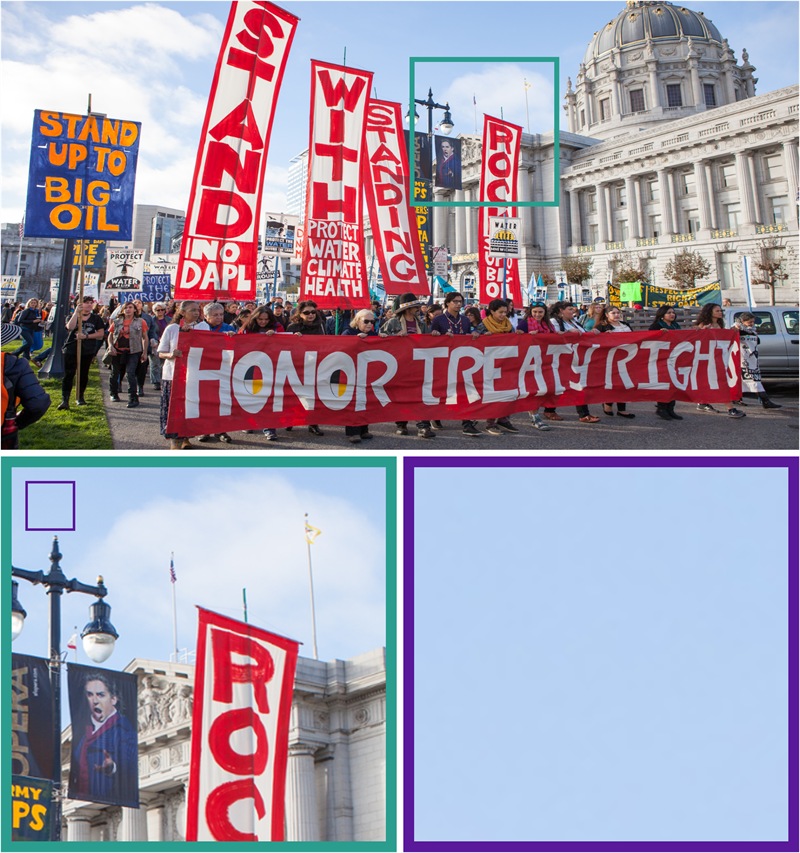} &
\includegraphics[width=\colw]{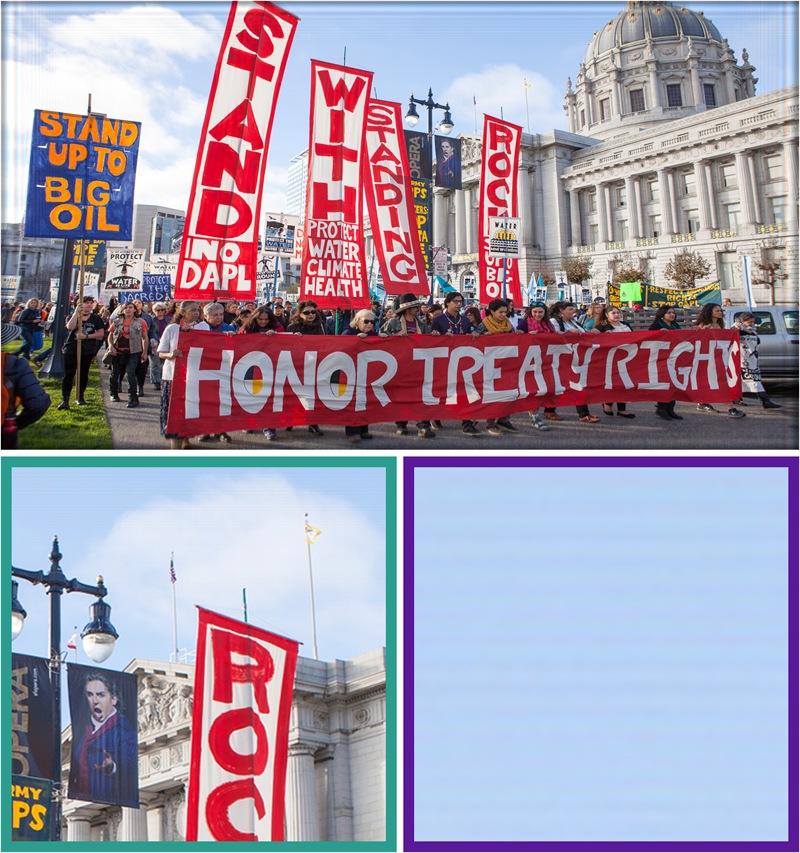} &
\includegraphics[width=\colw]{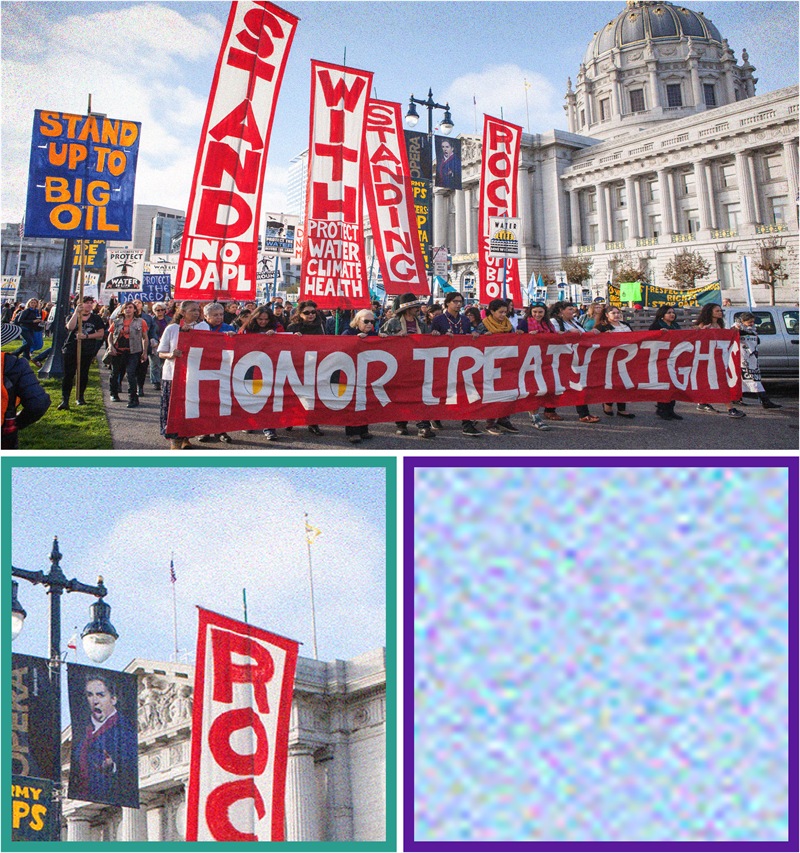} &
\includegraphics[width=\colw]{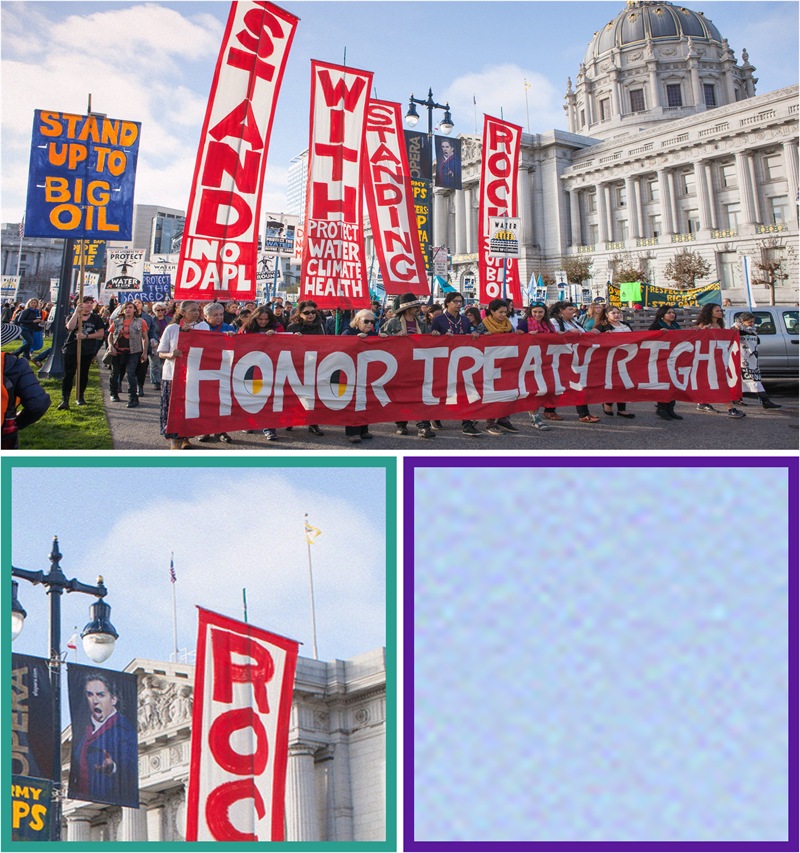} &
\includegraphics[width=\colw]{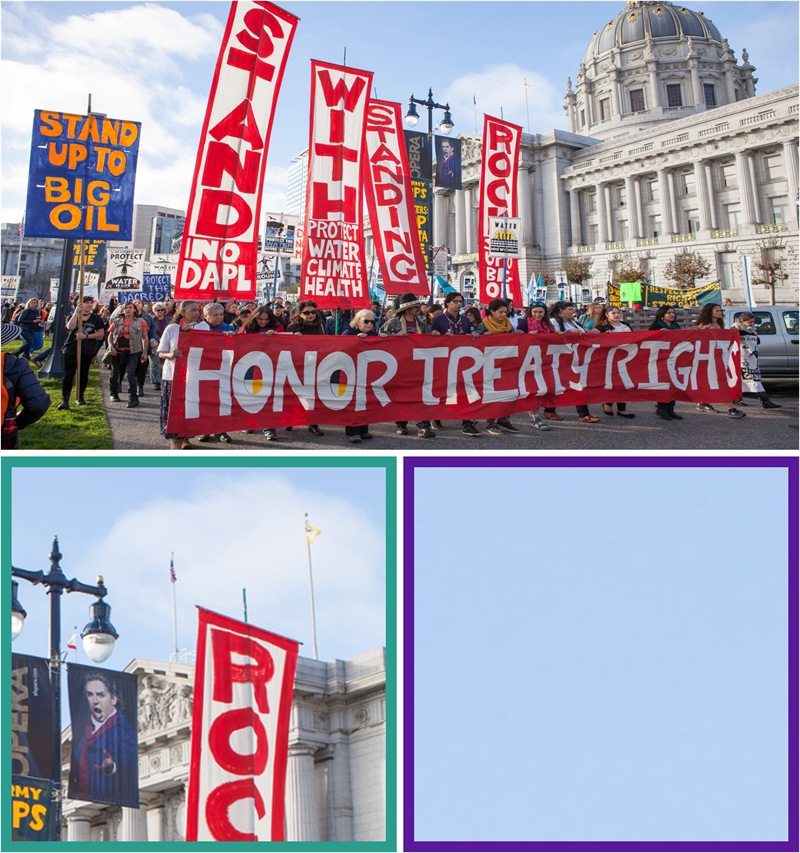} &
\includegraphics[width=\colw]{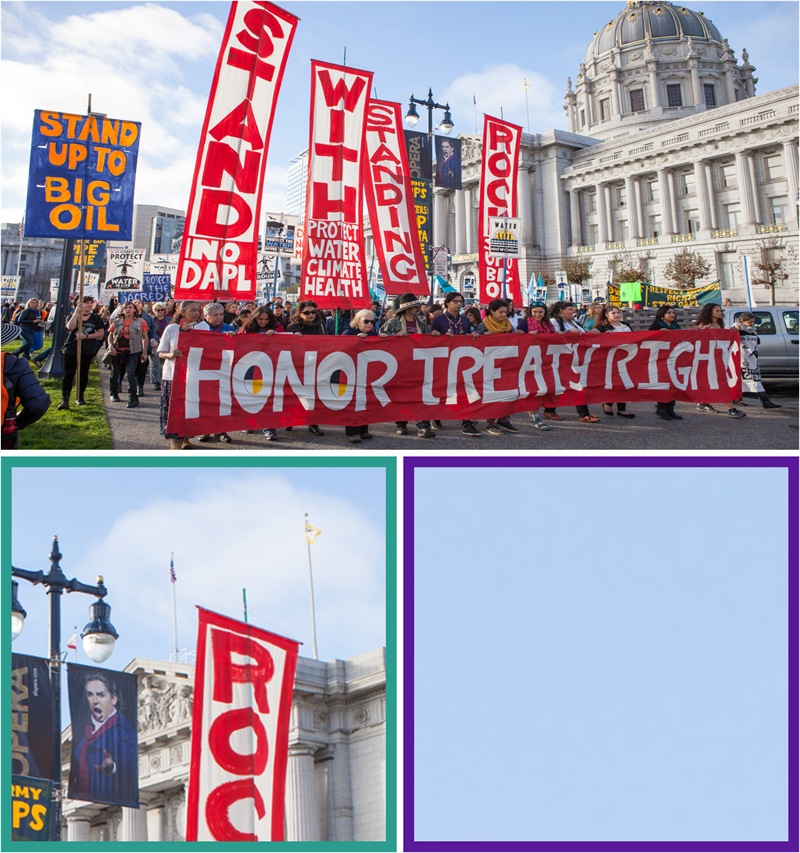} \\
[0.0ex]

\includegraphics[width=\colw]{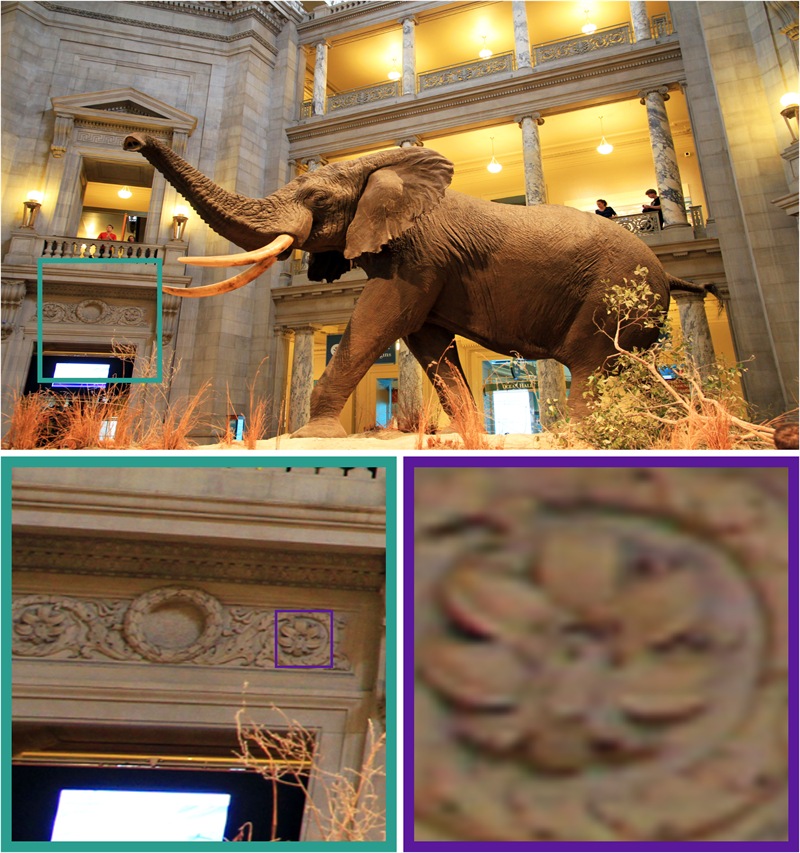} &
\includegraphics[width=\colw]{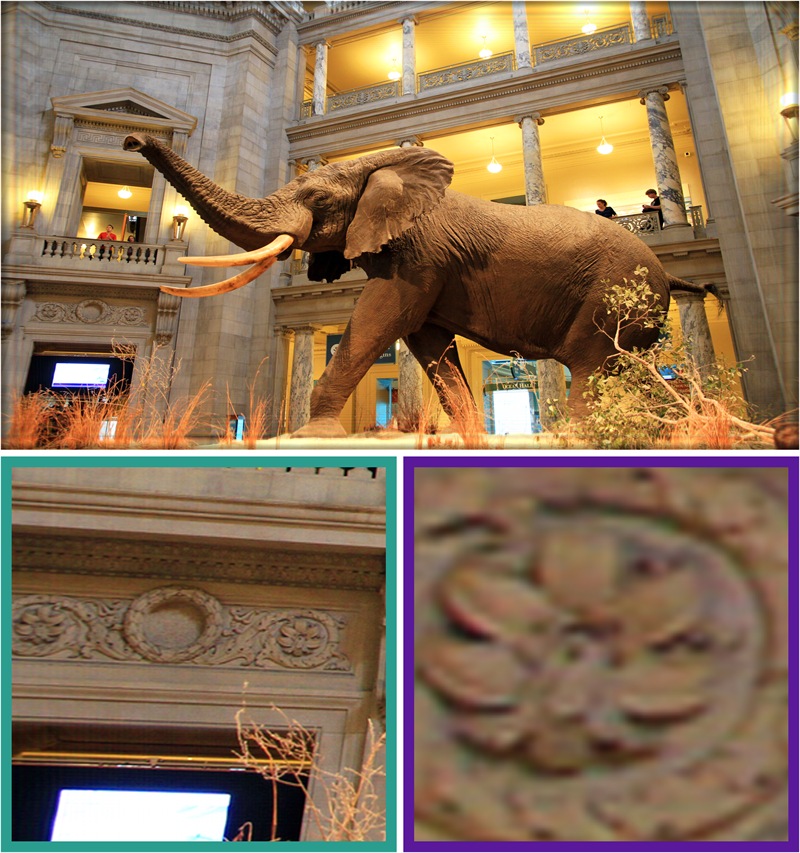} &
\includegraphics[width=\colw]{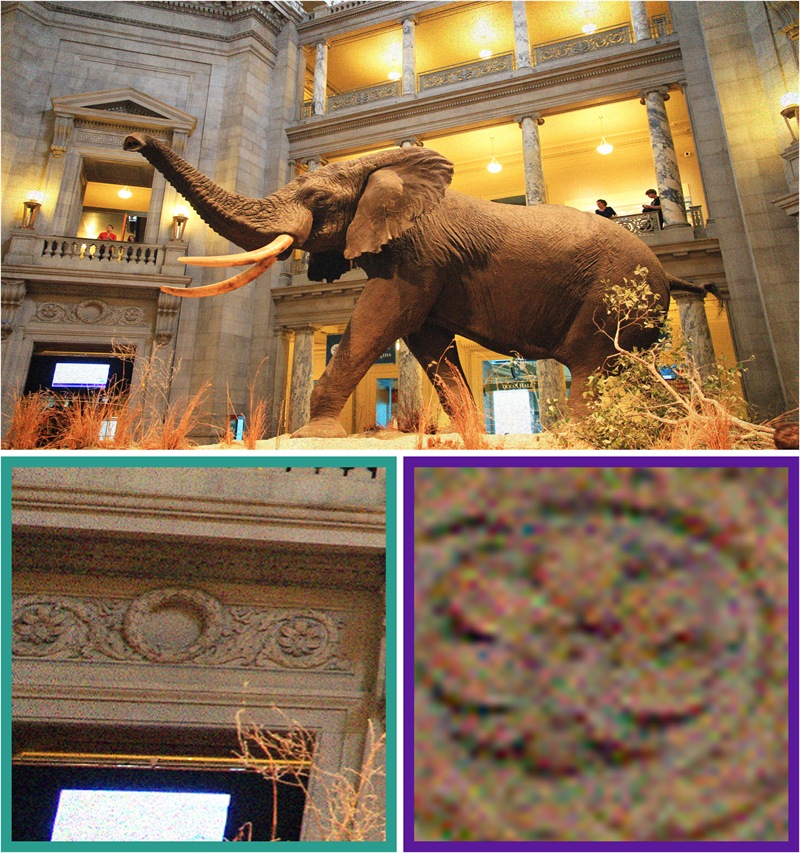} &
\includegraphics[width=\colw]{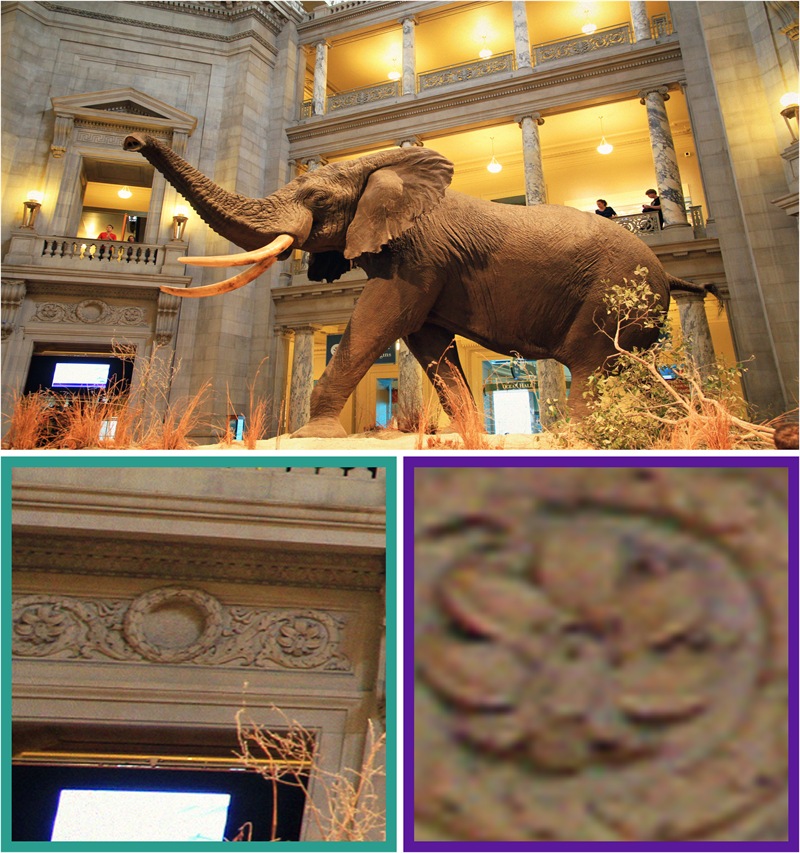} &
\includegraphics[width=\colw]{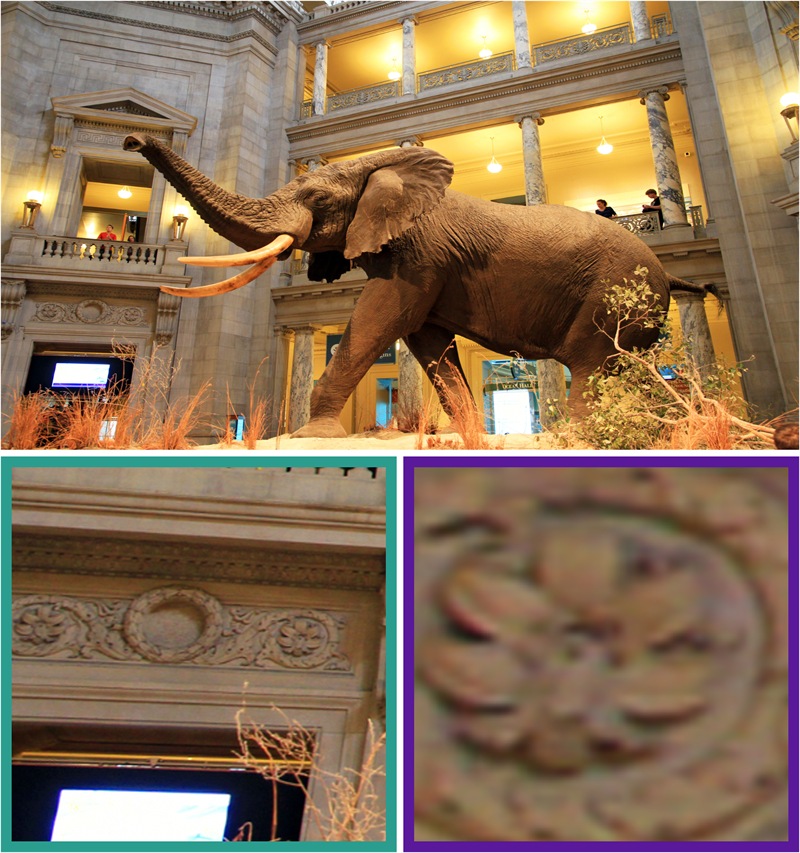} &
\includegraphics[width=\colw]{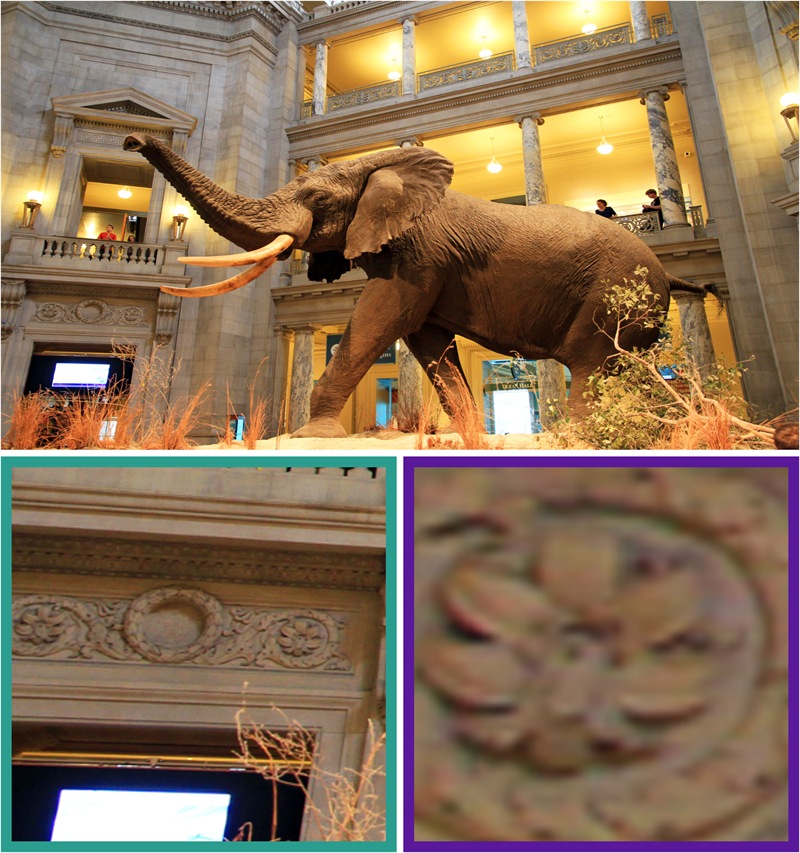} \\
[0.0ex]

\end{tabular}%
}
\vspace{-2mm}
    \caption{ 
   \textit{Synthetic Evaluation of 2D Holography Methods}. 
Our ellipsography approach demonstrates significantly improved image quality and speckle suppression compared to established smooth phase and random phase holography techniques, including depolarized holography (random phase), double phase amplitude coding (smooth phase) and neural holography SGD optimization (shallow random phase). 
Insets highlight key regions of interest, showing that ellipsography---both with global (GA) and per-pixel (PA) analyzers---consistently produces the sharpest reconstructions while effectively eliminating high-frequency speckle.
Insets show progressive zoom-ins. The left green inset is cropped from the main image, and the right purple inset is further cropped from the left inset.
    }
 \label{fig:sim_2d_results}
\end{figure*}

\setlength{\tabcolsep}{1.0pt}
\renewcommand{\arraystretch}{0.6}

\begin{figure*}[t]
\centering
\newcommand{\panelW}{0.155\textwidth} 
\newcommand{\errRatio}{0.46484375}    
\resizebox{\textwidth}{!}{%
\newcommand{\hdr}[2]{%
  \begin{tabular}[c]{@{}c@{}}
    \fontsize{8}{9}\selectfont #1\\[-0.1ex]
    \fontsize{8}{9}\selectfont #2
  \end{tabular}%
}
\begin{tabular}{@{}cccccc@{}}

\multicolumn{1}{c}{\hdr{Target}} &
\multicolumn{1}{c}{\hdr{Michelson Holo.}} &
\multicolumn{1}{c}{\hdr{Depolarized Holo.}} &
\multicolumn{1}{c}{\hdr{Complex Modulation}} &
\multicolumn{1}{c}{\hdr{Ellipsography (GA)}} &
\multicolumn{1}{c}{\hdr{Ellipsography (PA)}} \\[-1.6ex]

\includegraphics[width=\panelW]{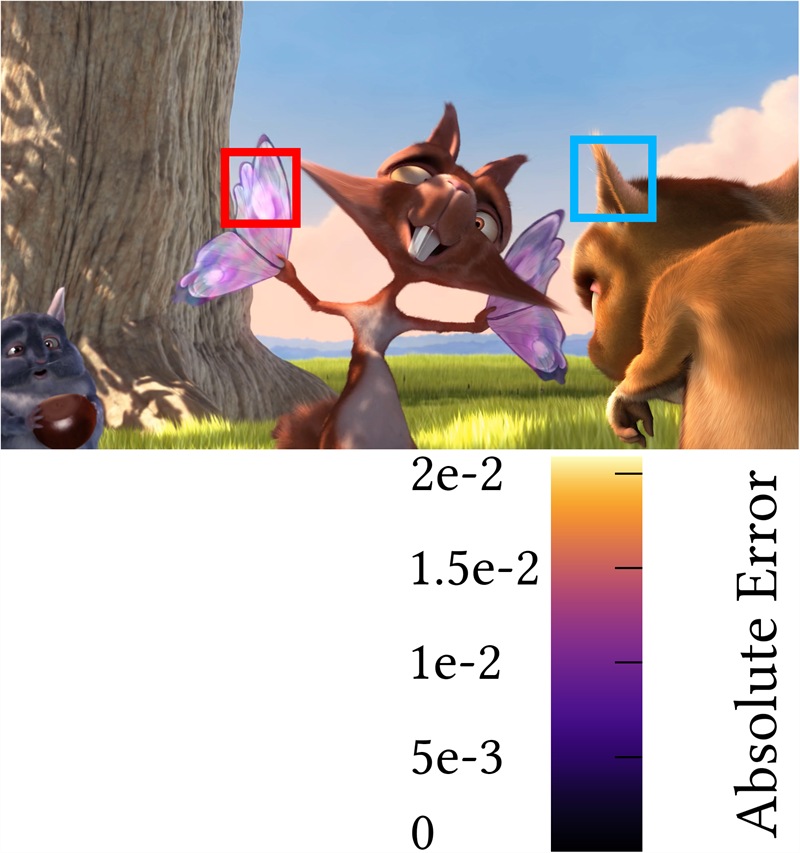} &
\includegraphics[width=\panelW]{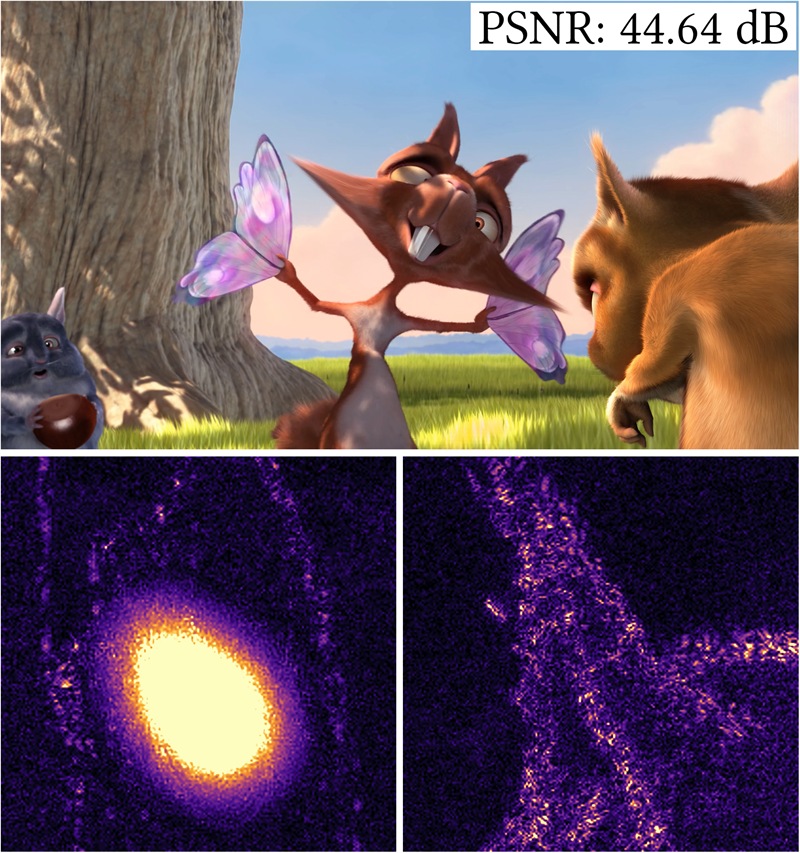} &
\includegraphics[width=\panelW]{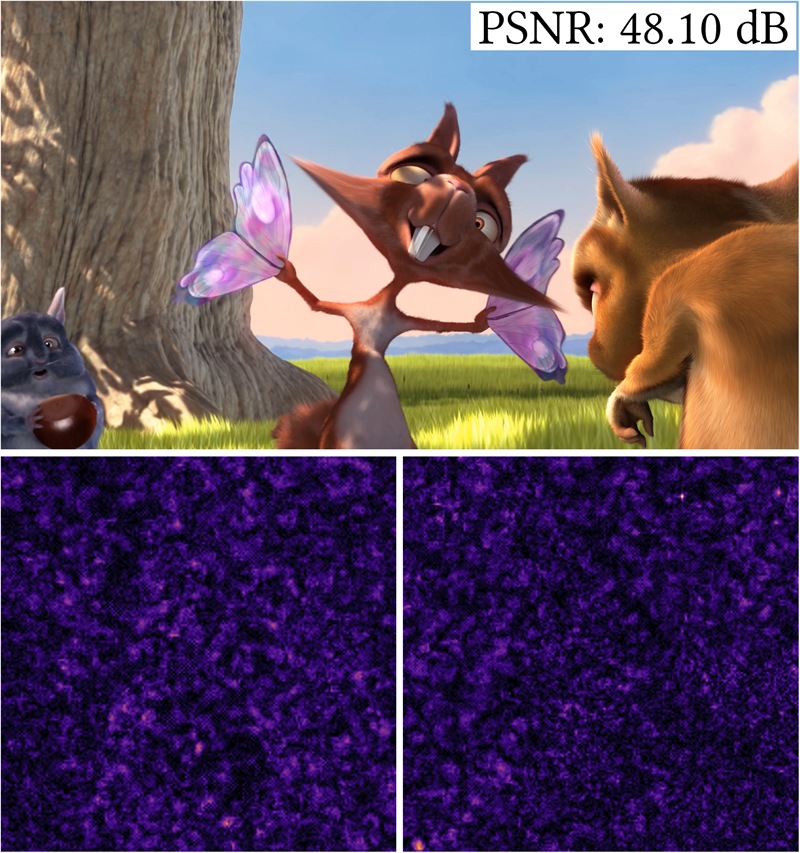} &
\includegraphics[width=\panelW]{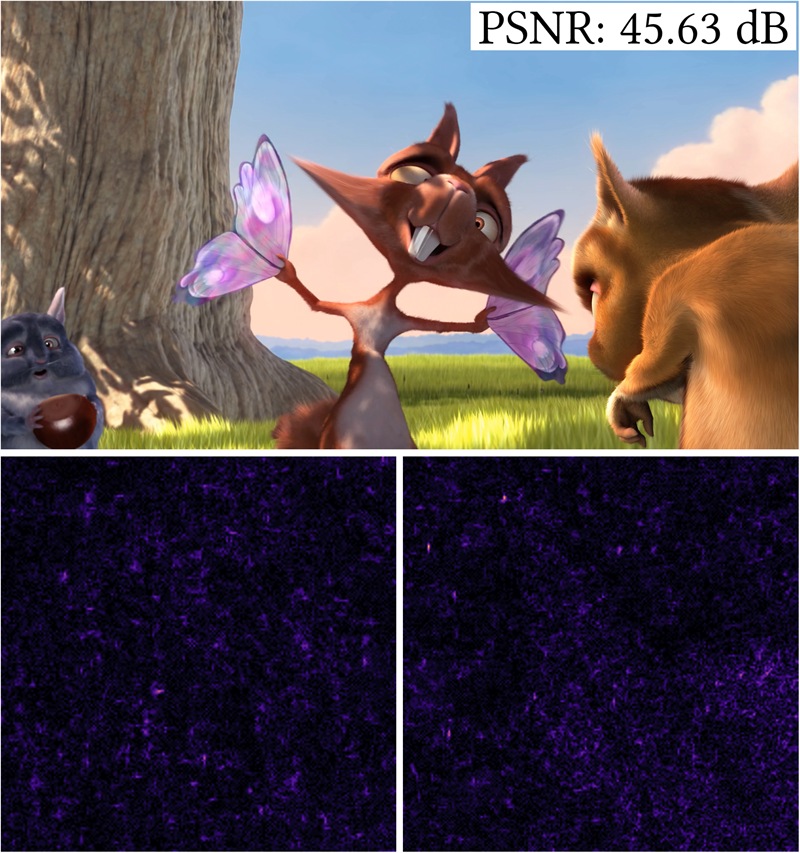} &
\includegraphics[width=\panelW]{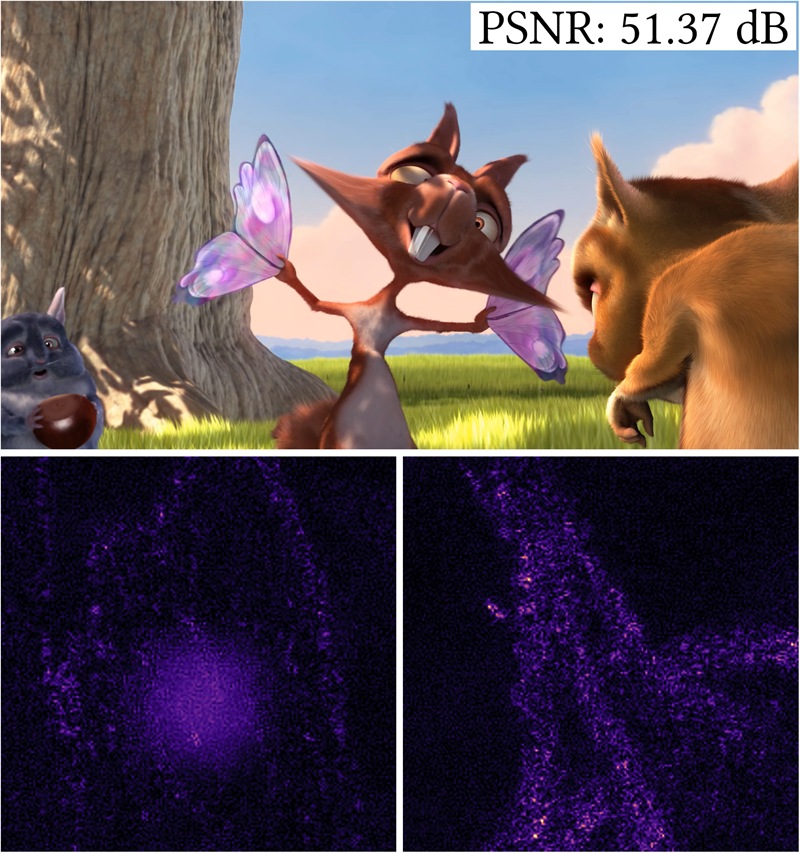} &
\includegraphics[width=\panelW]{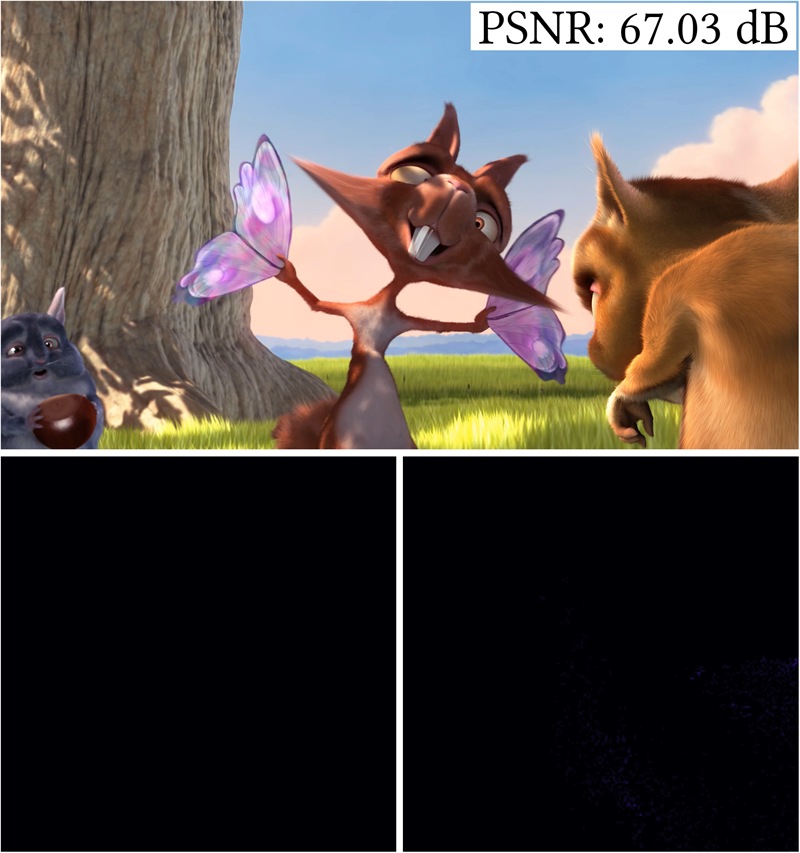} \\[-0.4ex]

\includegraphics[width=\panelW]{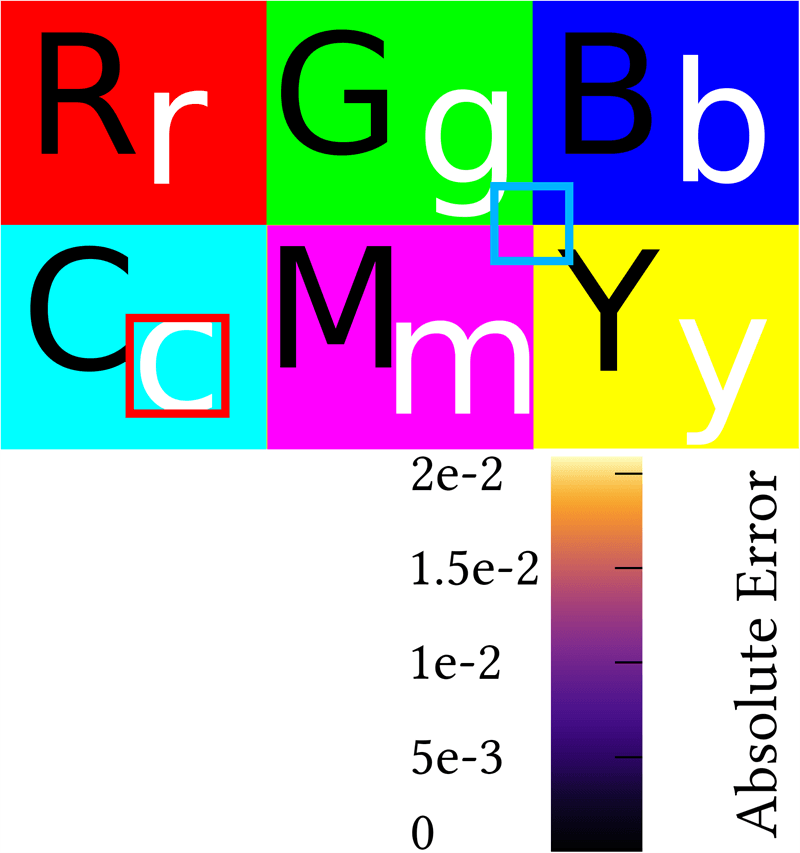} &
\includegraphics[width=\panelW]{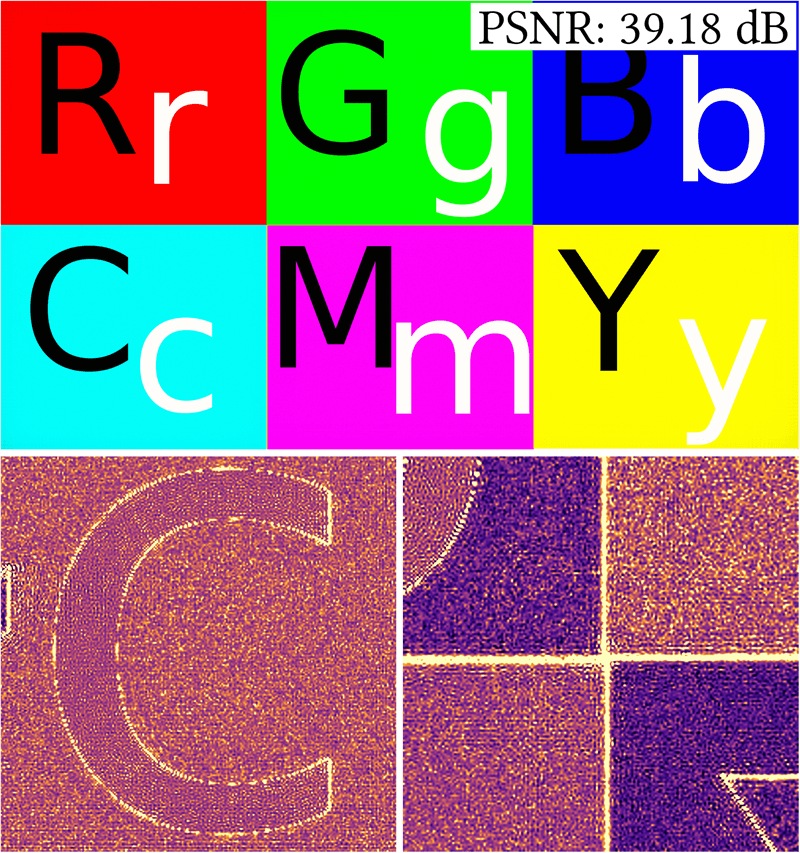} &
\includegraphics[width=\panelW]{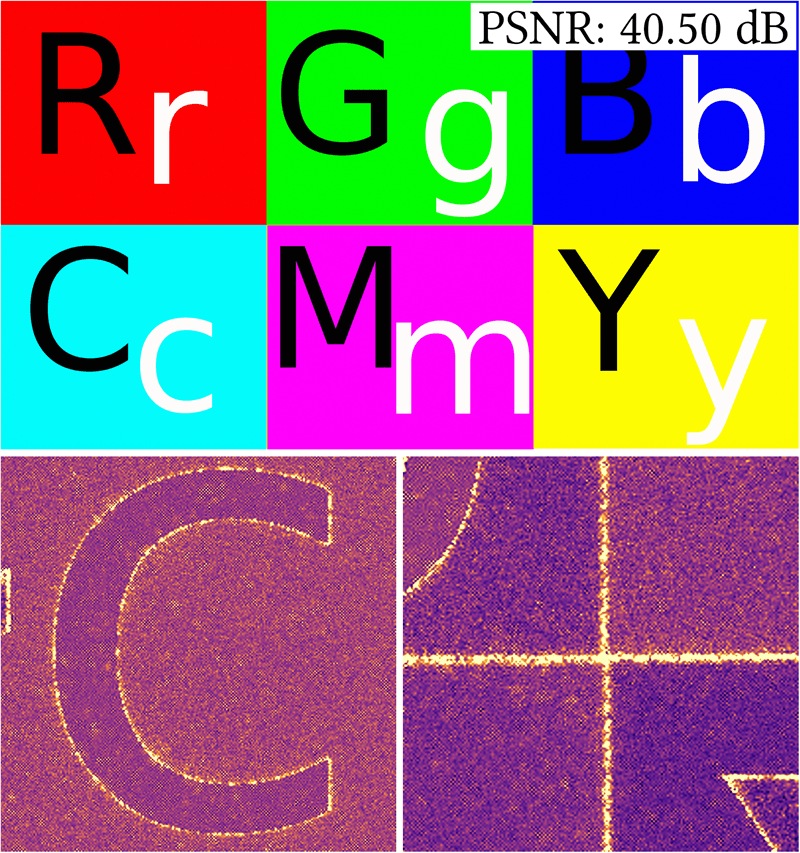} &
\includegraphics[width=\panelW]{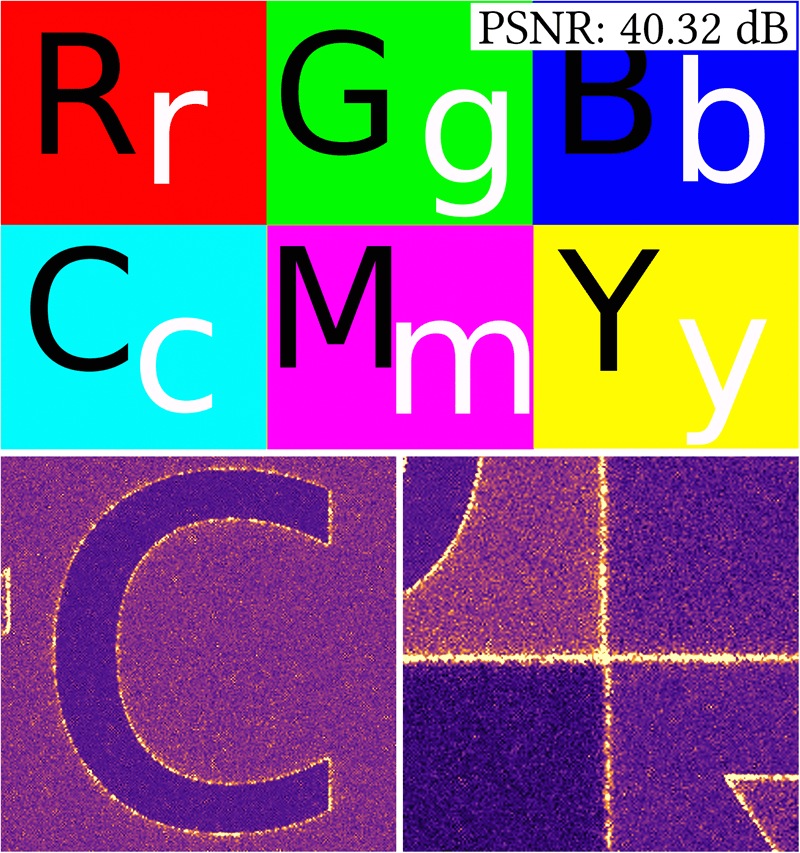} &
\includegraphics[width=\panelW]{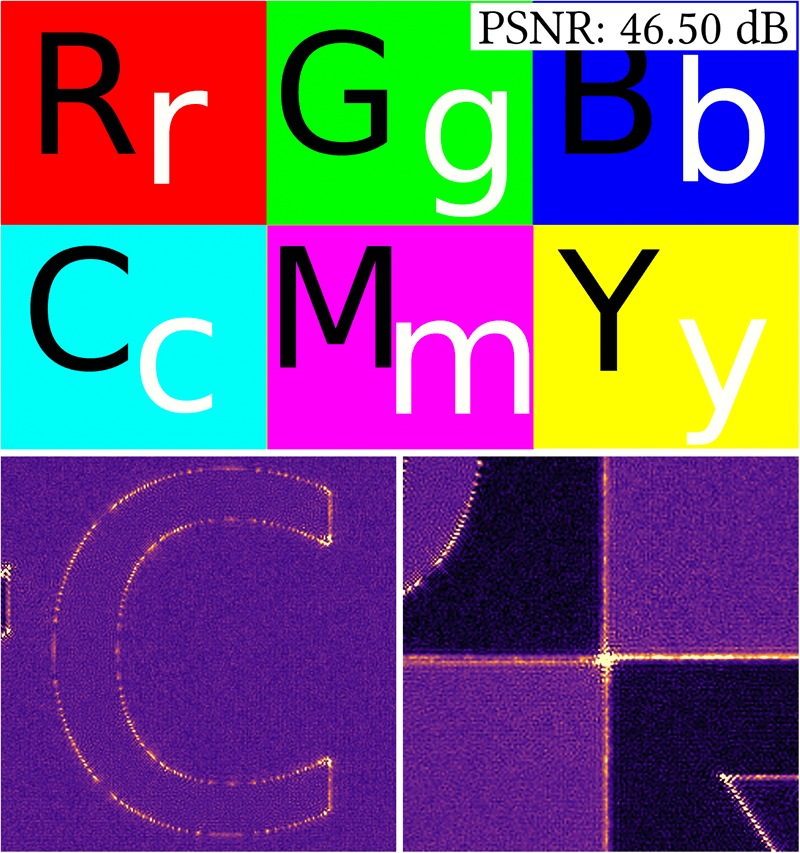} &
\includegraphics[width=\panelW]{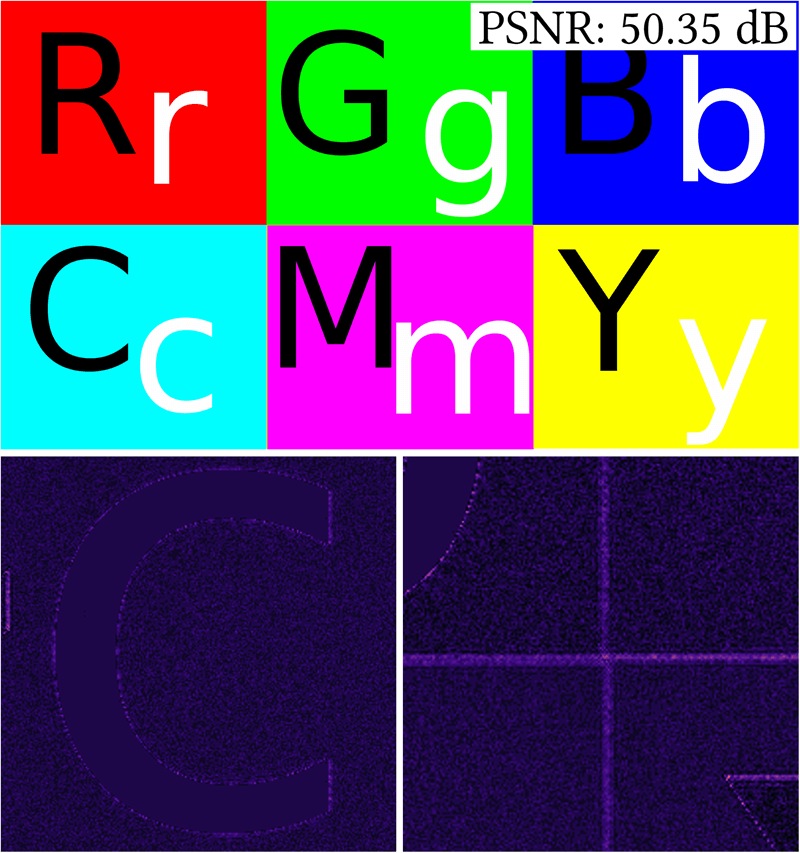}  \\

\end{tabular}%
}
\vspace{-2mm} 
		\caption{\textit{Synthetic Evaluation of Dual-Modulation Approaches}.
We compare ellipsography---both with global (GA) and per-pixel (PA) analyzers---against representative dual-modulation approaches with comparable degrees of freedom, including michelson holography, polarization
multiplexing, and combined amplitude--phase modulation.
All methods are optimized under the same simulation conditions.
Insets show zoomed regions of interest, together with corresponding error maps visualizing the residual reconstruction error relative to the target.
Despite having similar modulation capacity, competing methods exhibit
residual interference artifacts, whereas ellipsography significantly reduces visible speckle, approaching speckle-free reconstructions.
         }
\label{fig:sim_michelson_main}
\end{figure*}

\begin{figure*}[t]
\centering
\resizebox{\textwidth}{!}{%
\begin{tabular}{c p{\colw}p{\colw}p{\colw}p{\colw}p{\colw}p{\colw}}
\newcommand{\titlestrut}{\rule{0pt}{2.4ex}}
\newcommand{\coltitleOne}[1]{\coltitle{#1\\\vphantom{Holography}}}
\newcommand{\coltitleTwo}[2]{\coltitle{#1\\#2}}

\parbox[c][3.2ex][c]{\colw}{} & \parbox[c][5.2ex][c]{\colw}{%
    \centering\titlesize
    \raisebox{-0.6ex}{Target}%
  }
& \coltitle{Tensor Holo.}
& \coltitle{Neural 3D Holo.}
& \coltitle{Depolarized Holo.}
& \coltitle{Ellipsography (GA)}
& \coltitle{Ellipsography (PA)} \\[-1.3ex]

\rotatebox{90}{\parbox[c]{\colw}{\centering\titlesize Front Focus}} &
\includegraphics[width=\colw]{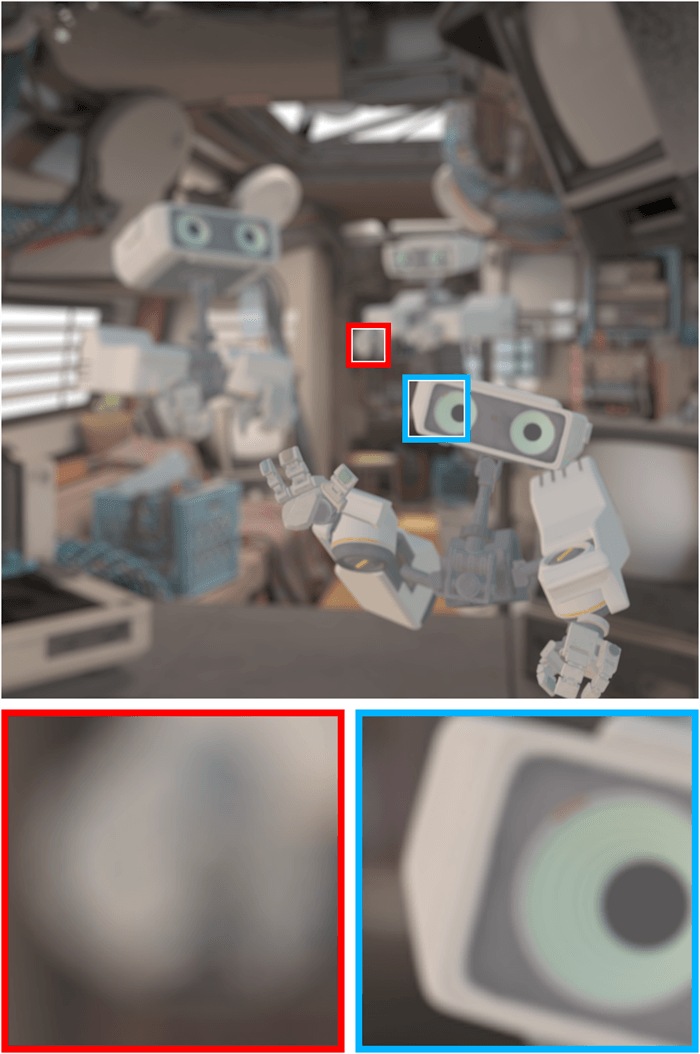} &
\includegraphics[width=\colw]{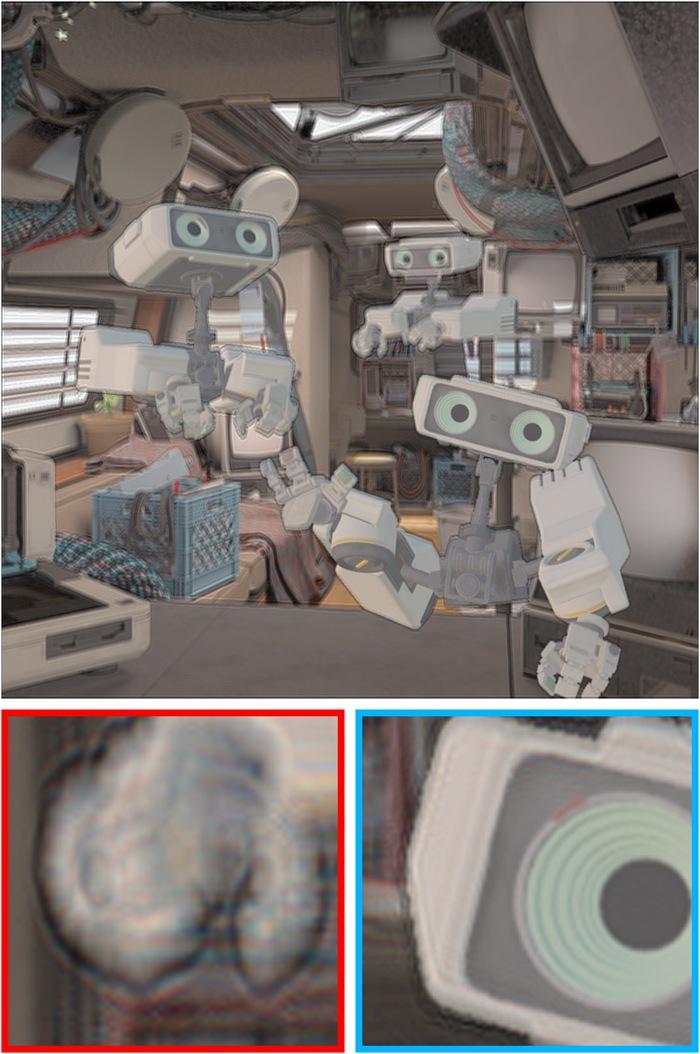} &
\includegraphics[width=\colw]{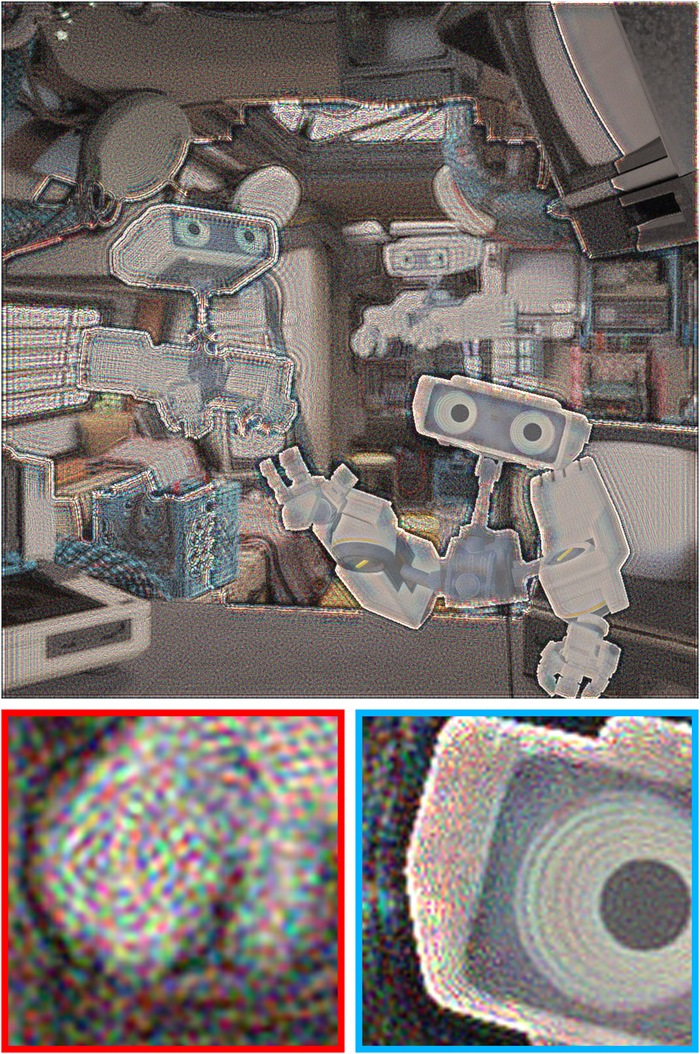} &
\includegraphics[width=\colw]{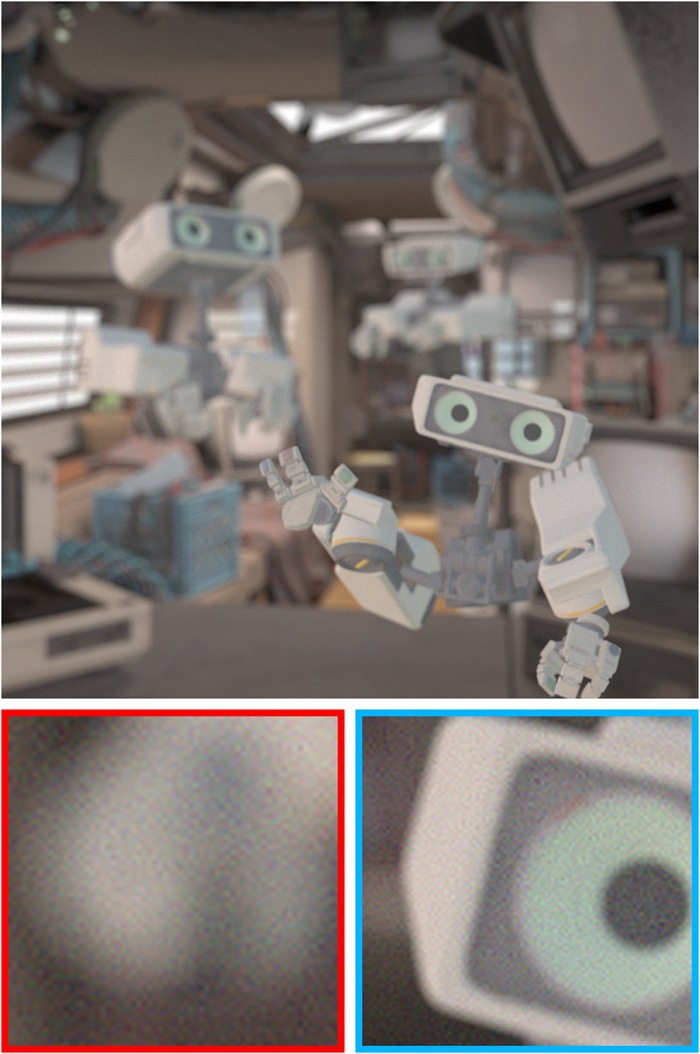} &
\includegraphics[width=\colw]{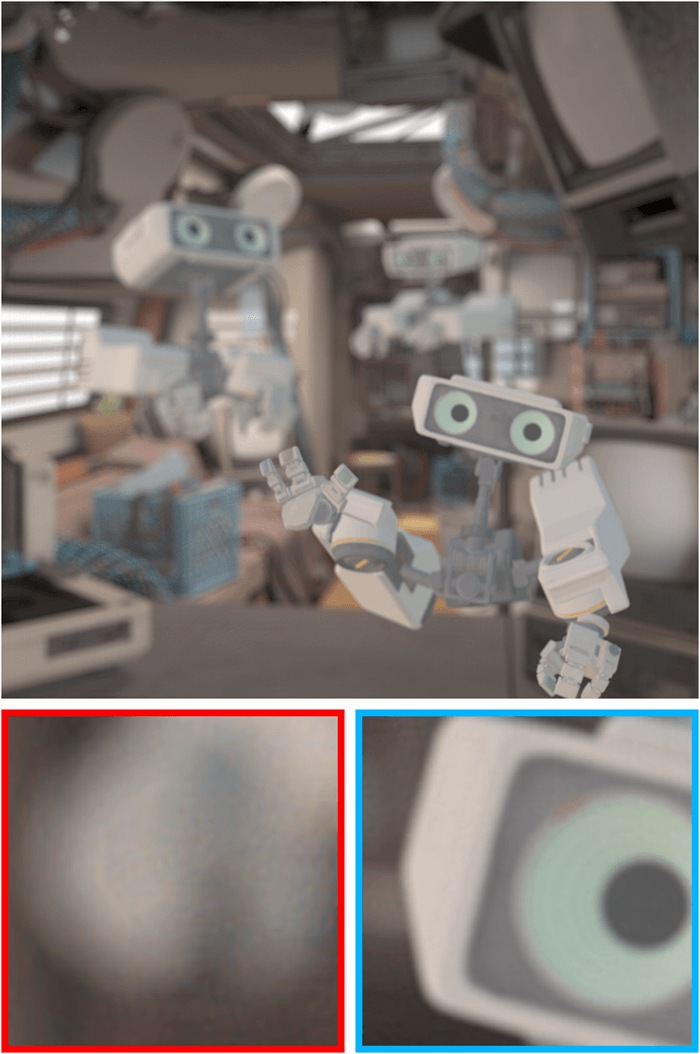} &
\includegraphics[width=\colw]{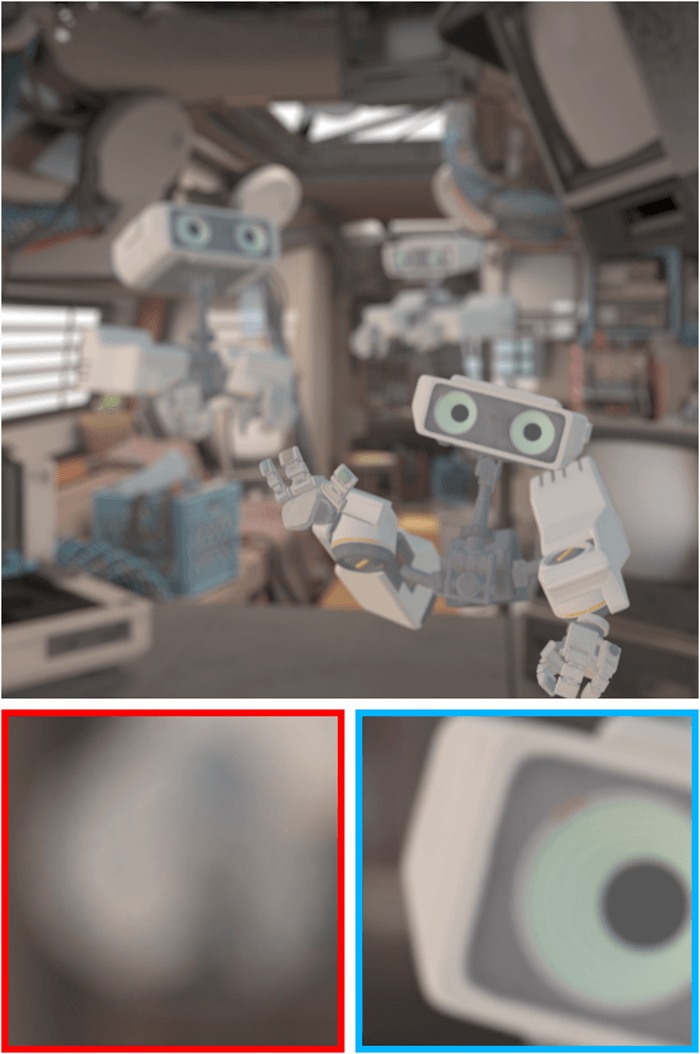} \\[0.0ex]

\rotatebox{90}{\parbox[c]{\colw}{\centering\titlesize Back Focus}} &
\includegraphics[width=\colw]{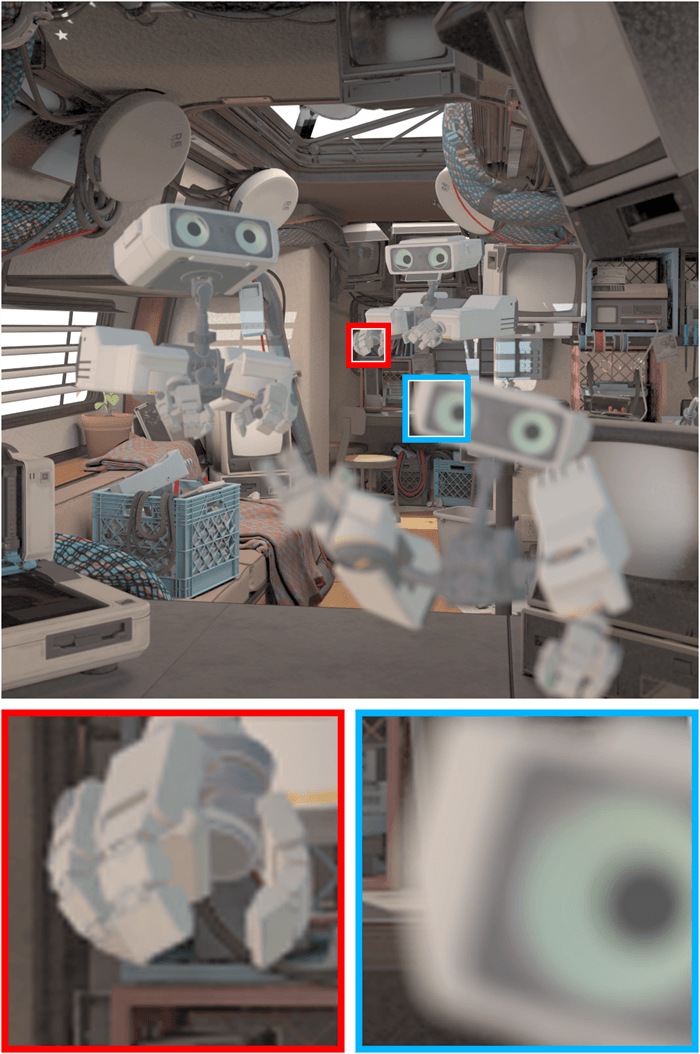} &
\includegraphics[width=\colw]{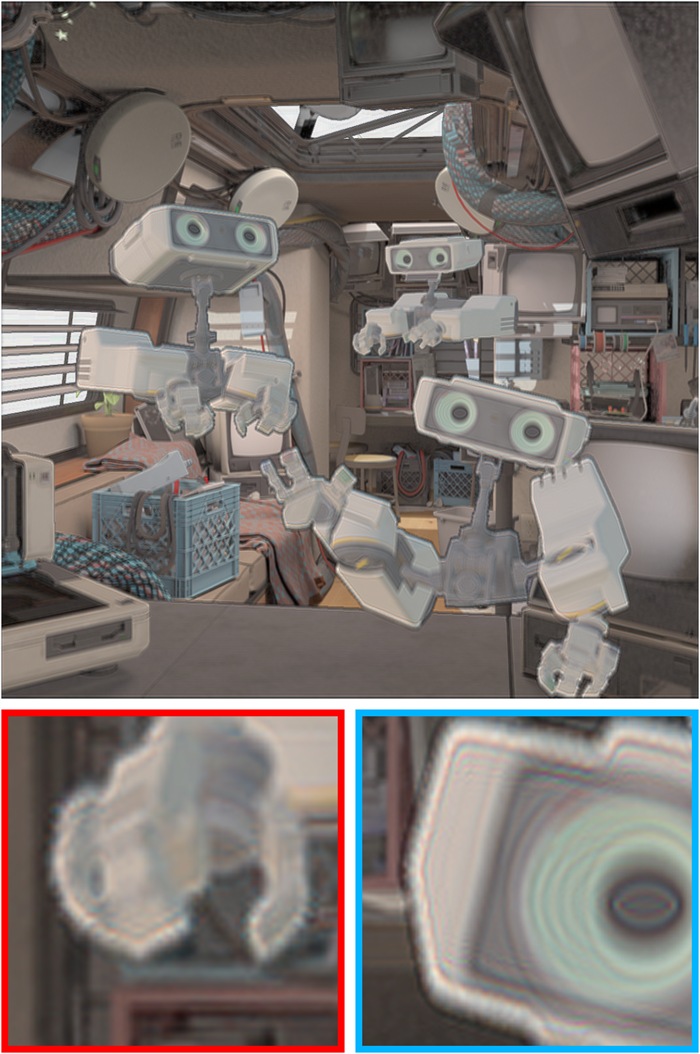} &
\includegraphics[width=\colw]{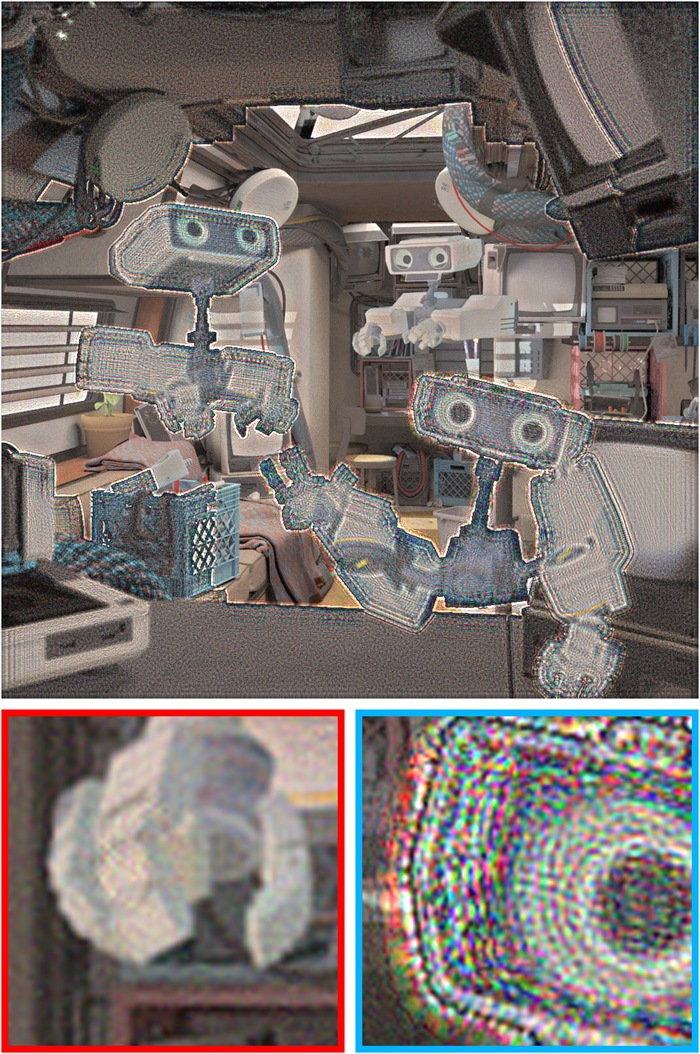} &
\includegraphics[width=\colw]{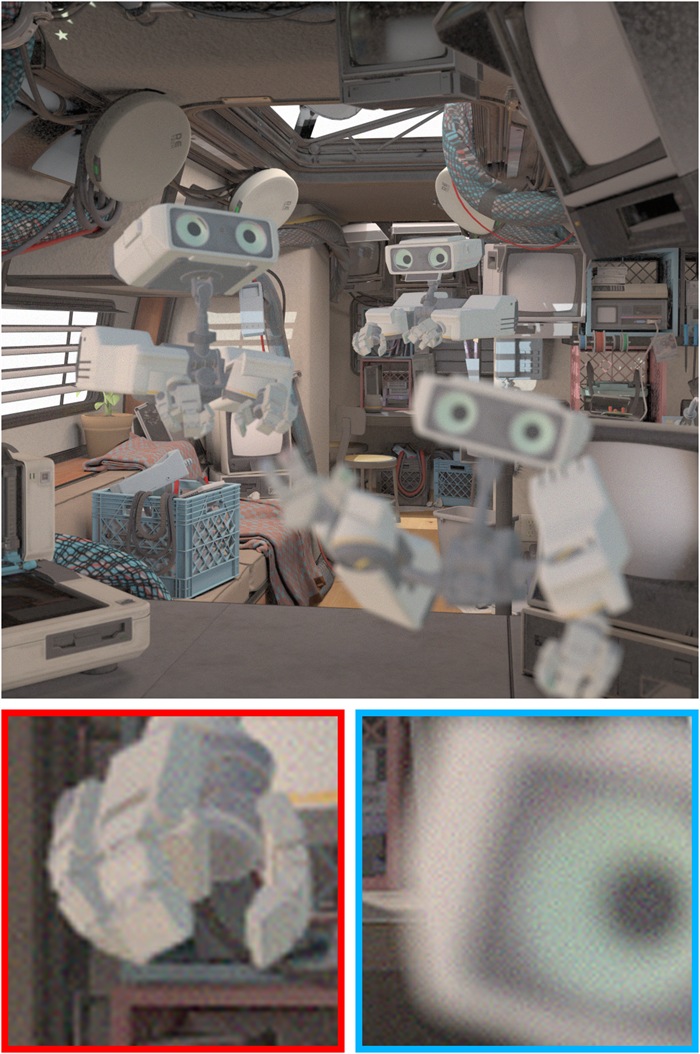} &
\includegraphics[width=\colw]{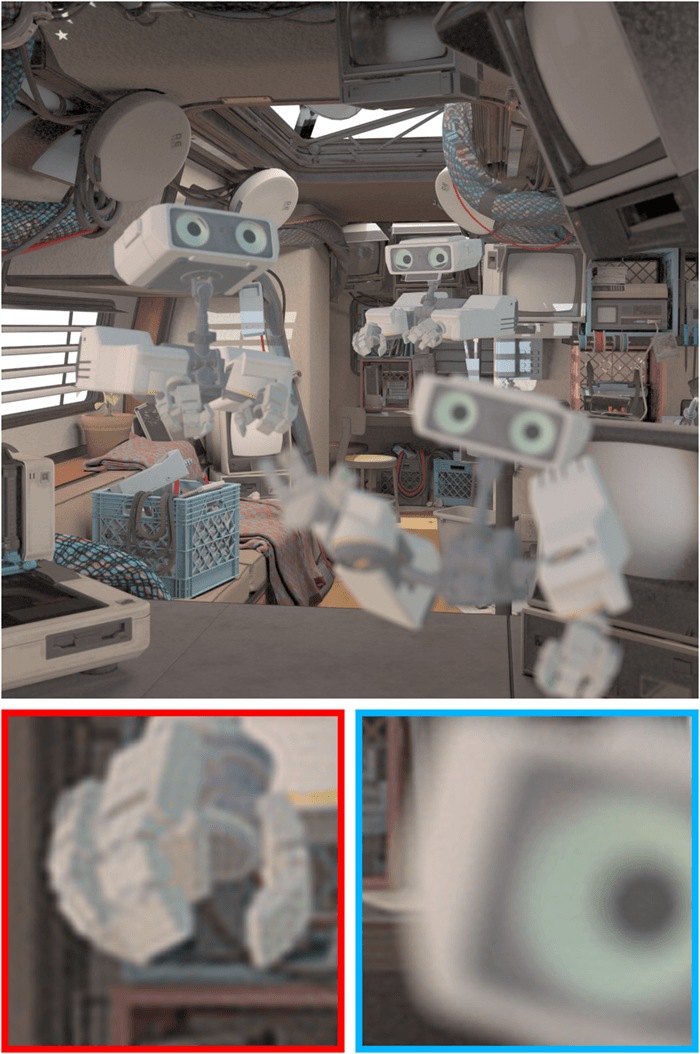} &
\includegraphics[width=\colw]{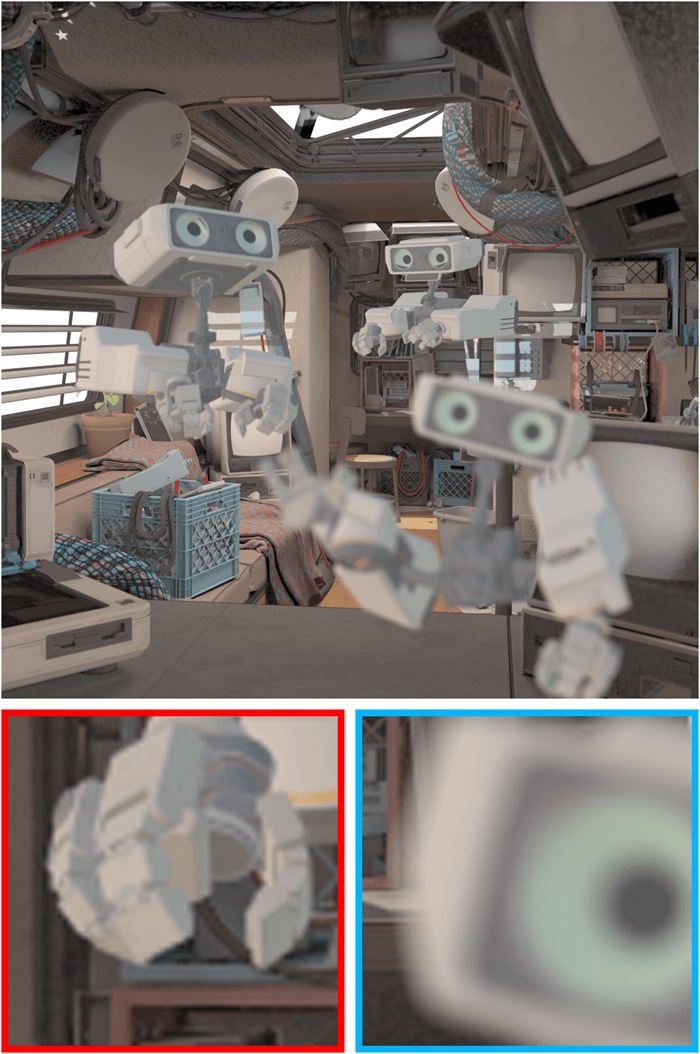} \\

\end{tabular}%
}
\vspace{-2mm}
    \caption{ 
    \textit{Synthetic Evaluation of 3D Holography Methods.}
    Compared to leading 3D holography techniques, including 
        Tensor holography \cite{shi2021towards}, neural 3D holography \cite{choi2021neural} and polarization multiplexing \cite{nam2023depolarized}, our ellipsography approach shows significant improvements in 3D reconstructions. 
    While Tensor Holography reduces speckle, it introduces unnatural defocus blur. Polarization multiplexing and Neural 3D Holography, both relying on random phase encoding, produce noisy in-focus reconstructions. In contrast, ellipsography---both with global (GA) and per-pixel (PA) analyzers---achieves clean, speckle-free reconstructions with sharp spatial details and natural defocus blur. Red and blue insets show zoomed-in regions corresponding to the front and back focal planes, respectively.
    }
    \label{fig:sim_3d_results}
\end{figure*}

\paragraph{Evaluation Metrics}
We use three metrics to evaluate the quality of reconstructed images from holograms: peak-signal-to-noise ratio (PSNR) as a per-pixel difference evaluator, the structural similarity index (SSIM) \cite{wang2004image} as a structural error metric and the learned perceptual image patch similarity (LPIPS) \cite{zhang2018perceptual} as a learned perceptual metric.
Specifically, LPIPS measures the perceptual difference between two images by computing distances in the feature space of a pre-trained VGG network \cite{simonyan2015very}, whose features are calibrated to reflect human perceptual similarity judgments.
A higher score is desired for PSNR and SSIM, whereas a lower score is desired for LPIPS. We evaluate our approach against existing 2D and 3D holography approaches, including both iterative and non-iterative methods. 
All metrics are computed on 100 randomly selected 2D images from the DIV2K dataset \cite{timofte2017ntire}, and 3D images from the Light field Saliency Dataset \cite{li2014saliency}.
Unlike prior approaches that rely on either region-of-interest (ROI) masking \cite{georgiou2008aspects,peng2020neural,chakravarthula2019wirtinger} or iris tuning \cite{chen2022off} to artificially suppress noise by dumping energy outside the target region or eyebox, we do not employ masking or post-processing tricks, thereby preserving full light efficiency in all reported evaluations.
Qualitative results for 2D and 3D reconstructions are shown in \Cref{fig:sim_2d_results,fig:sim_michelson_main,fig:sim_3d_results}, respectively.
Corresponding quantitative results are summarized in \Cref{tab:sim-metrics}.

\renewcommand{\arraystretch}{1.06}
\setlength{\tabcolsep}{1.2pt}
\begin{table}[ht]
\centering
\caption{Quantitative comparison of simulation results for 2D and 3D scenes using traditional methods and ellipsography. No tricks like region-of-interest (ROI) masking were applied to suppress image-plane noise.
}
\vspace{-2mm}
\label{tab:sim-metrics}
\begin{tabular}{|c|c|c|c|c|c|c|}

\hline
\multirow{2}{*}{Method} & \multicolumn{3}{c|}{2D}
 & \multicolumn{3}{c|}{3D} \\ \cline{2-7} 
 & PSNR$\uparrow$ & SSIM $\uparrow$& LPIPS $\downarrow$ & PSNR$\uparrow$ & SSIM $\uparrow$  & LPIPS $\downarrow$  \\ \hline
\begin{tabular}{c}
     Traditional Holo.   \\
     (Random Phase) 
\end{tabular} & 35.59  & 0.98 & 0.09 & 28.35 & 0.94 & 0.14 \\ \hline
 \begin{tabular}{c}
     Traditional Holo.   \\
     (Smooth Phase) 
\end{tabular} & 29.84 & 0.97&  0.19& 24.48 & 0.96 & 0.43 \\ \hline
\rowcolor{green!20}
\shortstack{
 \rule{0pt}{2.0ex}Ellipsography \\ (Global Analyzer)} & \textbf{52.63} & \textbf{0.99} & \textbf{1e-3} & \textbf{47.32} & \textbf{0.98} & \textbf{1e-3} \\ \hline
 \rowcolor{green!20}
 \shortstack{
\rule{0pt}{2.0ex}Ellipsography \\ (Per-pixel)} &  
\textbf{67.42} & \textbf{0.99} & \textbf{1.5e-4} & \textbf{52.94} & \textbf{0.99} & \textbf{1e-3} \\ \hline
\end{tabular}
\vspace{-5mm}
\end{table}

\paragraph{Results}
\Cref{tab:sim-metrics} presents the quantitative evaluation results using PSNR, SSIM, and LPIPS. The proposed ellipsography method achieves a substantial performance gain, outperforming all baseline methods by over 30 dB for 2D holograms and 25 dB for 3D holograms in PSNR. 
This reflects a significant reduction in reconstruction error, exceeding at least \textit{two orders of magnitude} over conventional scalar holography techniques. 
Most importantly, even ellipsography with a global-analyzer shows improvements of over 20 dB PSNR for both 2D and 3D reconstructions.
In addition to pixel-wise accuracy, ellipsography also shows superior performance on perceptual metrics. Both SSIM and LPIPS consistently indicate that ellipsography yields reconstructions with higher structural fidelity and closer perceptual similarity to the ground truth. 

\Cref{fig:sim_2d_results} shows simulation results comparing ellipsography to existing baselines, including the double phase amplitude coding \cite{maimone2017holographic}, neural holography \cite{peng2020neural}, and depolarized holography \cite{nam2023depolarized}. 
As highlighted in the purple-bordered insets, ellipsography is the only method that produces speckle-free reconstructions with accurate texture details across all methods. All baseline methods exhibit residual interference artifacts, reflecting the inherent limitation of scalar or incoherently summed polarization channels.

\Cref{fig:sim_michelson_main} compares ellipsography against dual-modulation methods with comparable degrees of freedom, including the michelson holography (with two phase-only modulators) \cite{choi2021optimizing}, polarization multiplexing (with phase-only and metasurface modulators) \cite{nam2023depolarized} and complex modulation (with combined amplitude and phase modulators) \cite{kabuli2025high}. 
These methods continue to exhibit structured interference artifacts, as shown in the first row error map insets. In contrast, ellipsography with a global analyzer achieves substantial error reduction, while the per-pixel analyzer configuration yields near-perfect, speckle-free reconstruction.
This result demonstrates that the performance gains of ellipsography arises from explicitly structuring the interference. 
As shown in the second row, targets like color squares amplify interference artifacts in existing approaches including complex (joint amplitude and phase) modulation, whereas ellipsography consistently reconstructs uniform intensity fields with minimal variance, confirming its robustness to coherent noise beyond baselines.

\Cref{fig:sim_3d_results} further evaluates depth-dependent reconstruction quality for 3D holography, with zoomed-in regions corresponding to the front focal plane (red box) and the back focal plane (blue box).
State-of-the-art techniques such as neural 3D holography \cite{choi2021neural} and depolarized holography \cite{nam2023depolarized} introduce noticeable speckle noise in both in-focus and out-of-focus regions. 
Particularly, these methods exhibit pronounced high-frequency intensity fluctuations and granular artifacts that obscure details (see the circular features of the robot's eye) and depict incorrect focus cues.
Tensor holography \cite{shi2021towards}, which employs smooth-phase heuristics, reduces speckle but suffers from visibly unnatural defocus characteristics and ringing artifacts.
In contrast, ellipsography delivers speckle-free reconstructions with high spatial fidelity and perceptually realistic defocus blur. 
This is evident in the blur insets below each reconstruction: ellipsography produces smooth, symmetric defocus profiles, whereas other methods exhibit noisy or irregular patterns indicative of residual coherent interference.
These results confirm that 
our method consistently achieves significantly higher reconstruction fidelity, both visually and quantitatively. 

\subsection{Speckle Analysis}
\label{sec:sc}

\begin{figure*}[h!]
	\centering
        \includegraphics[width=1.0\linewidth]{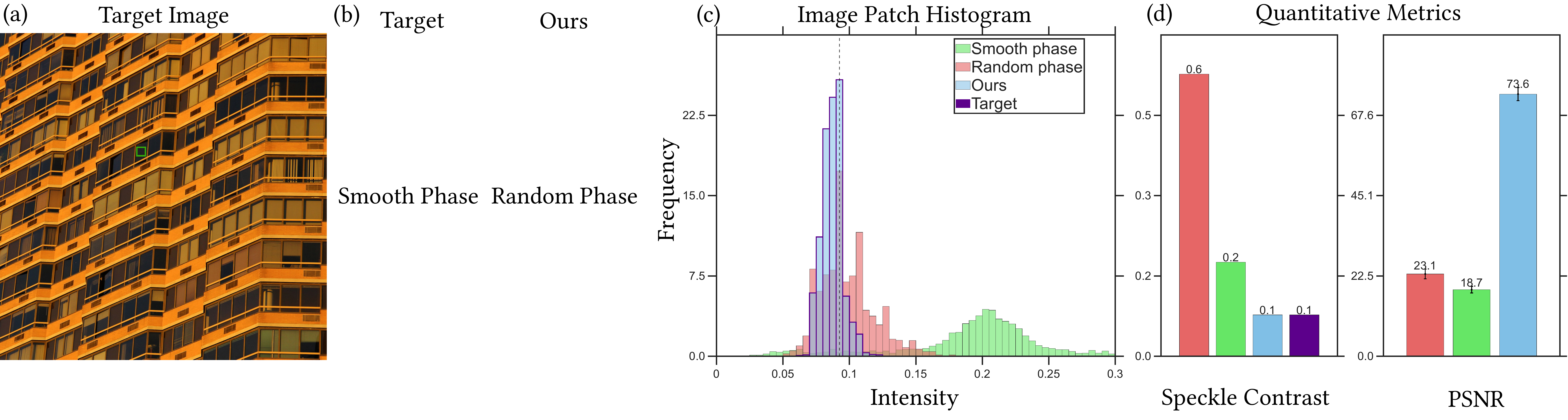}
        \vspace{-6mm}
 		\caption{\textit{Synthetic Validation of Speckle Suppression.} We compare traditional holography with smooth and random phase, and our proposed ellipsography method using random phase. The green box indicates the region used for computing speckle contrast and histogram statistics. While smooth-phase holography reduces speckle at the cost of image sharpness, and random-phase holography preserves detail but suffers from strong noise, ellipsography achieves both low speckle contrast and high fidelity. The histogram (middle) shows that only ellipsography closely matches the target intensity distribution, with the most frequent intensity of the ground truth image marked by
a dashed black line. Speckle contrast and PSNR metrics confirm the superior reconstruction quality and noise suppression enabled by polarization-guided interference control. 
        }
	\label{fig:speckle_contrast_sim}
\end{figure*}

\paragraph{Evaluation Metric}
To quantitatively assess speckle suppression, we use speckle contrast (SC)---a standard metric that captures the relative magnitude of intensity fluctuations in a speckle pattern.
It is defined as the ratio of the standard deviation ($\sigma_I$) to the mean intensity ($\langle I \rangle$) of the speckle:
\begin{equation}
    \textit{SC}=\frac{\sigma_{I}}{\langle I \rangle}.
\end{equation}
SC values range from 0 to 1, with higher values indicating stronger speckle visibility (\eg fully developed speckle) and lower values indicating more effective noise suppression. 
To avoid structural bias, SC is computed over a uniform-intensity region of the image, as indicated by the green box in \Cref{fig:speckle_contrast_sim}(a).

\begin{figure*}[t]
	\centering
		\includegraphics[width=0.97\linewidth]{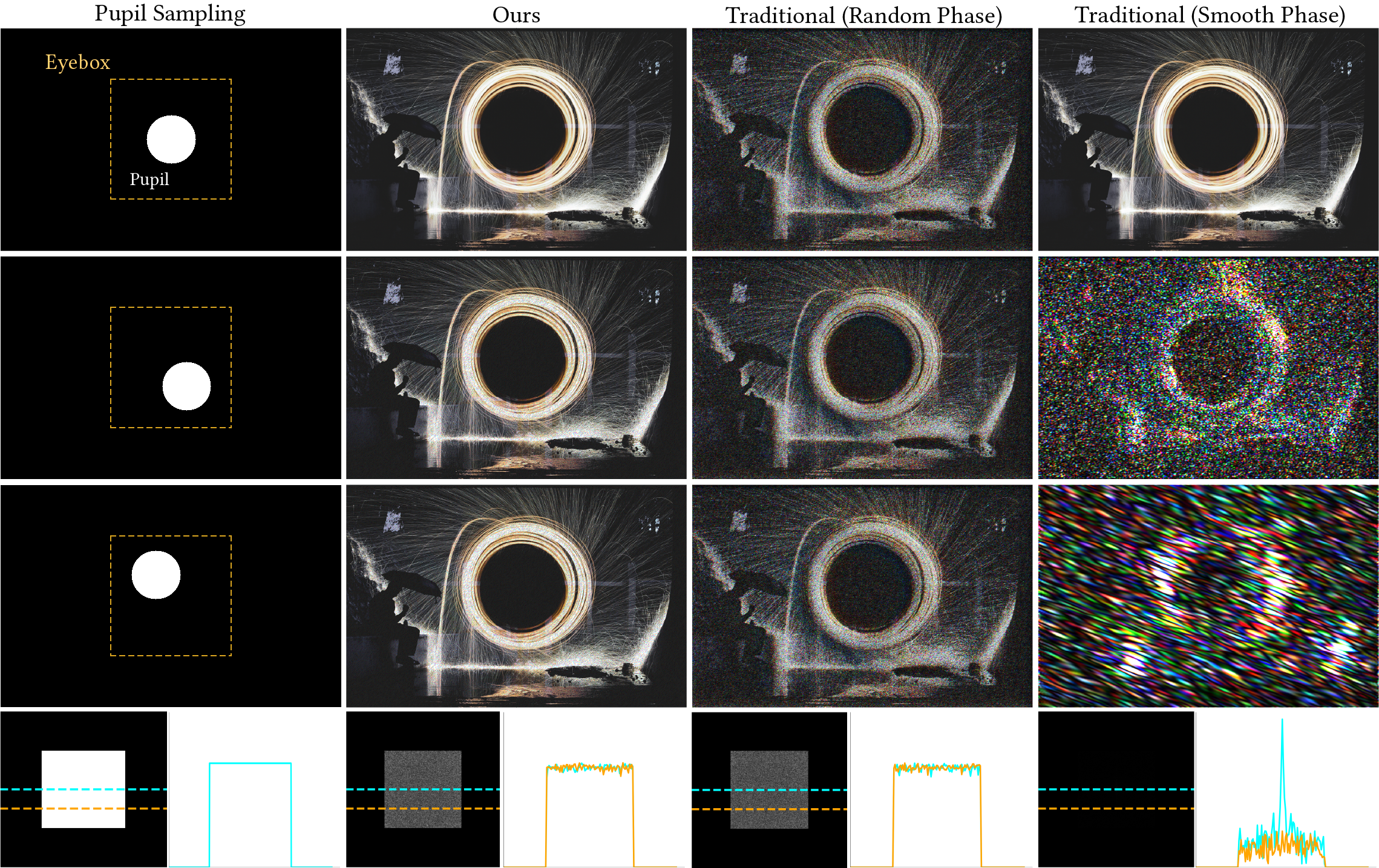}
        \vspace{-2mm}
		\caption{\textit{Pupil Invariance.} We present reconstructions at three distinct pupil positions within the eyebox in simulation. Smooth phase holograms produce high quality images when the pupil is centered; however, image quality deteriorates significantly 
        when pupil shifts away from the center. Random phase holograms exhibit uniform energy distribution across the eyebox but suffer from speckle artifacts, reducing overall image clarity. 
        Our ellipsography method achieves high quality image reconstruction that remains invariant to pupil position, making it a more desirable solution for practical displays. 
        } 
		\label{fig:pupil-sampling}
\end{figure*}

\setlength{\tabcolsep}{1.0pt}
\renewcommand{\arraystretch}{0.6}

\begin{figure*}[t]
\centering
\resizebox{\textwidth}{!}{%
\begin{tabular}{ccccc}

\multicolumn{1}{c}{\fontsize{8}{9}\selectfont Target} &
\multicolumn{1}{c}{\fontsize{8}{9}\selectfont DPAC} &
\multicolumn{1}{c}{\fontsize{8}{9}\selectfont CITL Optimization} &
\multicolumn{1}{c}{\fontsize{8}{9}\selectfont Ellipsography (GA)} &
\multicolumn{1}{c}{\fontsize{8}{9}\selectfont Ellipsography (PA)} \\
[0.1ex]

\includegraphics[width=3.6cm]{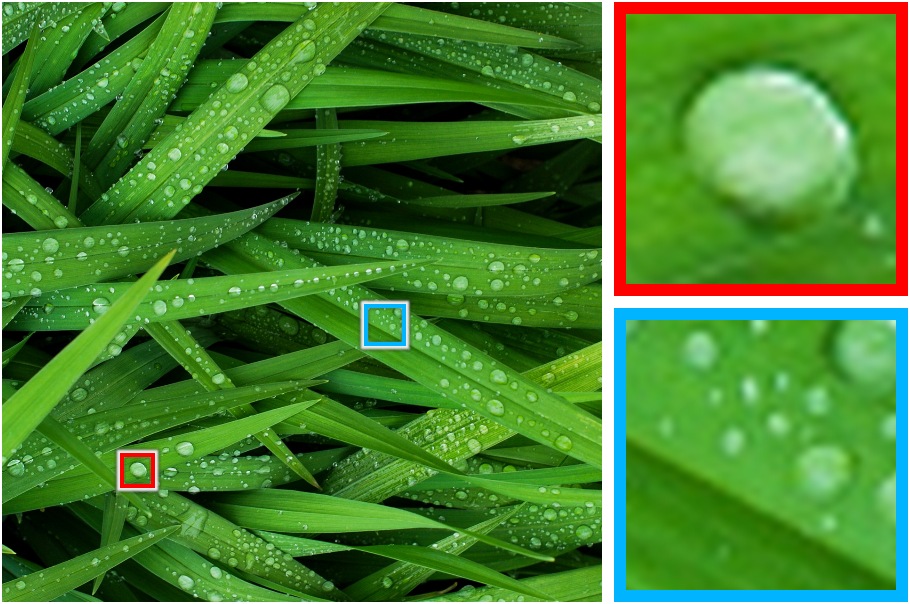} &
\includegraphics[width=3.6cm]{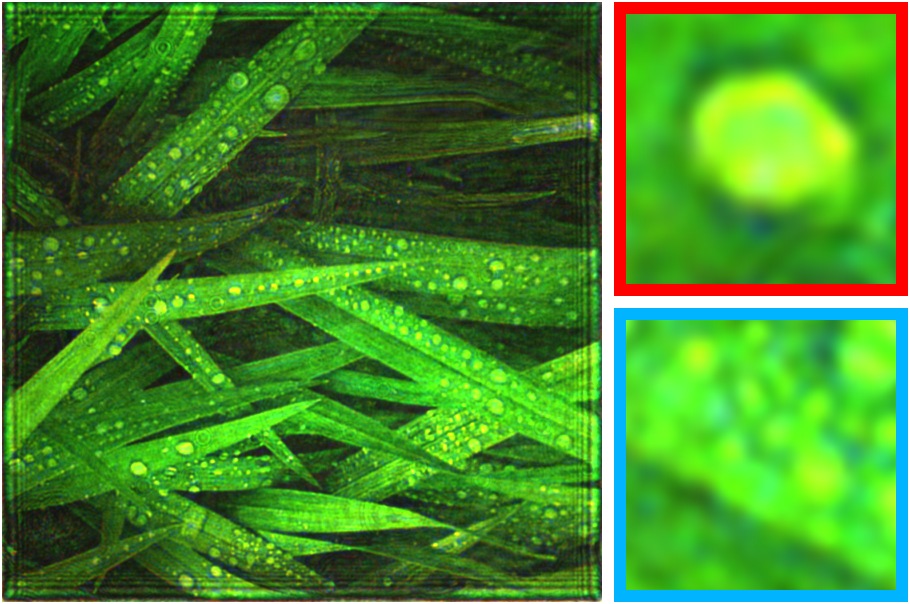} &
\includegraphics[width=3.6cm]{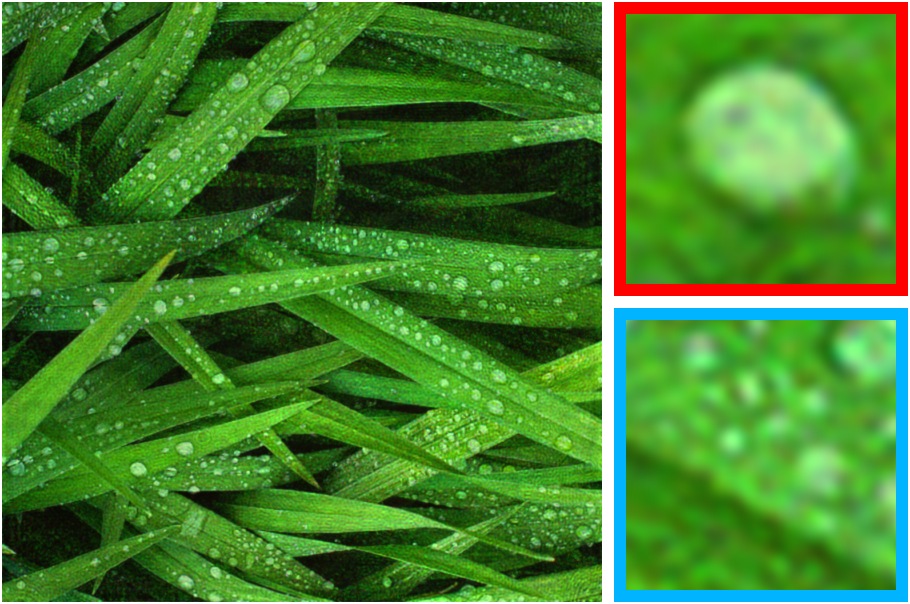} &
\includegraphics[width=3.6cm]{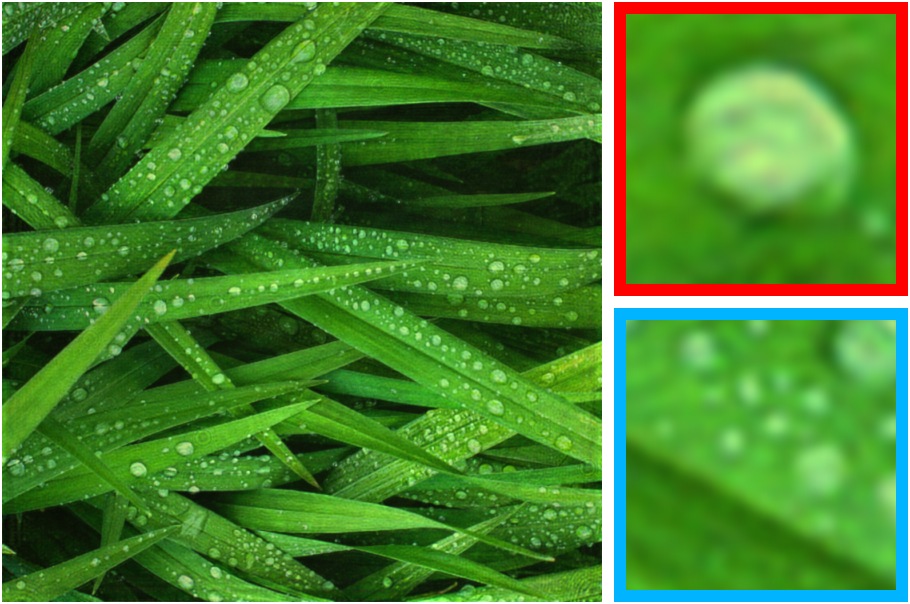} &
\includegraphics[width=3.6cm]{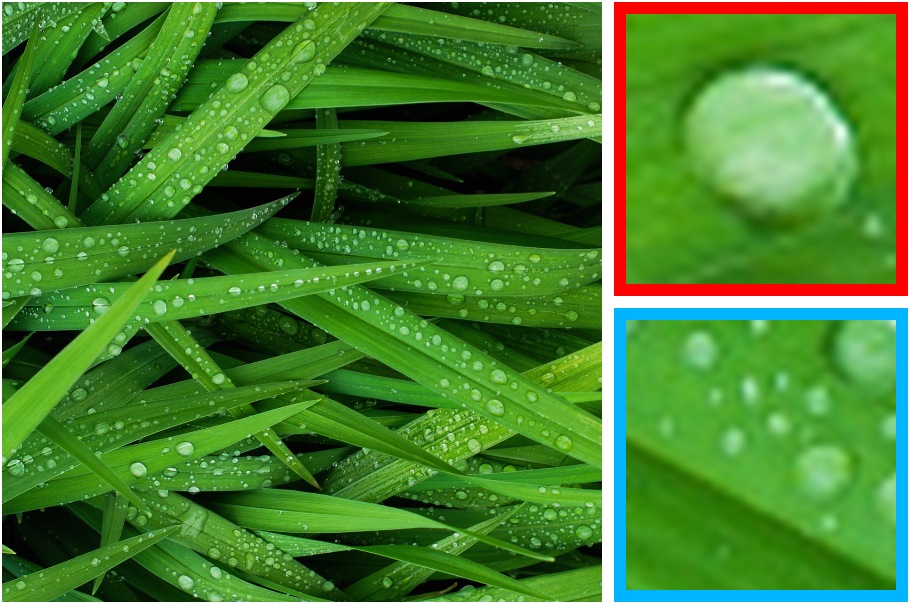} \\
[0.0ex]

\includegraphics[width=3.6cm]{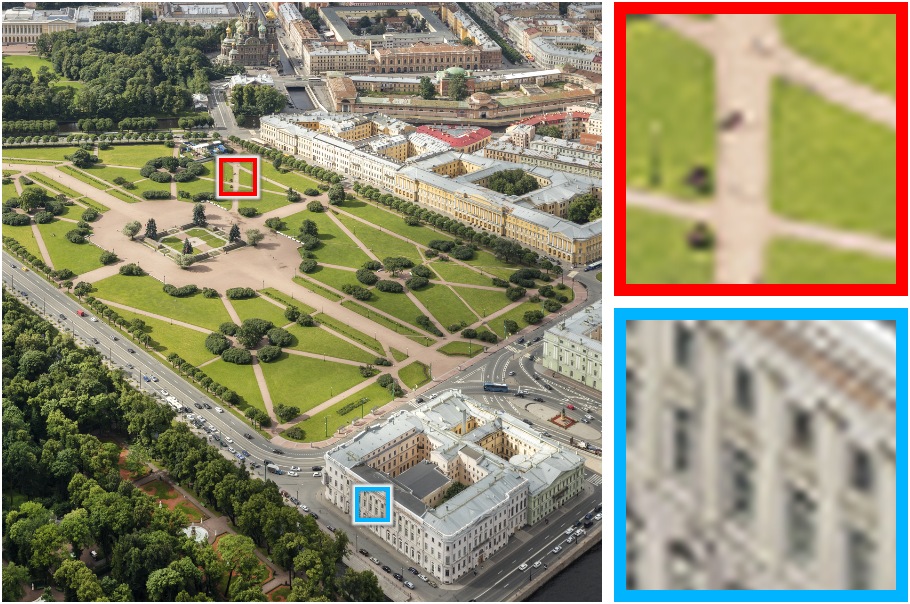} &
\includegraphics[width=3.6cm]{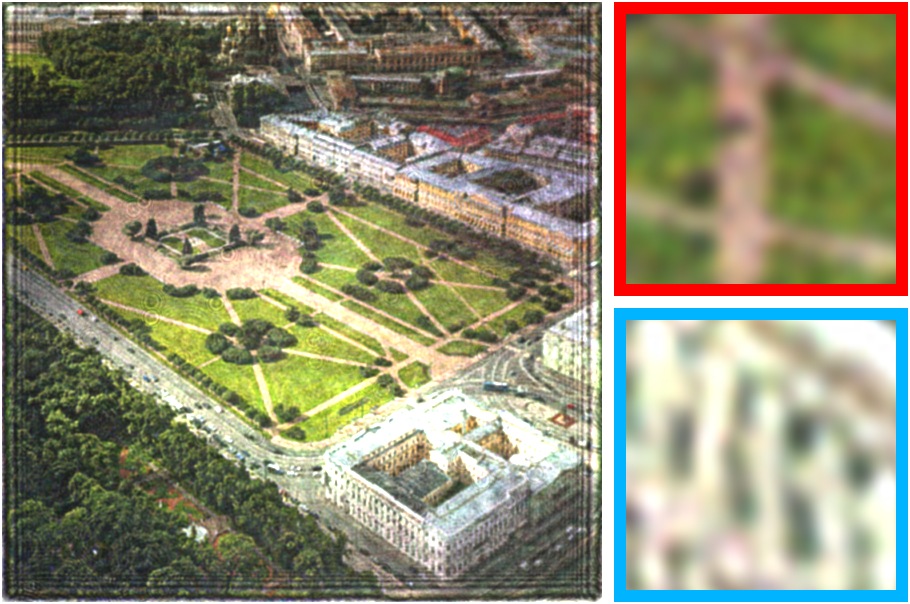} &
\includegraphics[width=3.6cm]{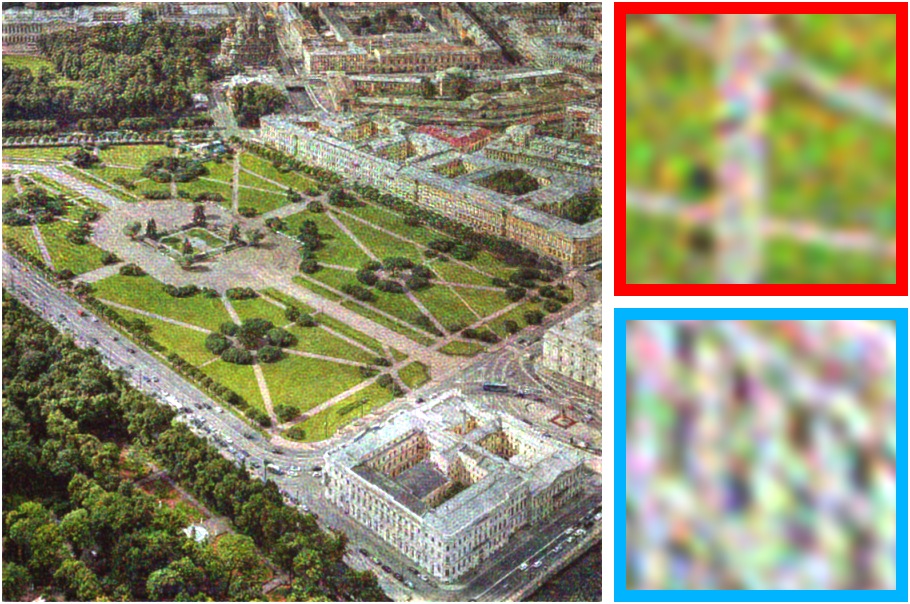} &
\includegraphics[width=3.6cm]{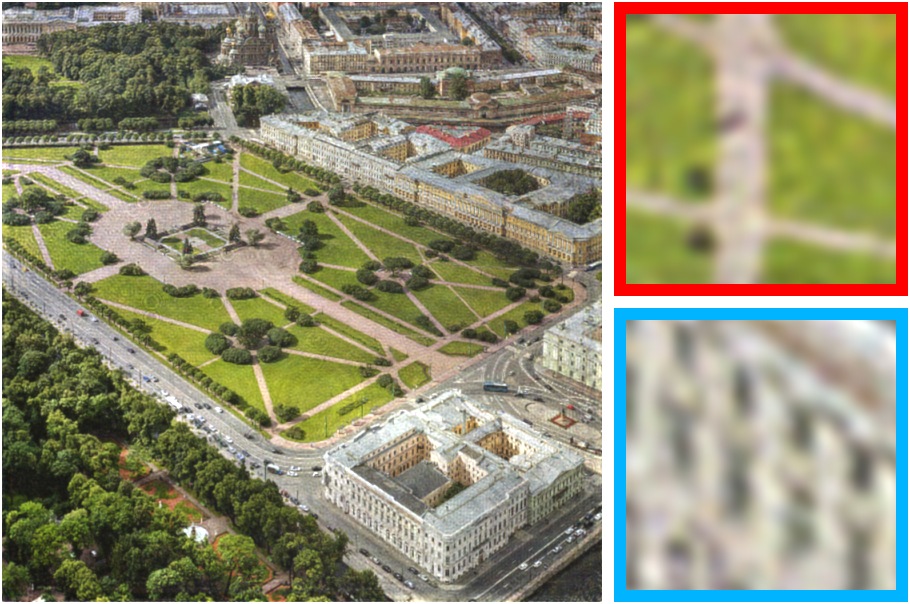} &
\includegraphics[width=3.6cm]{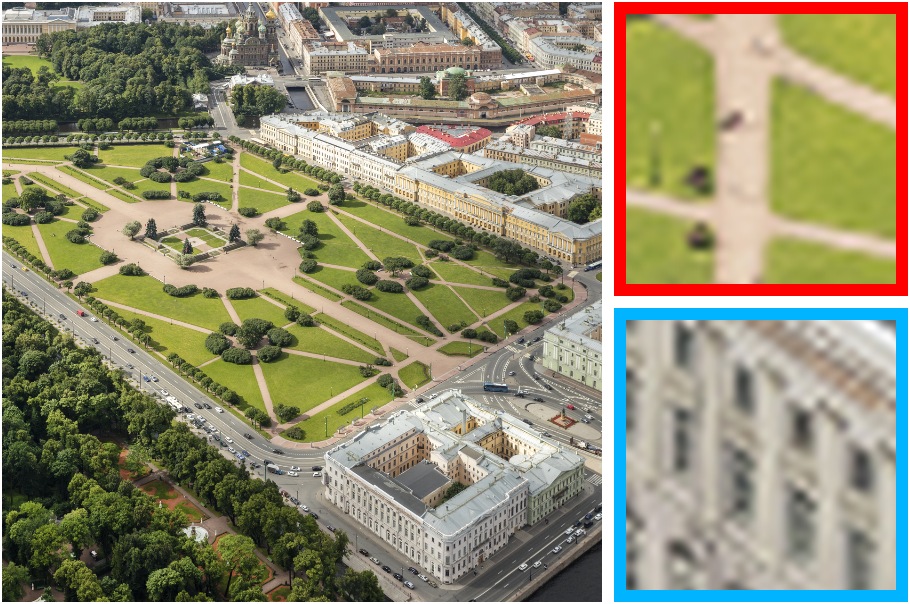} \\
[0.0ex]

\includegraphics[width=3.6cm]{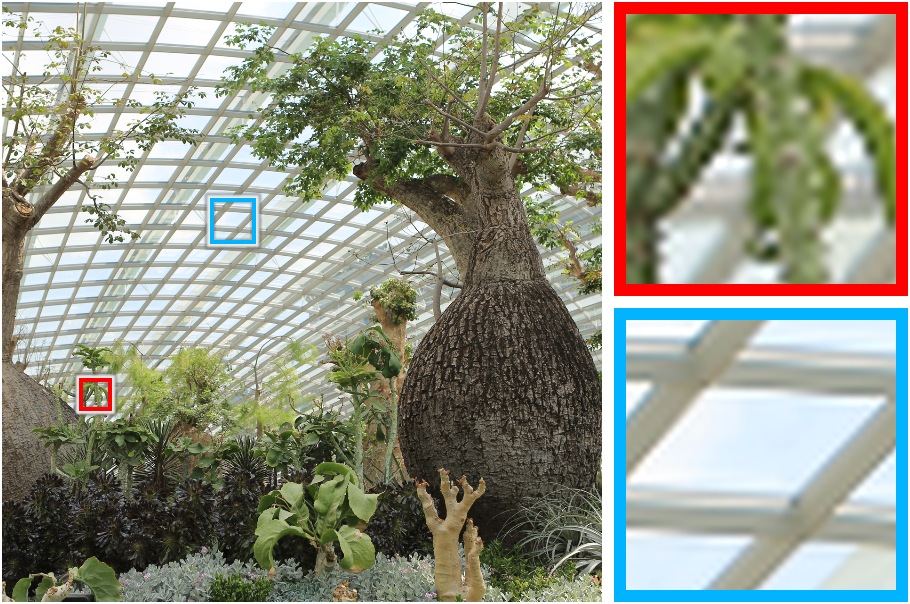} &
\includegraphics[width=3.6cm]{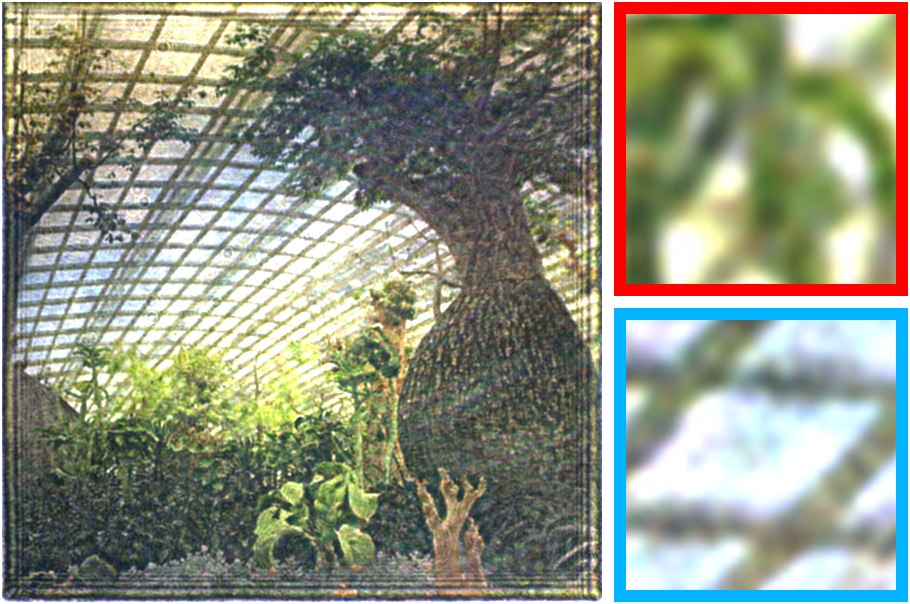} &
\includegraphics[width=3.6cm]{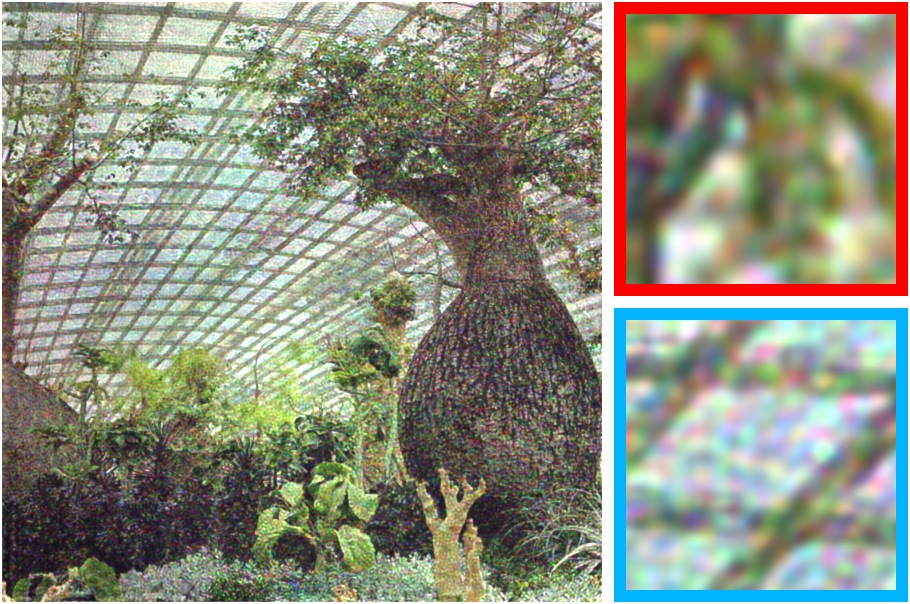} &
\includegraphics[width=3.6cm]{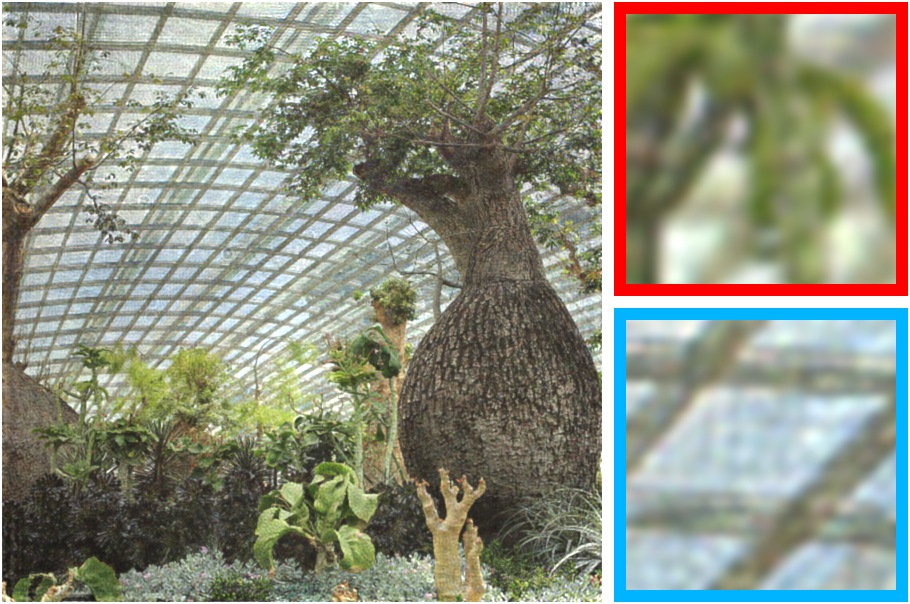} &
\includegraphics[width=3.6cm]{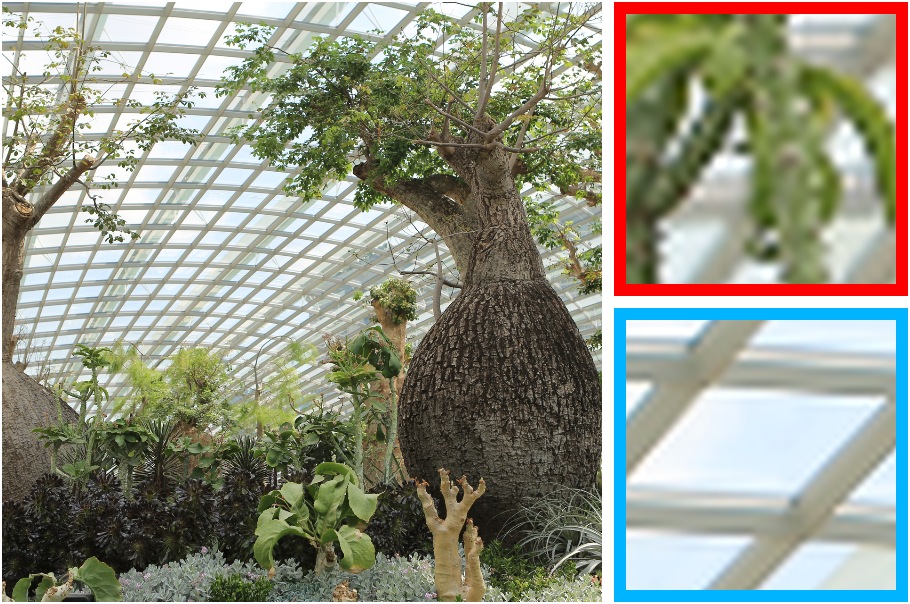} \\
[0.0ex]

\end{tabular}%
}
\vspace{-2mm}
		\caption{\textit{Experimental Evaluation of 2D Holography}.
Our ellipsography approach---both with global (GA) and per-pixel (PA) analyzers---produces high-resolution speckle-free images from a single random phase hologram.
Traditional holograms generated using camera-in-the-loop (CITL) optimization \cite{peng2020neural} exhibit visible speckle artifacts due to their inherent reliance on random phase, while smooth phase holograms from double phase amplitude coding (DPAC) \cite{maimone2017holographic} reduce noise at the cost of spatial detail.
In contrast, ellipsography simultaneously preserves fine image features and suppresses noise, without requiring iterative refinement or multi-frame averaging. 
        } 
		\label{fig:2d_captures}
\label{fig:2d_exp_main}
\end{figure*}

\paragraph{Results and Discussion}
\Cref{fig:speckle_contrast_sim} presents both qualitative and quantitative comparisons between traditional holography with both smooth and random phase, and ellipsography under random-phase conditions.
In traditional holography, smooth phase holograms reduce speckle more effectively than random phase holograms, but at the cost of resolution and natural focus cues. 
In contrast, ellipsography delivers superior speckle suppression without sacrificing phase randomness (see zoom-ins in \Cref{fig:speckle_contrast_sim}(b)), overcoming the longstanding trade-off between random phase modulation and image quality. 
To further evaluate reconstruction fidelity, we compare histograms computed from the region specified by the green box in the target and reconstructed images. 
As shown in \Cref{fig:speckle_contrast_sim}(c), both smooth and random phase holograms yield intensity histograms that deviate substantially from the ground truth. 
In comparison, ellipsography produces the sharpest and most accurate intensity distribution, closely matching the target. 
We indicate the peak of the target distribution with a dashed line, for reference. 
This trend is further supported by PSNR measurements in \Cref{fig:speckle_contrast_sim}(d), where ellipsography outperforms baselines, achieving an improvement of up to 50 dB. 
Together, the SC values, PSNR scores, and histogram alignment validate
the effectiveness of ellipsography in structuring vectorial interference for speckle suppression.

\subsection{Pupil Sampling and Invariance}
\label{subsec:pupil-invariance}

For holographic displays to be practical, they must support a large eyebox that exceeds the typical pupil size in order to accommodate natural, unrestricted eye movements without visual dropout. However, prior work has shown that when the pupil samples only a subset of the eyebox, significant image degradation and visual artifacts can arise \cite{chakravarthula2022pupil}.

\Cref{fig:pupil-sampling} presents pupil sampling experiments for both traditional holography and ellipsography. Importantly, no pupil-aware loss functions \cite{chakravarthula2022pupil} or adaptive sampling strategies \cite{schiffers2023stochastic} were employed in any of the experiments. 
Our simulations demonstrate that ellipsography maintains uniform image quality across the entire eyebox without requiring pupil localization or any form of pupil-aware optimization. This robustness effectively mitigates the eyebox-dependent artifacts commonly observed in conventional holographic displays.

Traditional smooth-phase holograms produce clean, low-noise images when the pupil samples the central region of the eyebox, where optical energy is most concentrated. 
As the pupil moves away from the center, however, image brightness and quality degrade rapidly, resulting in dim, noisy, and incomplete reconstructions (see the last column of \Cref{fig:pupil-sampling}). 
In contrast, random-phase holograms distribute optical energy more uniformly across the eyebox but suffer from persistent speckle noise regardless of pupil position.

Ellipsography overcomes both limitations, producing high-fidelity reconstructions across the full eyebox independent of pupil location. \textit{These results highlight the potential of ellipsography to enable practical, large {\'e}tendue holographic displays that support natural eye movements without compromising image quality.}

\section{Experimental Evaluation}
\label{sec:exp-eval}

Here, we experimentally validate the performance of ellipsography using a benchtop display prototype. We evaluate key performance characteristics, including image quality, speckle suppression, and comparisons against existing state-of-the-art baseline methods.

\subsection{Image Quality Assessment}

\setlength{\tabcolsep}{1.5pt}
\renewcommand{\arraystretch}{0.6}
\begin{figure*}[t]
\begin{center}
\resizebox{\textwidth}{!}{%
\begin{tabular}{cccccccc}

\multicolumn{2}{c}{\fontsize{11}{11}\selectfont Target} &
\multicolumn{2}{c}{\fontsize{11}{11}\selectfont Traditional Holography} &
\multicolumn{2}{c}{\fontsize{11}{11}\selectfont Ellipsography (GA)} &
\multicolumn{2}{c}{\fontsize{11}{11}\selectfont Ellipsography (PA)} \\[0.5ex]

\includegraphics[width=2.92cm]{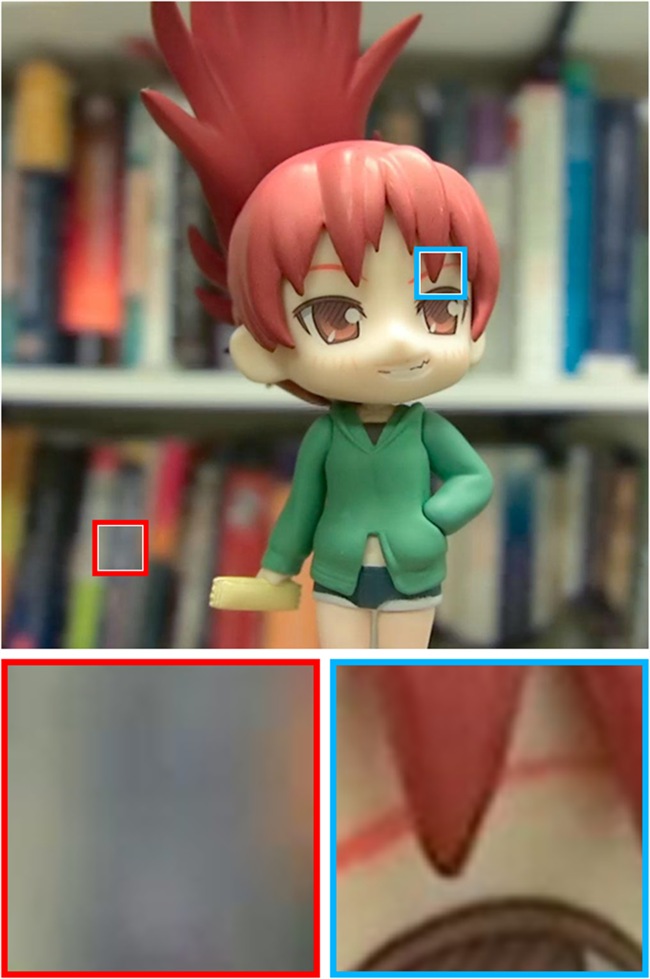} &
\includegraphics[width=2.92cm]{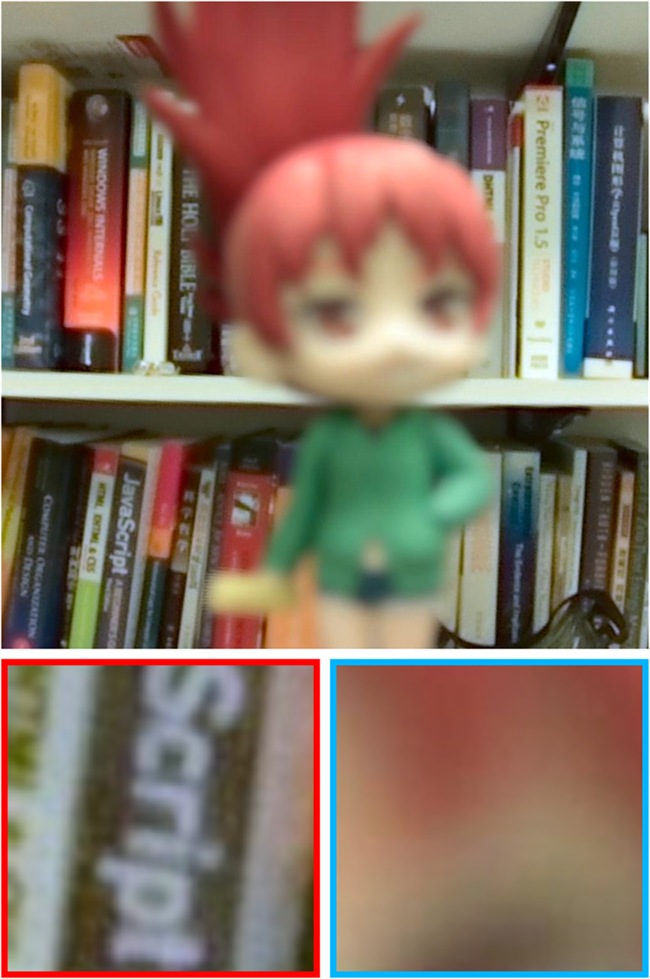} &
\includegraphics[width=2.92cm]{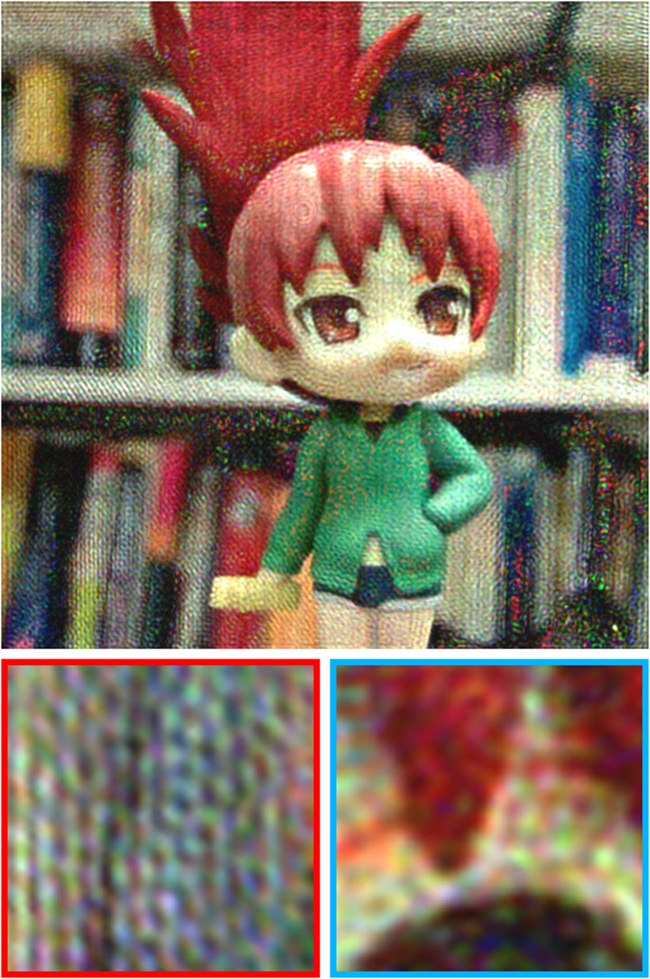} &
\includegraphics[width=2.92cm]{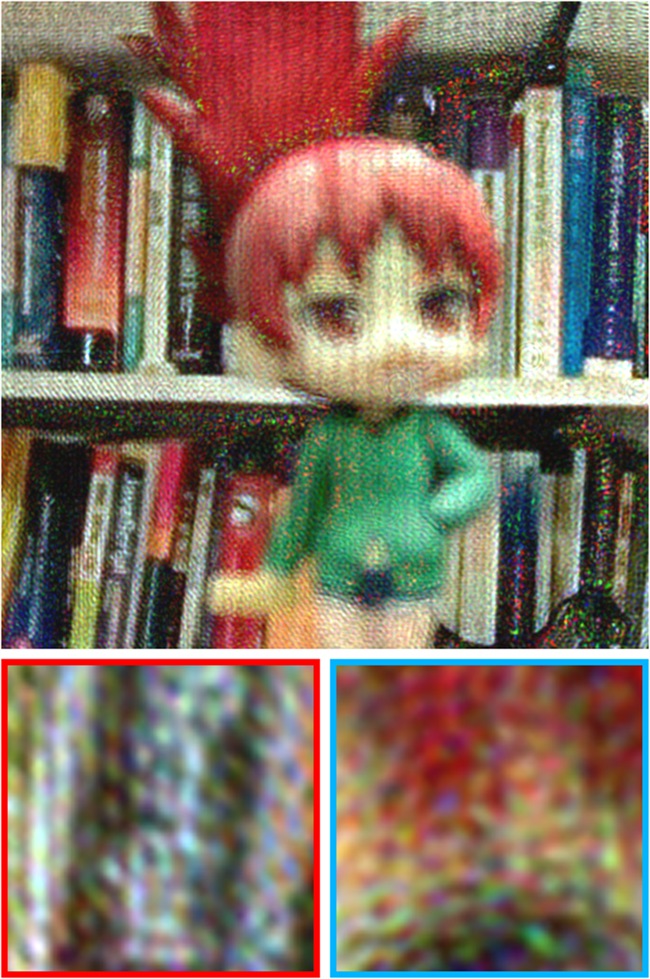} &
\includegraphics[width=2.92cm]{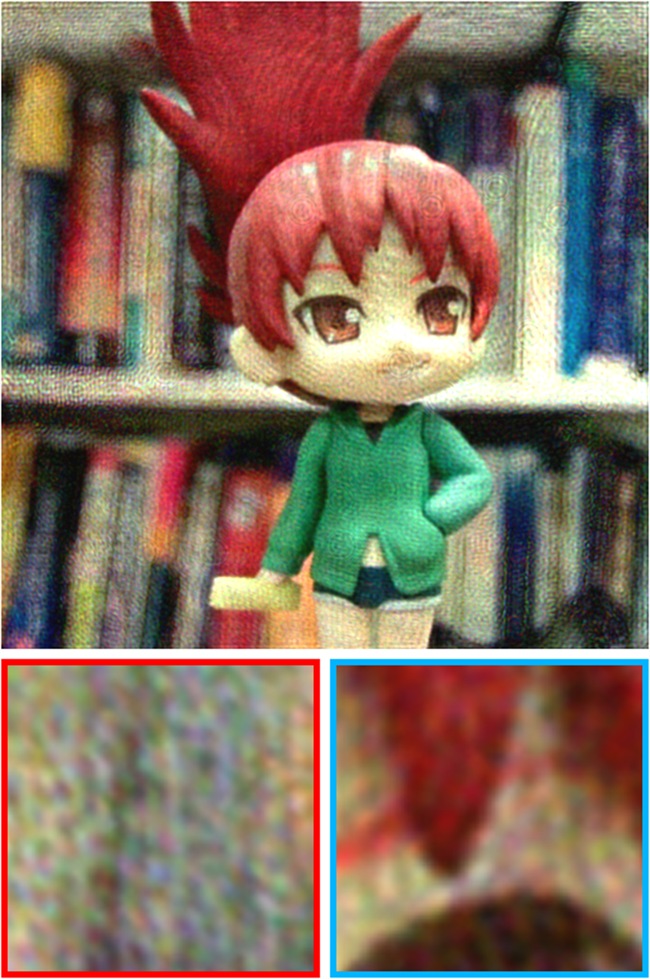} &
\includegraphics[width=2.92cm]{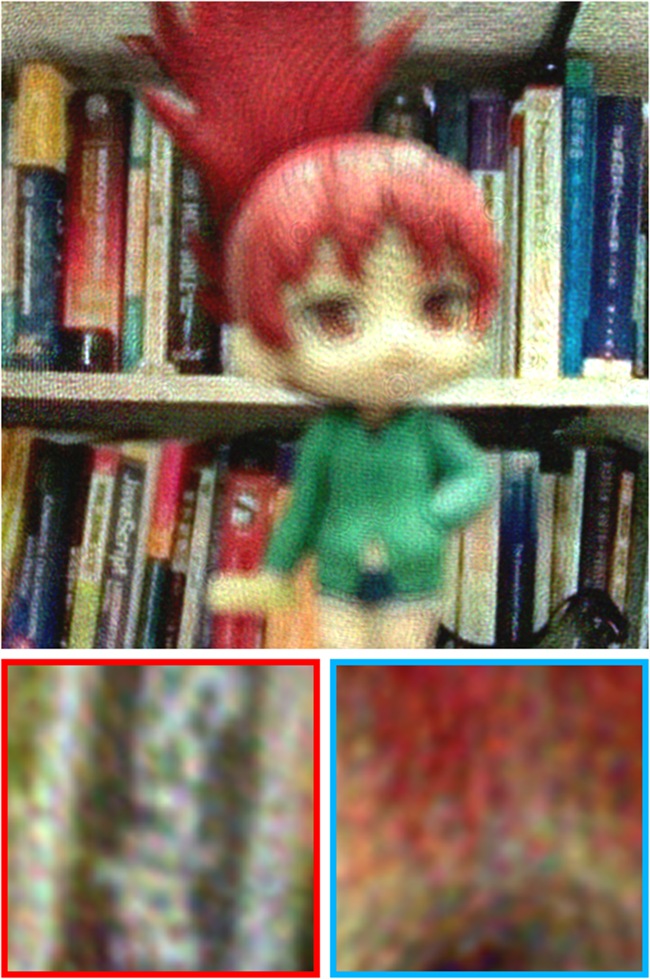} &
\includegraphics[width=2.92cm]{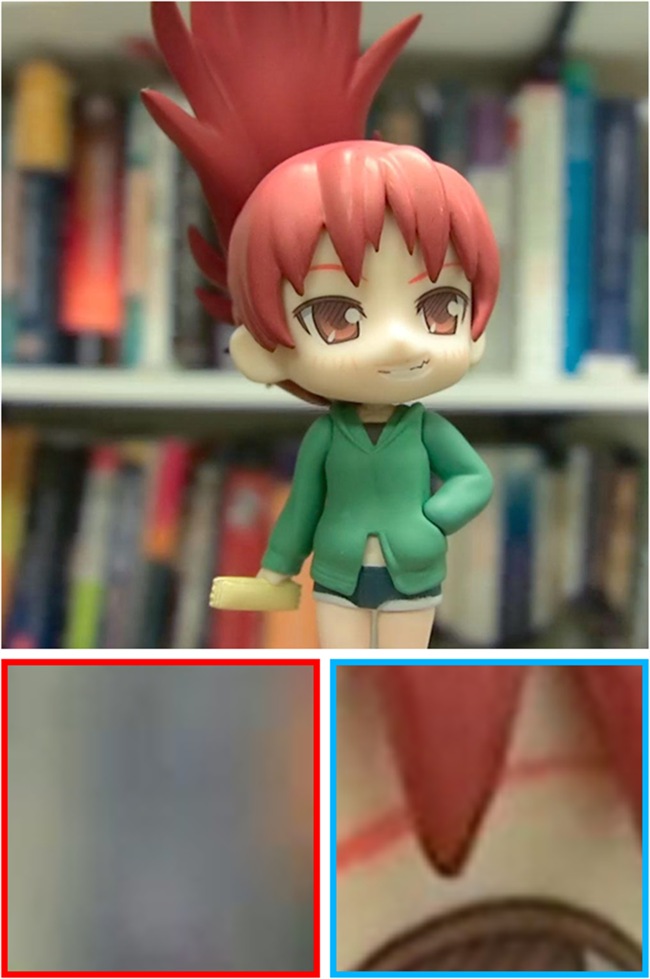} &
\includegraphics[width=2.92cm]{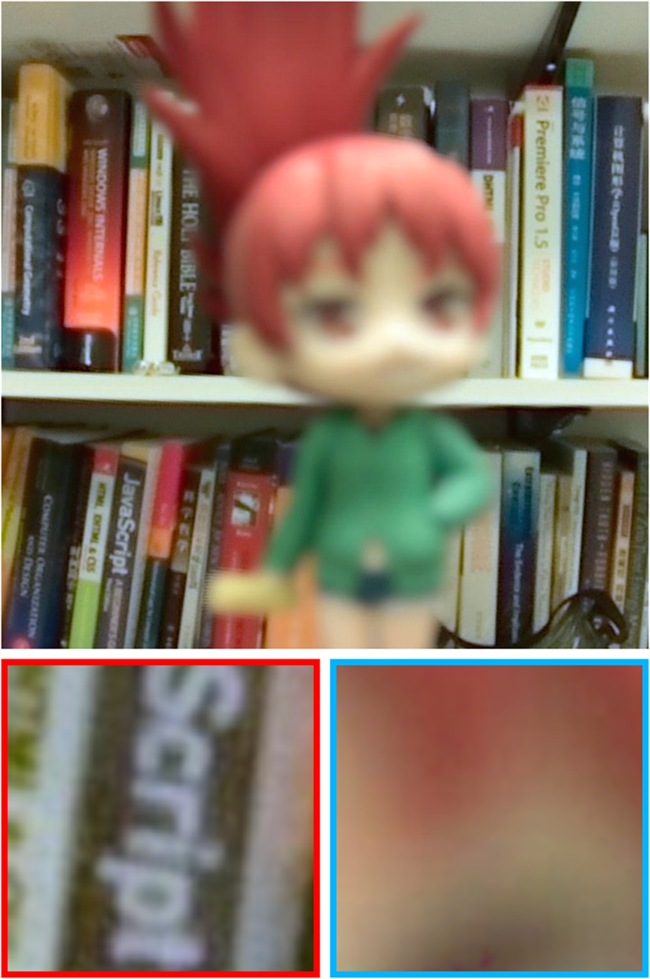} \\[0.8ex]

\includegraphics[width=2.92cm]{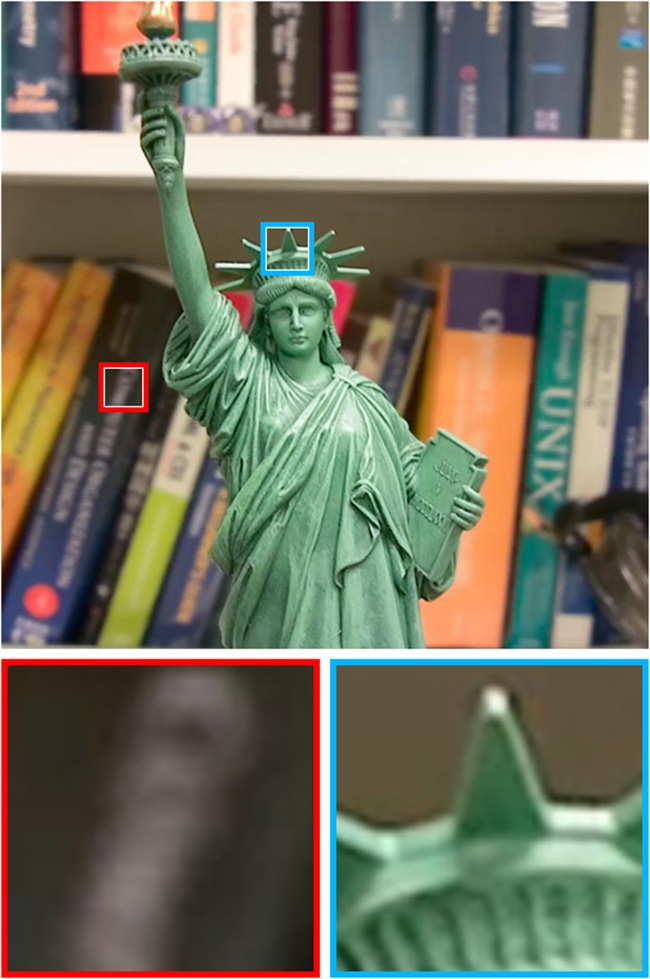} &
\includegraphics[width=2.92cm]{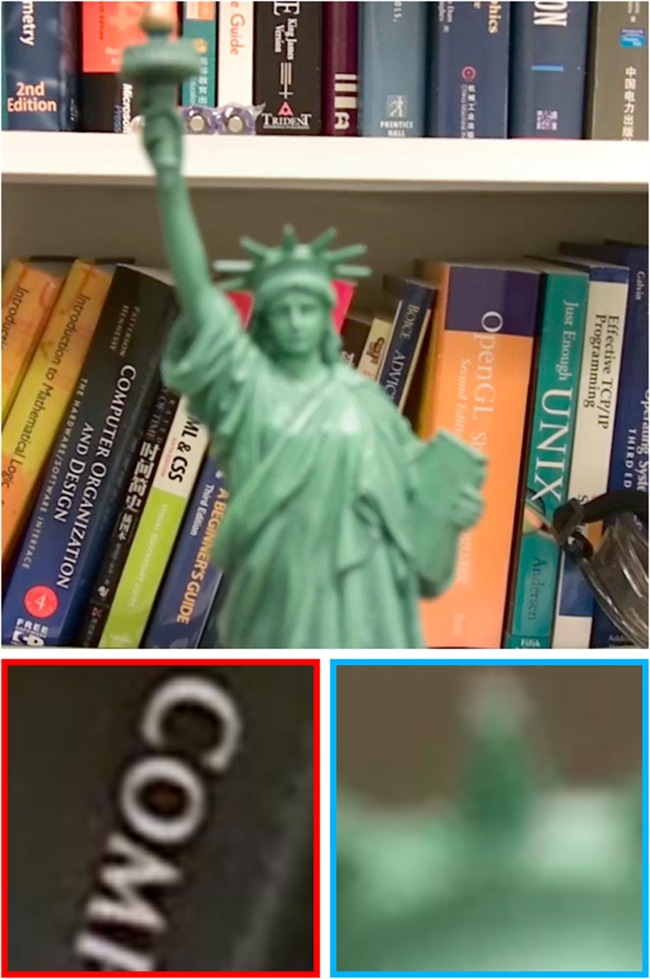} &
\includegraphics[width=2.92cm]{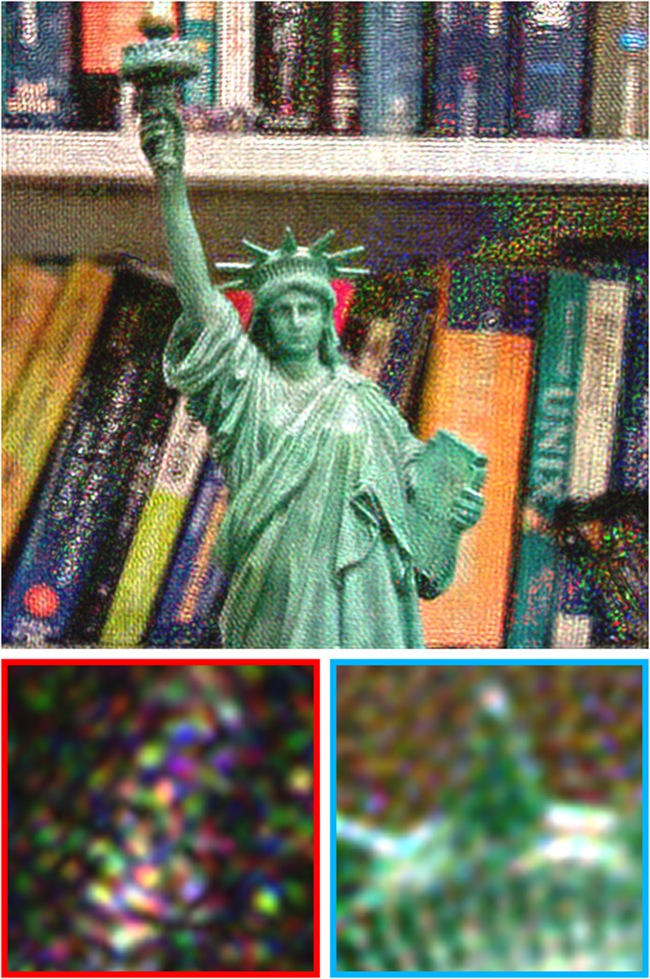} &
\includegraphics[width=2.92cm]{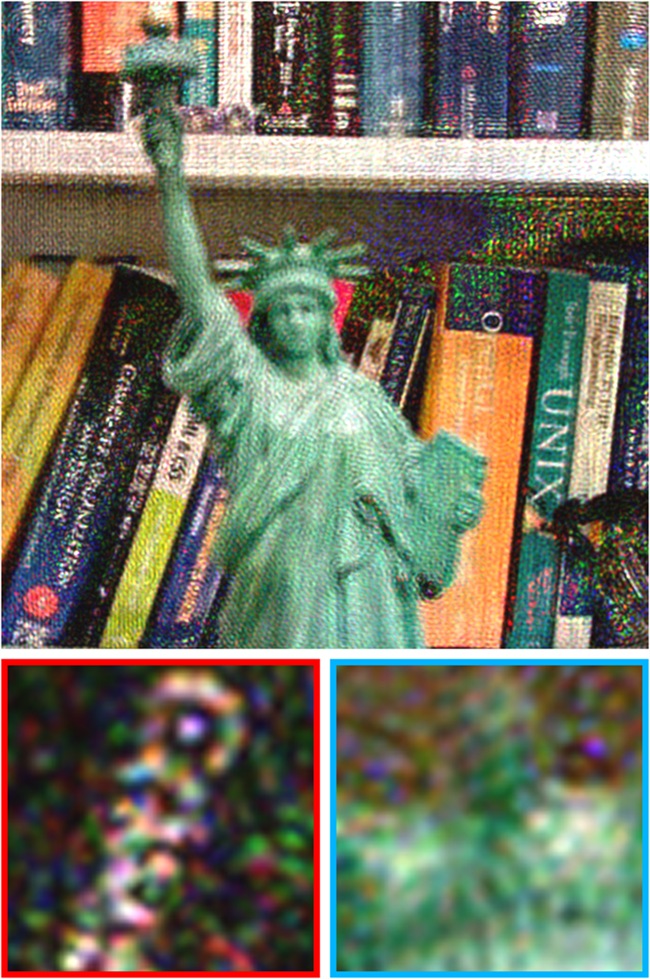} &
\includegraphics[width=2.92cm]{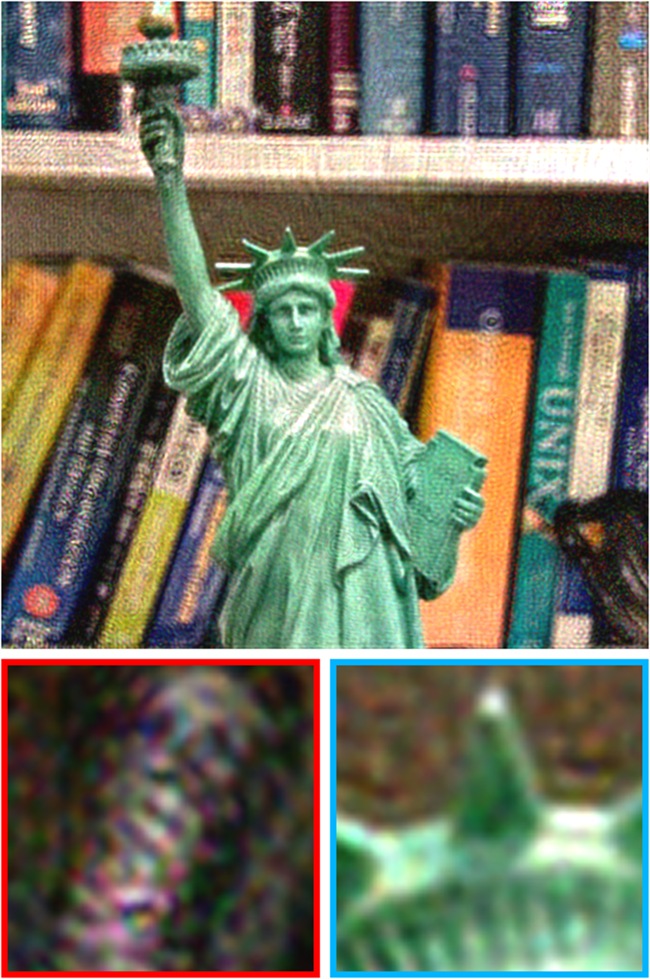} &
\includegraphics[width=2.92cm]{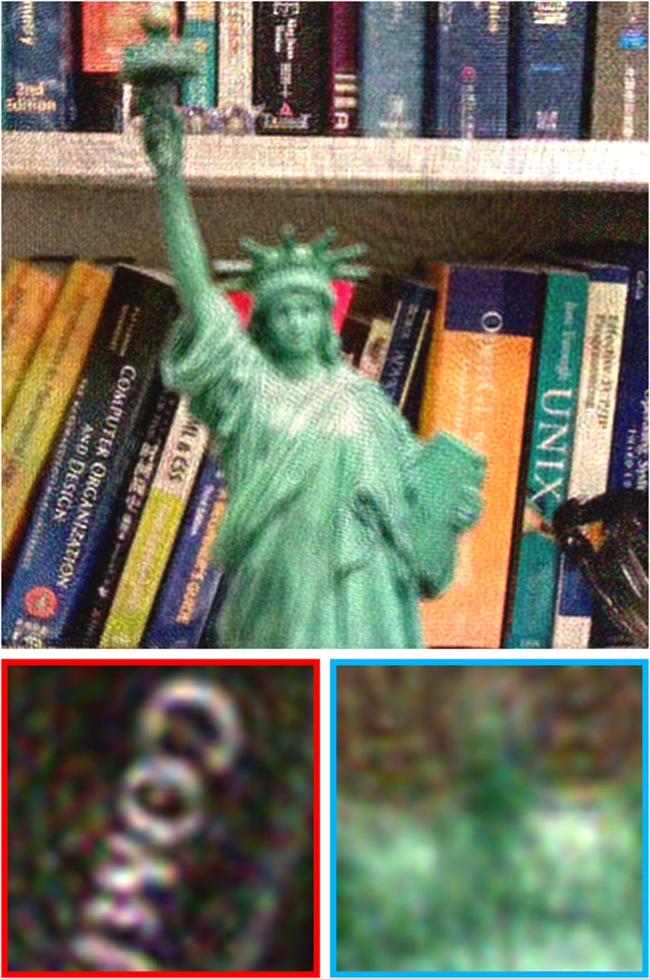} &
\includegraphics[width=2.92cm]{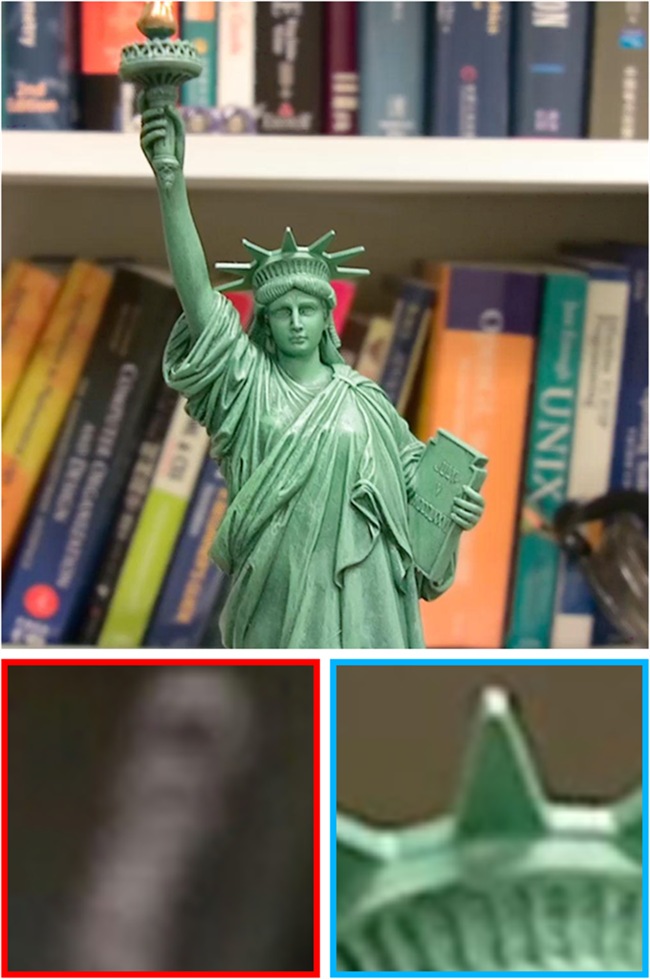} &
\includegraphics[width=2.92cm]{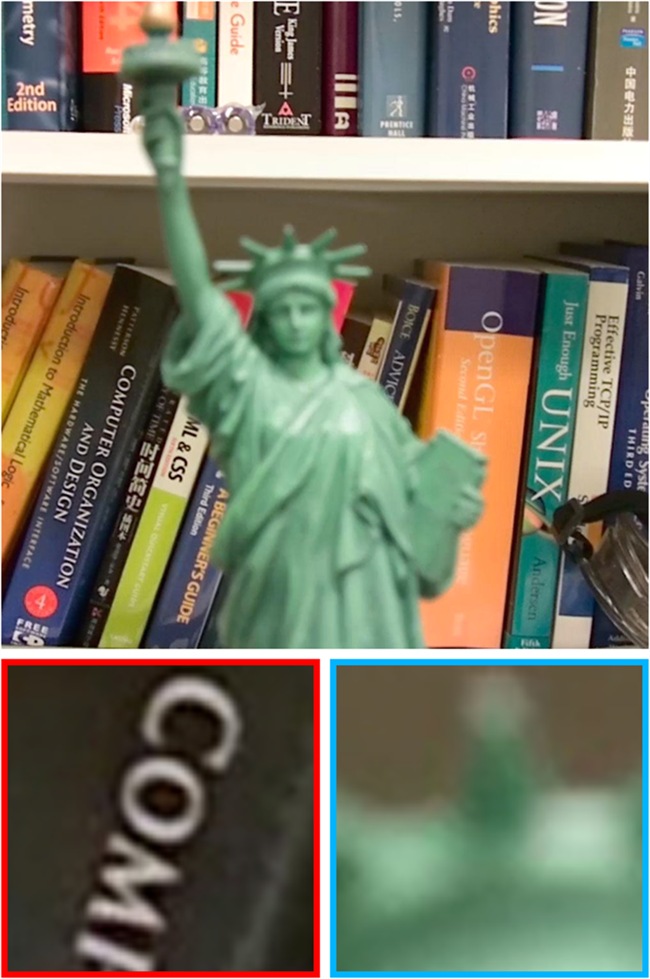} \\[0.8ex]

 Front Focus & Rear Focus &
 Front Focus & Rear Focus &
 Front Focus & Rear Focus &
Front Focus & Rear Focus \\
\end{tabular}
}
\end{center}

\vspace{-2mm}
\caption{
\textit{ Experimental Evaluation of 3D Holography}.
Our ellipsography approach---both with global (GA) and per-pixel (PA) analyzers---achieves speckle-suppressed volumetric reconstructions with sharper edges, natural defocus blur, and consistent image quality across focal depths.
In contrast, 3D holograms generated using traditional methods \cite{choi2021neural} exhibit visible artifacts, particularly around occlusion boundaries and high-frequency regions such as text, resulting in degraded detail and depth perception.
These results validate ellipsography's ability to preserve phase randomness while effectively suppressing interference noise from a single hologram frame.
}
\label{figure:3d_figure}
\end{figure*}

\paragraph{Qualitative Analysis of 2D Reconstructions}
\Cref{fig:2d_captures} presents experimental results for 2D holographic reconstructions generated using the proposed ellipsography method, using both per-pixel and global analyzers, compared against traditional random and smooth phase holography.
Specifically, we benchmark against camera-in-the-loop (CITL) optimization \cite{peng2020neural} and double phase amplitude coding (DPAC) \cite{maimone2017holographic, shi2021towards} which are shown to produce previous state-of-the-art results.
CITL methods suppress artifacts by iteratively refining phase patterns using active camera feedback, but inherently rely on random phase, resulting in visible speckle.
DPAC eliminates random interference by enforcing smooth image-plane phase, reducing noise at the cost of lower spatial resolution and contrast.
This trend can be observed from the zoomed insets of \Cref{fig:2d_captures}.
In contrast, ellipsography consistently outperforms both approaches, producing speckle-free, high-contrast reconstructions with sharp spatial features, closely approaching the ground truth target images; compare first and last columns of \Cref{fig:2d_captures}.
Importantly, even ellipsography with a global analyzer outperforms all the baseline methods.
These results validate that ellipsography combines the advantages of both smooth and random phase holograms: it maintains phase randomness for energy uniformity while achieving speckle suppression via vectorial interference shaping, all in a single hologram frame without active camera-based refinement or frame averaging.

\paragraph{Qualitative Analysis of 3D Reconstructions}
We further evaluate volumetric reconstruction performance of ellipsography using 3D holograms captured across multiple depths. 
\Cref{figure:3d_figure} presents experimental comparisons between CITL optimized holography \cite{choi2021neural} and our method, tested with both per-pixel and a global analyzers.
We omit comparisons with smooth phase holography, as prior work has demonstrated its inability to generate natural defocus cues or drive accommodation responses effectively \cite{chakravarthula2022hogel,kim2022accommodative}.
As shown in the insets of \Cref{figure:3d_figure}, traditional random phase holography suffers from pronounced speckle, particularly around occlusion boundaries and in high-frequency regions, where rapid phase variations lead to strong interference noise.
These artifacts degrade perceptual depth cues and obscure fine structural details across the focal stack.
In contrast, ellipsography produces reconstructions with significant speckle suppression, sharper spatial detail, and smoother, perceptually consistent defocus blur.

\paragraph{Quantitative Analysis}

\renewcommand{\arraystretch}{1.06}
\setlength{\tabcolsep}{1.2pt}
\begin{table}[ht]
\centering
\caption{Quantitative comparison of experimental results for 2D and 3D scenes using traditional and proposed methods. No region-of-interest (ROI) masking was used to suppress noise in the image plane. }
\label{tab:exp-metrics}
\begin{tabular}{|c|c|c|c|c|c|c|}

\hline
\multirow{2}{*}{Method} & \multicolumn{3}{c|}{2D} & \multicolumn{3}{c|}{3D} \\ \cline{2-7}
 & PSNR$\uparrow$ & SSIM$\uparrow$ & LPIPS$\downarrow$ & PSNR$\uparrow$ & SSIM$\uparrow$ & LPIPS$\downarrow$ \\ \hline
\begin{tabular}[c]{@{}c@{}}Smooth Phase Holo. \\ (DPAC)\end{tabular} & 18.49 & 0.70 & 0.51 & -- & -- & -- \\ \hline
\begin{tabular}[c]{@{}c@{}}Random Phase Holo. \\ (CITL)\end{tabular} & 18.74 & 0.69 & 0.43 & 17.53 & 0.55 & 0.51 \\ \hline
\rowcolor{green!20}
\begin{tabular}[c]{@{}c@{}}Ellipsography \\ (Global Analyzer)\end{tabular} & \textbf{23.05} & \textbf{0.81} & \textbf{0.31} & \textbf{24.42} & \textbf{0.62} & \textbf{0.47} \\ \hline
\rowcolor{green!20}
\begin{tabular}[c]{@{}c@{}}Ellipsography \\ (Per-pixel Analyzer)\end{tabular} & \textbf{28.66} & \textbf{0.82} & \textbf{0.21} & \textbf{28.12} & \textbf{0.78} & \textbf{0.32} \\ \hline
\end{tabular}
\end{table}

\Cref{tab:exp-metrics} presents quantitative evaluations of our experimental reconstructions using PSNR, SSIM, and LPIPS.
Consistent with simulation results, per-pixel ellipsography achieves an order of magnitude improvement over baseline methods for both 2D and 3D holograms. 
Even ellipsography with a global analyzer outperforms the baseline methods by at least 5 dB in PSNR, indicating a substantial reduction in reconstruction error compared to conventional scalar holography.

\begin{figure}[t]
\centering
\setlength{\tabcolsep}{1pt}
\renewcommand{\arraystretch}{0.6}

\begin{tabular}{ccc}
\makecell[c]{\small CITL Optimization} &
\makecell[c]{\small Ellipsography (GA)} &
\makecell[c]{\small Ellipsography (PA)} \\[-0.2ex]

\includegraphics[width=\columnwidth/3]{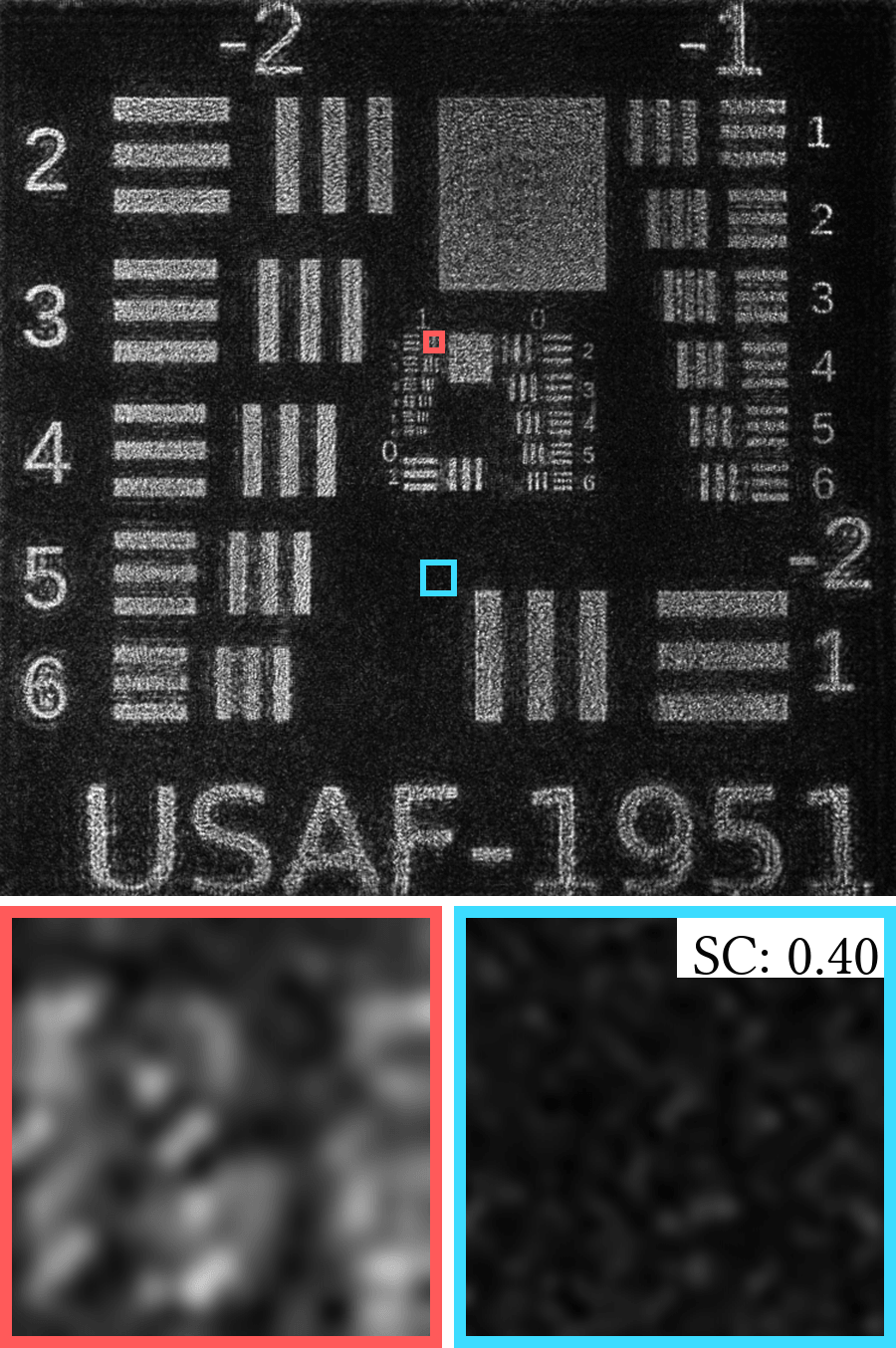} &
\includegraphics[width=\columnwidth/3]{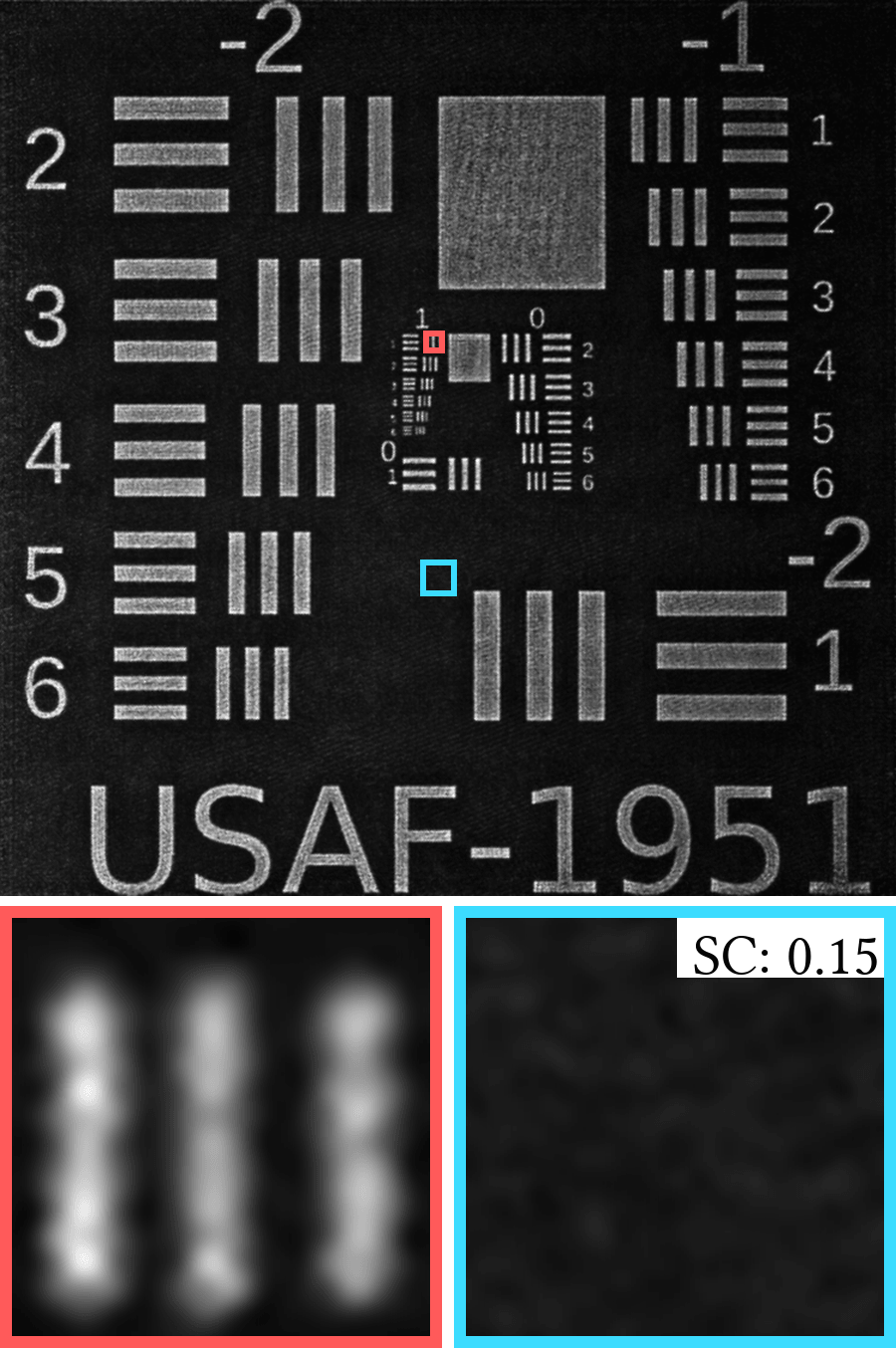} &
\includegraphics[width=\columnwidth/3]{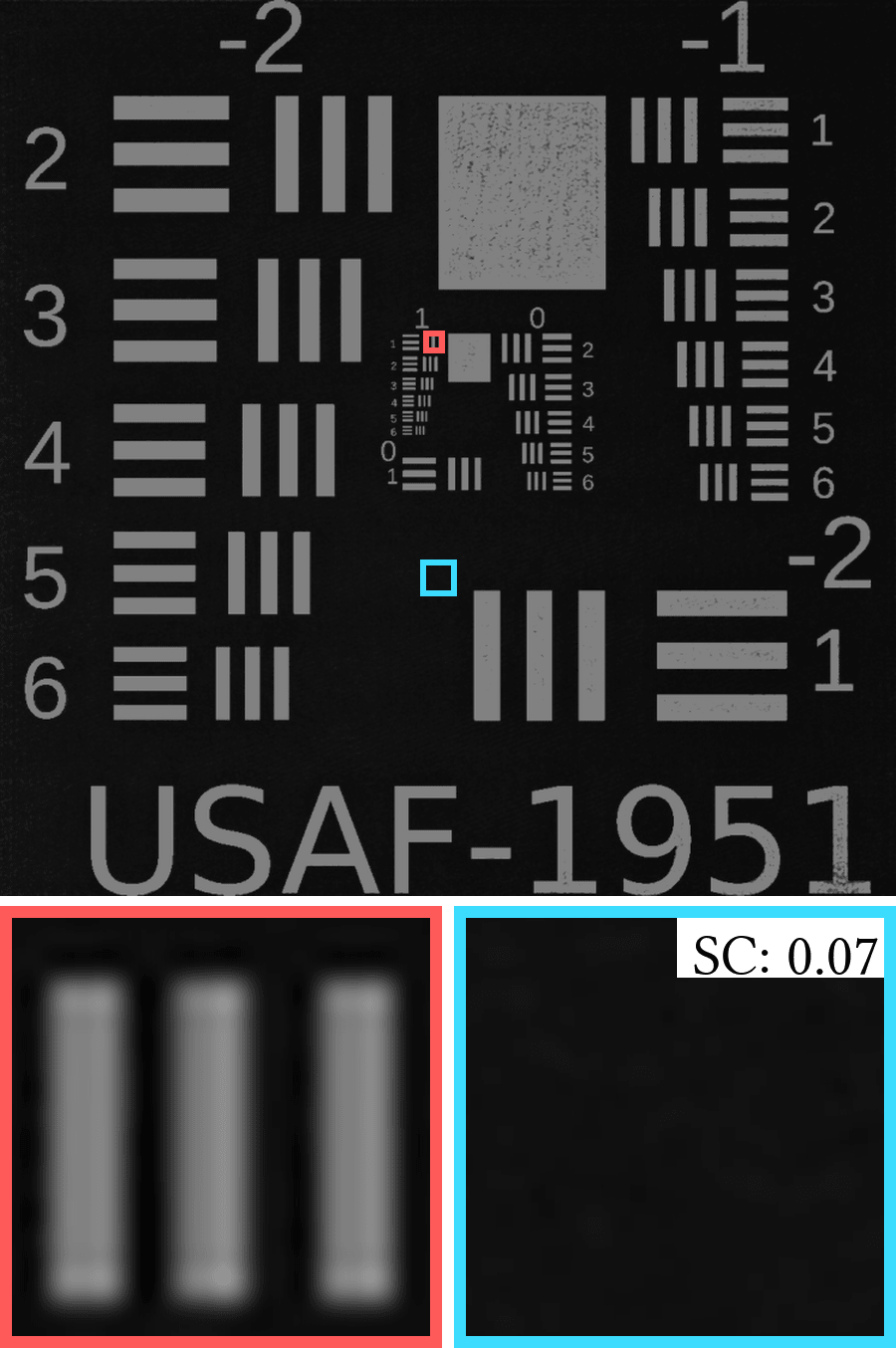}
\end{tabular}

\vspace{-1.5mm}
  \caption{
  \textit{Experimental Evaluation of Speckle Contrast.} 
  Compared to CITL optimization, the proposed ellipsography method---both with global (GA) and per-pixel (PA) analyzers---achieves substantial image quality improvements and significant speckle reduction, as validated by the speckle contrast.
  } 
  \label{fig:exp-speckle}
\end{figure}

In addition to pixel-wise accuracy, ellipsography also shows significant improvements in perceptual metrics, demonstrating closer perceptual similarity to the ground truth compared to competing methods. 
While the absolute performance in physical experiments is slightly lower than in simulation, this discrepancy is not due to limitations of the framework itself, but from practical hardware imperfections such as optical misalignments, lens aberrations, and non-ideal polarization filter characteristics.
These real-world factors are inherent to optical systems and do not diminish the fundamental advantages of ellipsography. We discuss these implementation considerations further in \Cref{sec:discussion}.

\subsection{Speckle and Resolution Analysis}

\setlength{\tabcolsep}{1.0pt}
\renewcommand{\arraystretch}{0.6}

\begin{figure*}[t]
\centering
\resizebox{\textwidth}{!}{%
\begin{tabular}{cccccc}

\multicolumn{1}{c}{\fontsize{8}{9}\selectfont Target} &
\multicolumn{1}{c}{\fontsize{8}{9}\selectfont CITL Optimization} &
\multicolumn{1}{c}{\fontsize{8}{9}\selectfont Polarization Multiplexing} &
\multicolumn{1}{c}{\fontsize{8}{9}\selectfont 64-Frame Average} &
\multicolumn{1}{c}{\fontsize{8}{9}\selectfont Ellipsography (GA)} &
\multicolumn{1}{c}{\fontsize{8}{9}\selectfont Ellipsography (PA)} \\
[0.1ex]

\includegraphics[width=3.1cm]{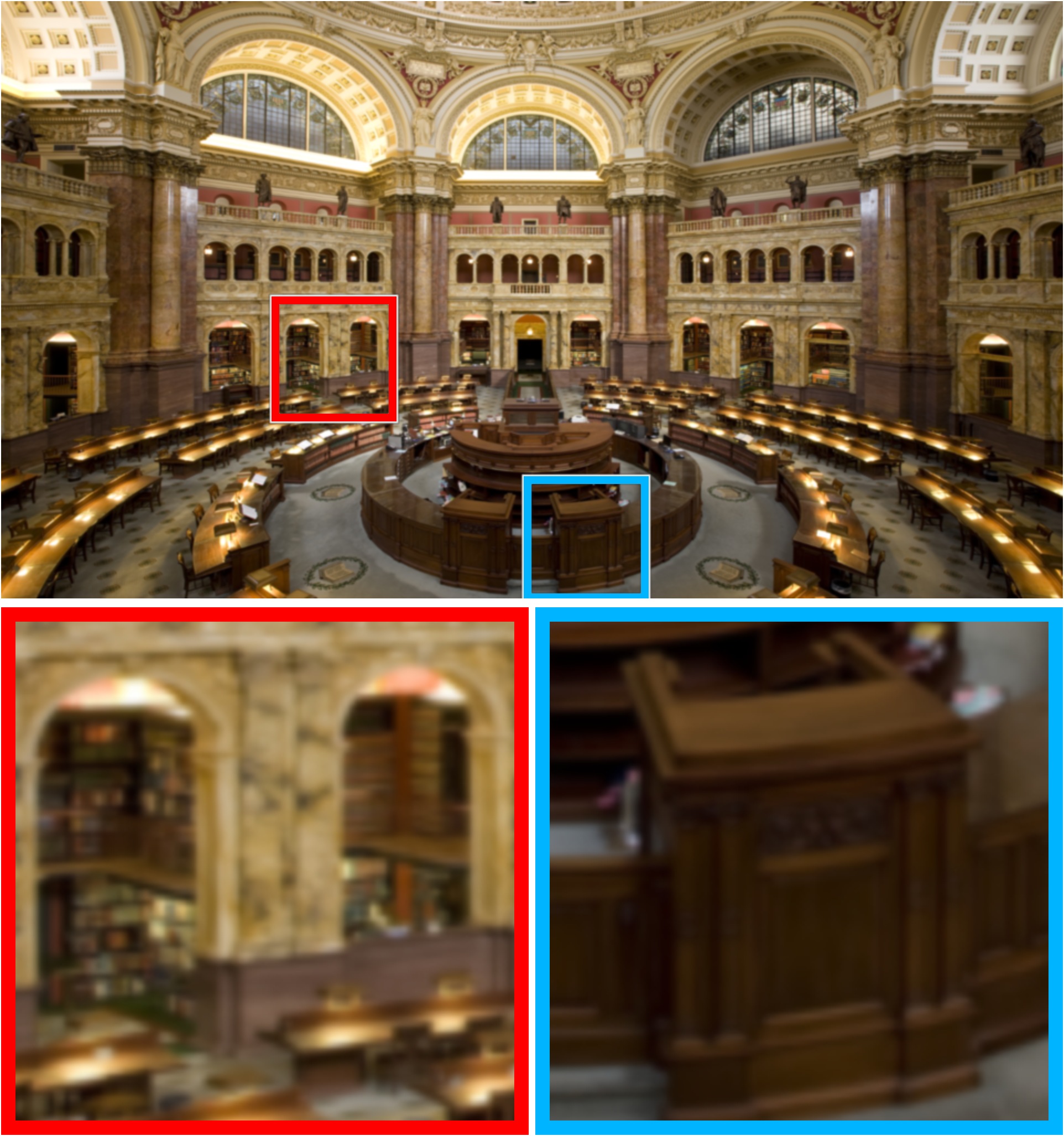} &
\includegraphics[width=3.1cm]{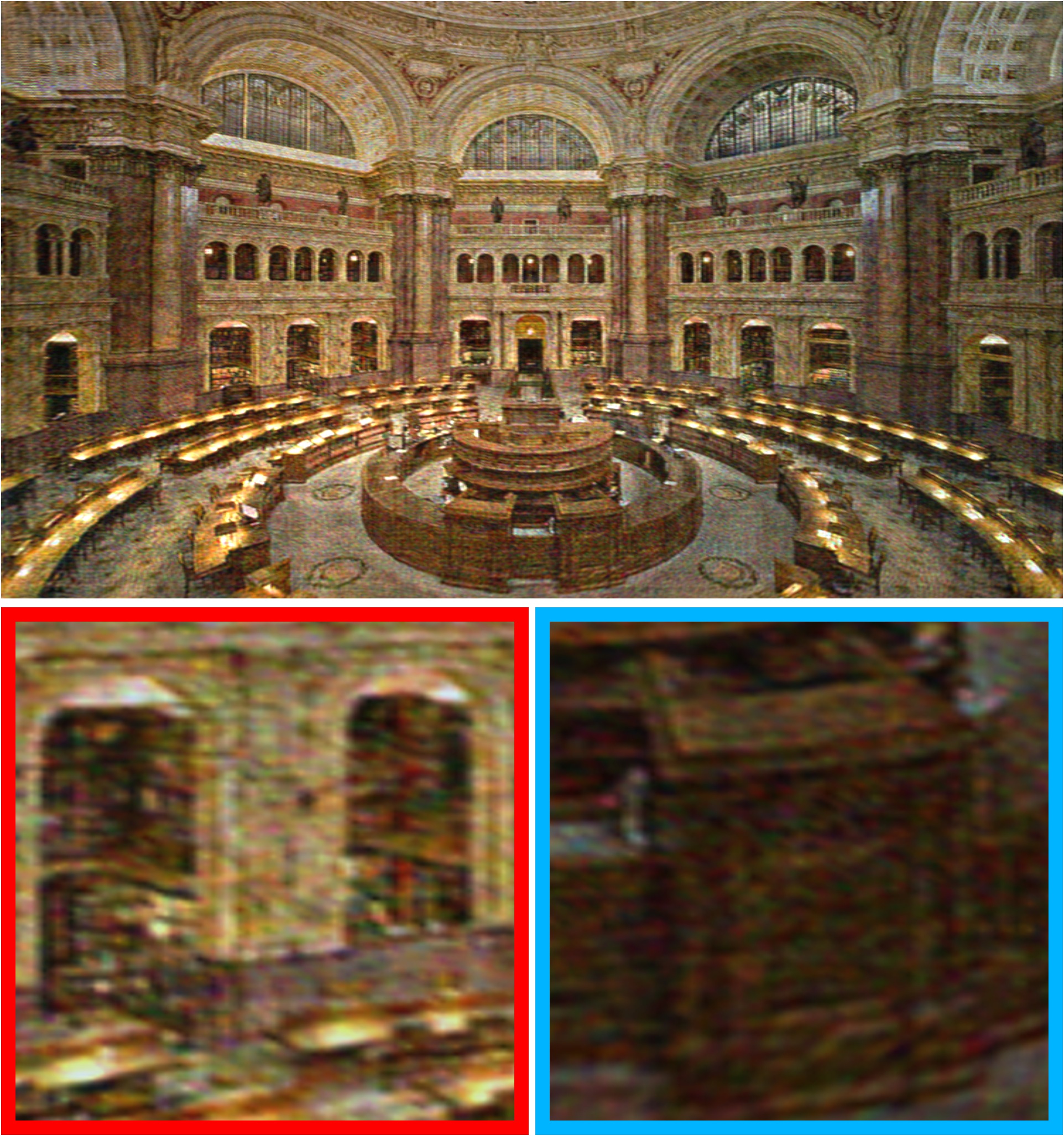} &
\includegraphics[width=3.1cm]{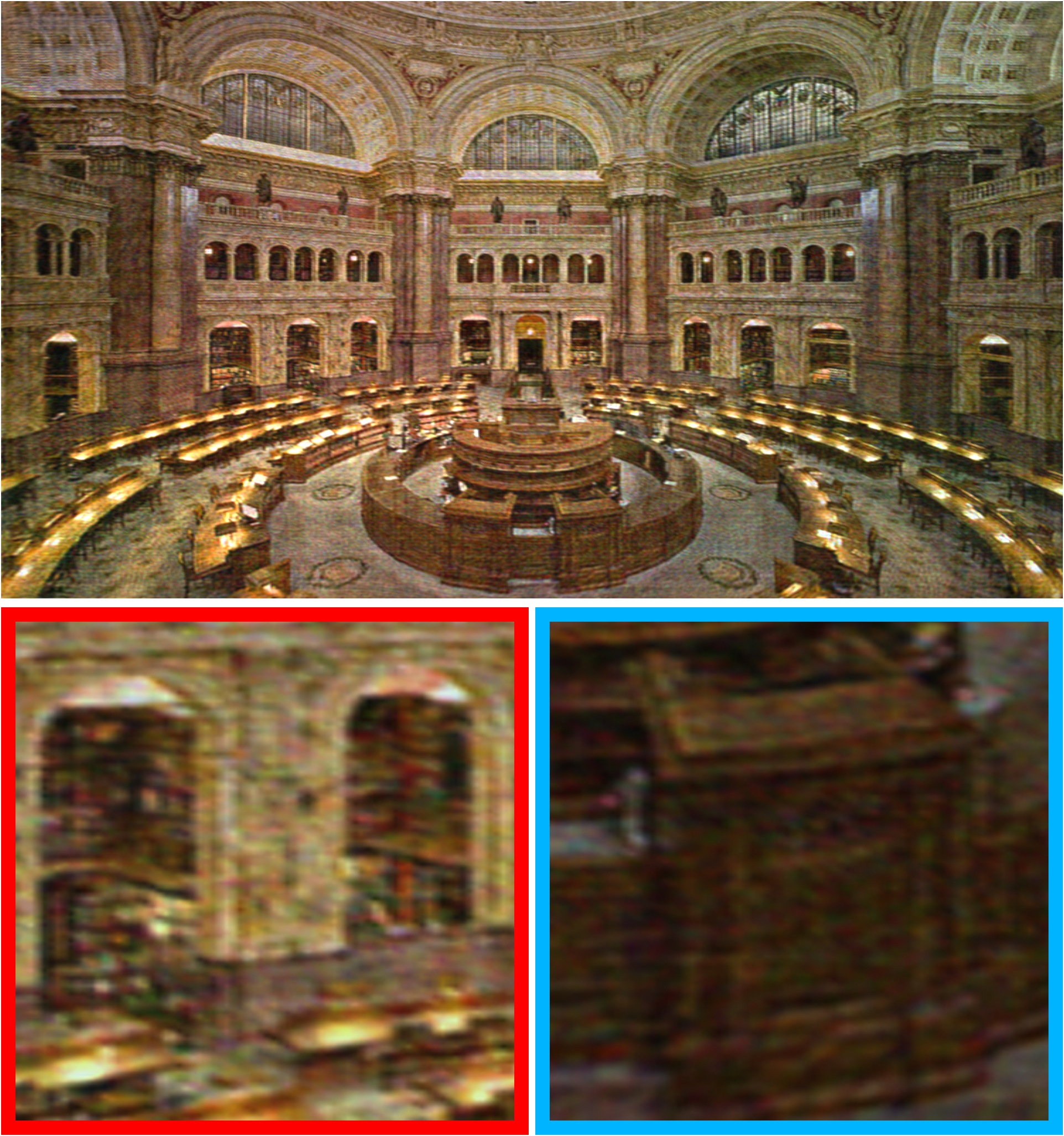} &
\includegraphics[width=3.1cm]{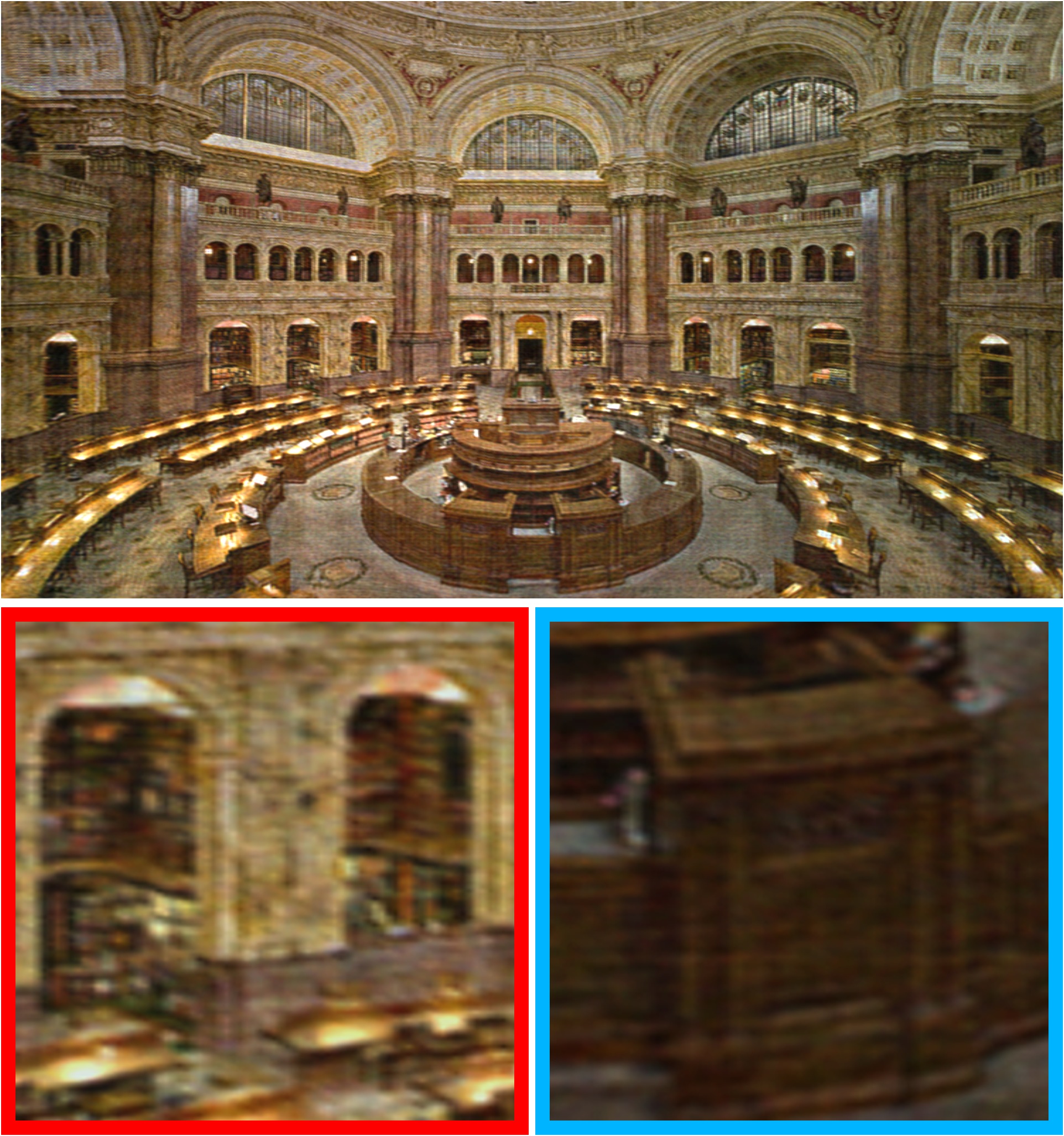} &
\includegraphics[width=3.1cm]{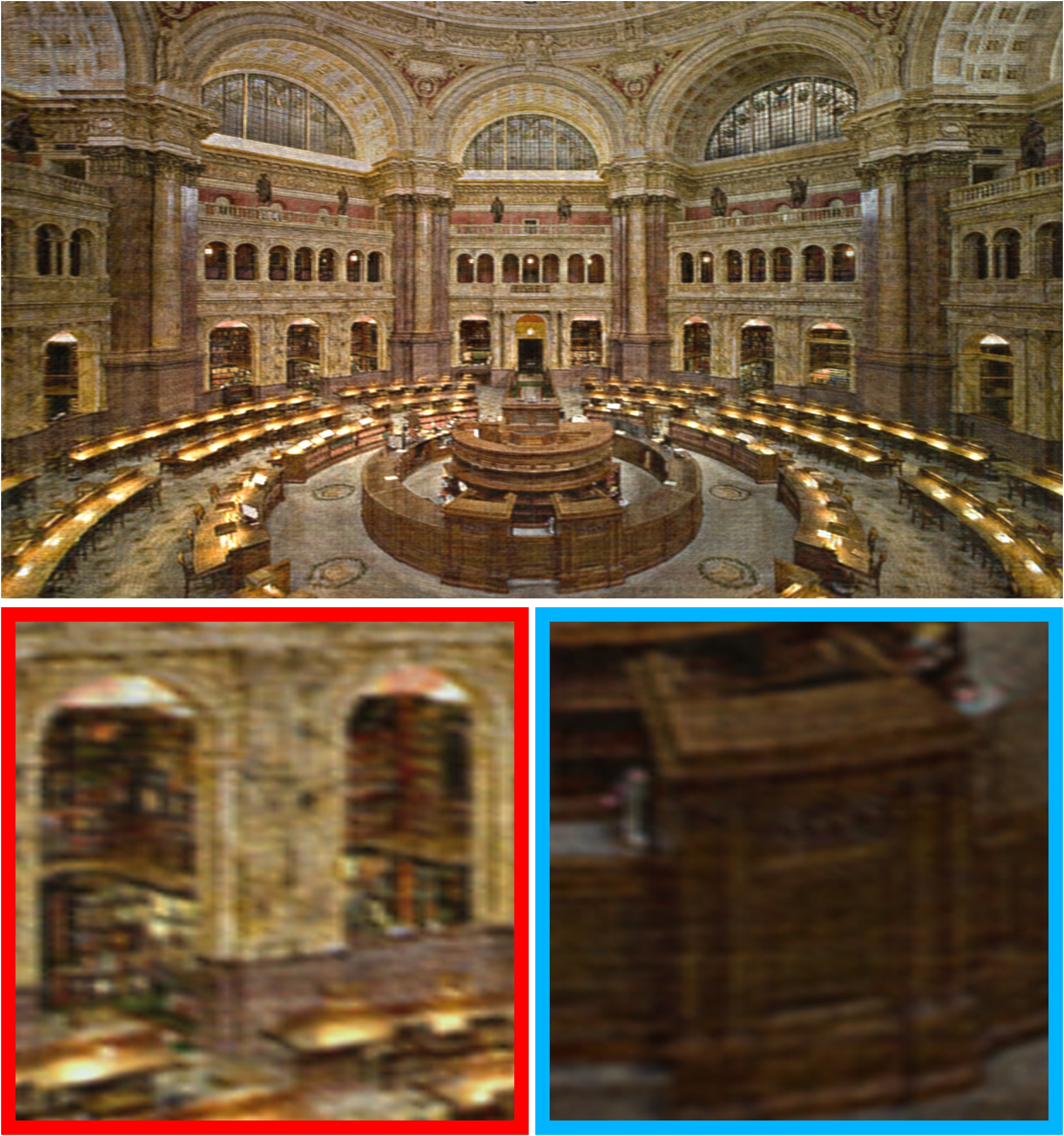} &
\includegraphics[width=3.1cm]{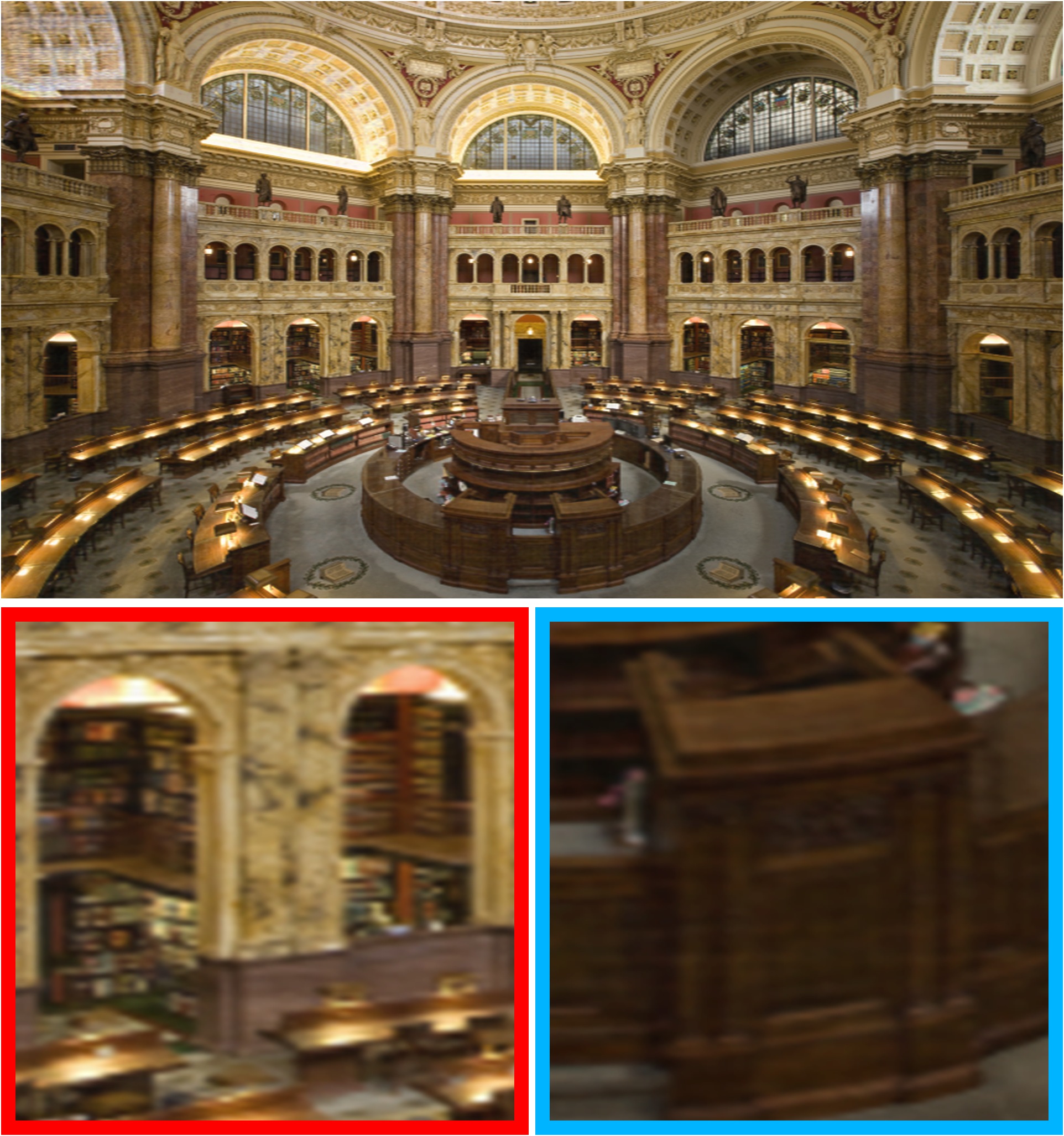} \\
[0.0ex]

\includegraphics[width=3.1cm]{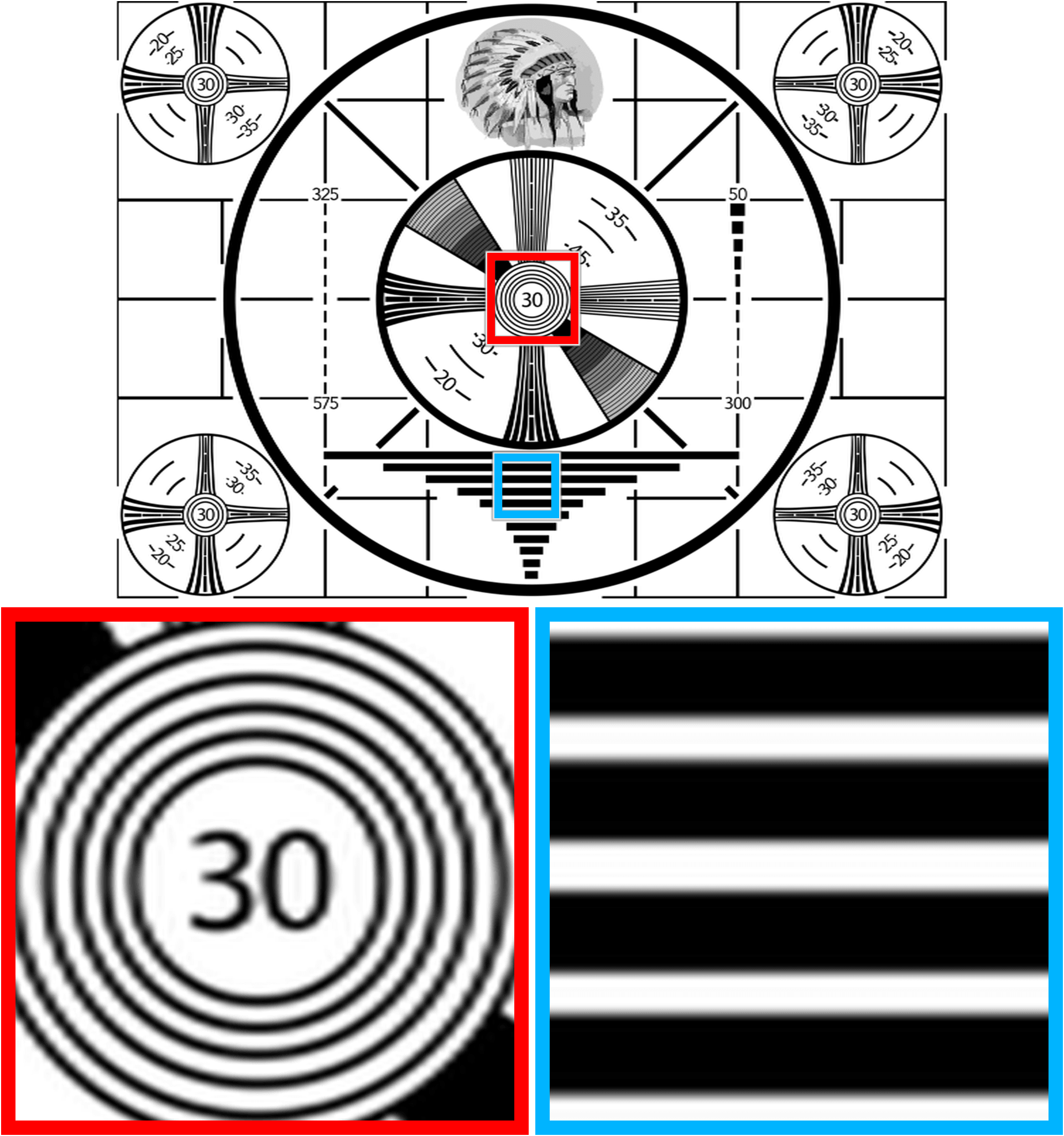} &
\includegraphics[width=3.1cm]{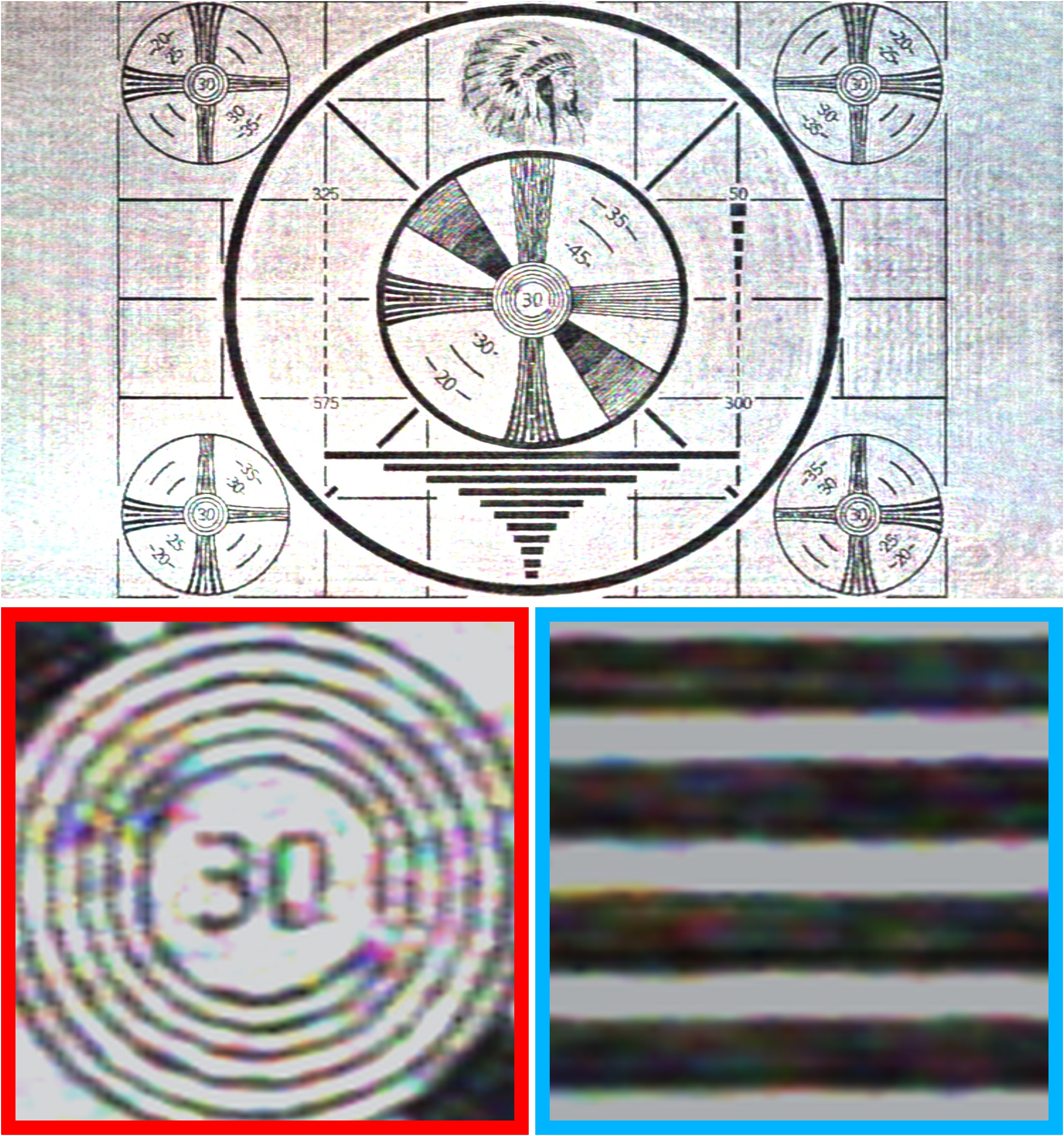} &
\includegraphics[width=3.1cm]{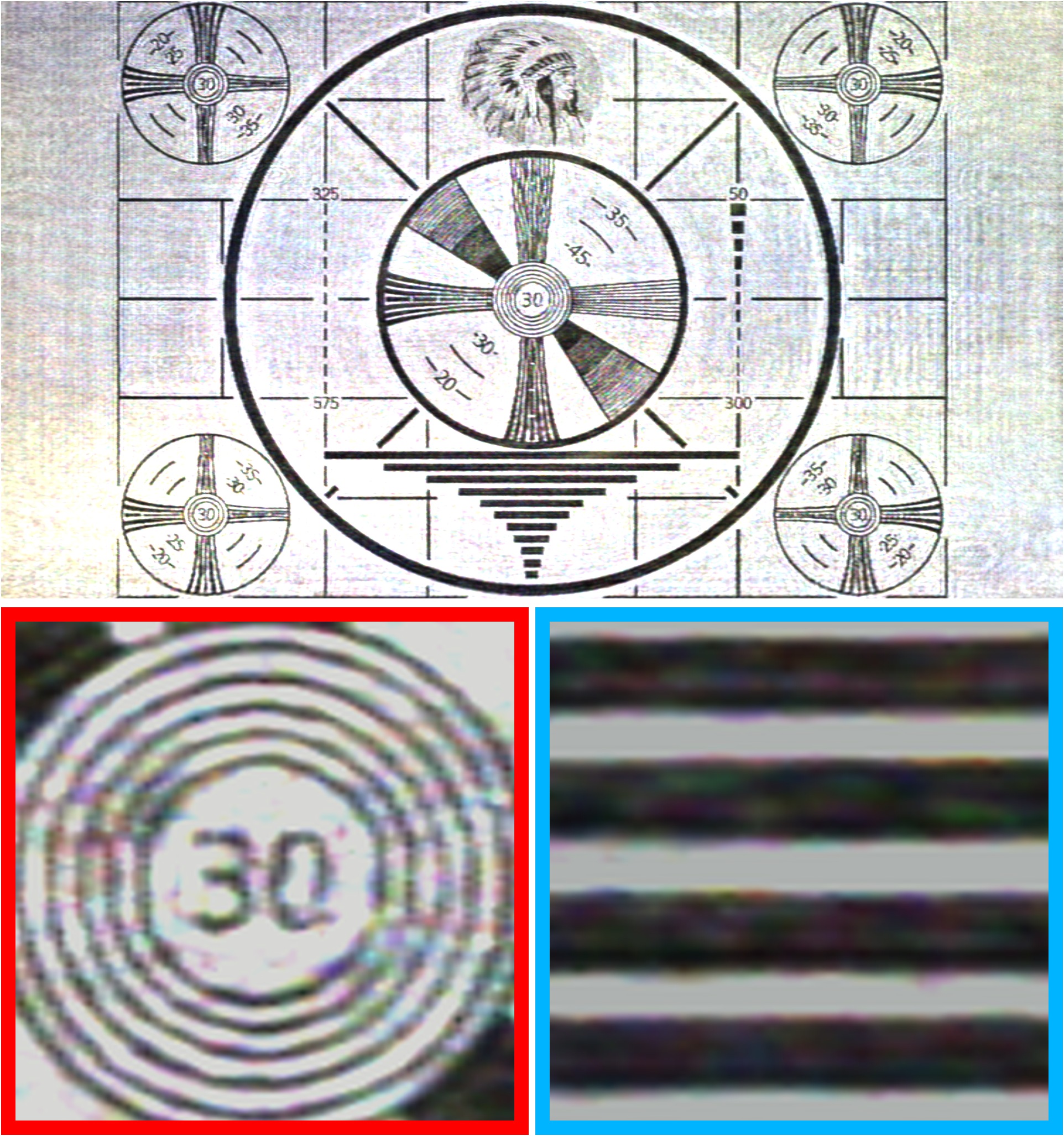} &
\includegraphics[width=3.1cm]{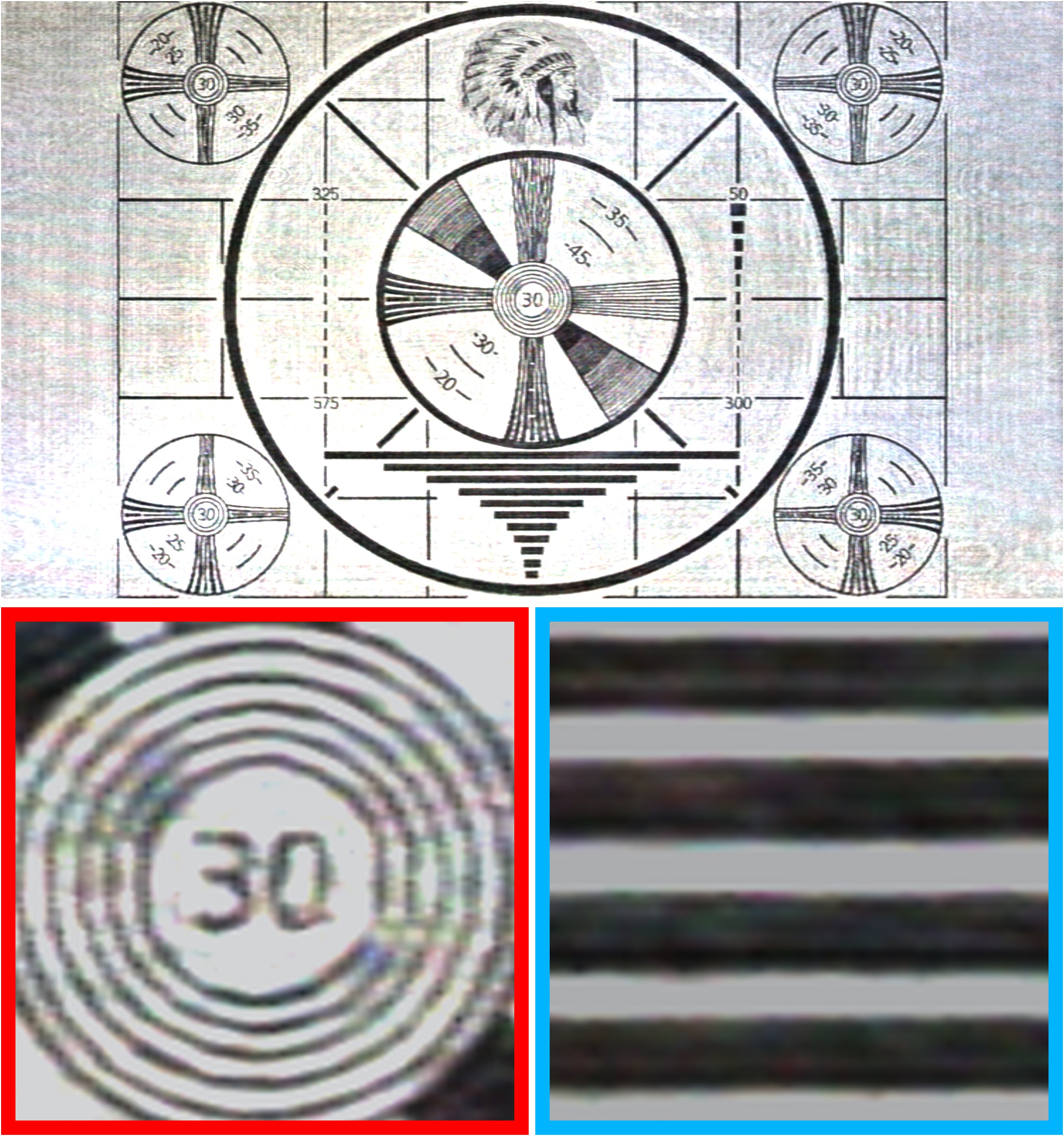} &
\includegraphics[width=3.1cm]{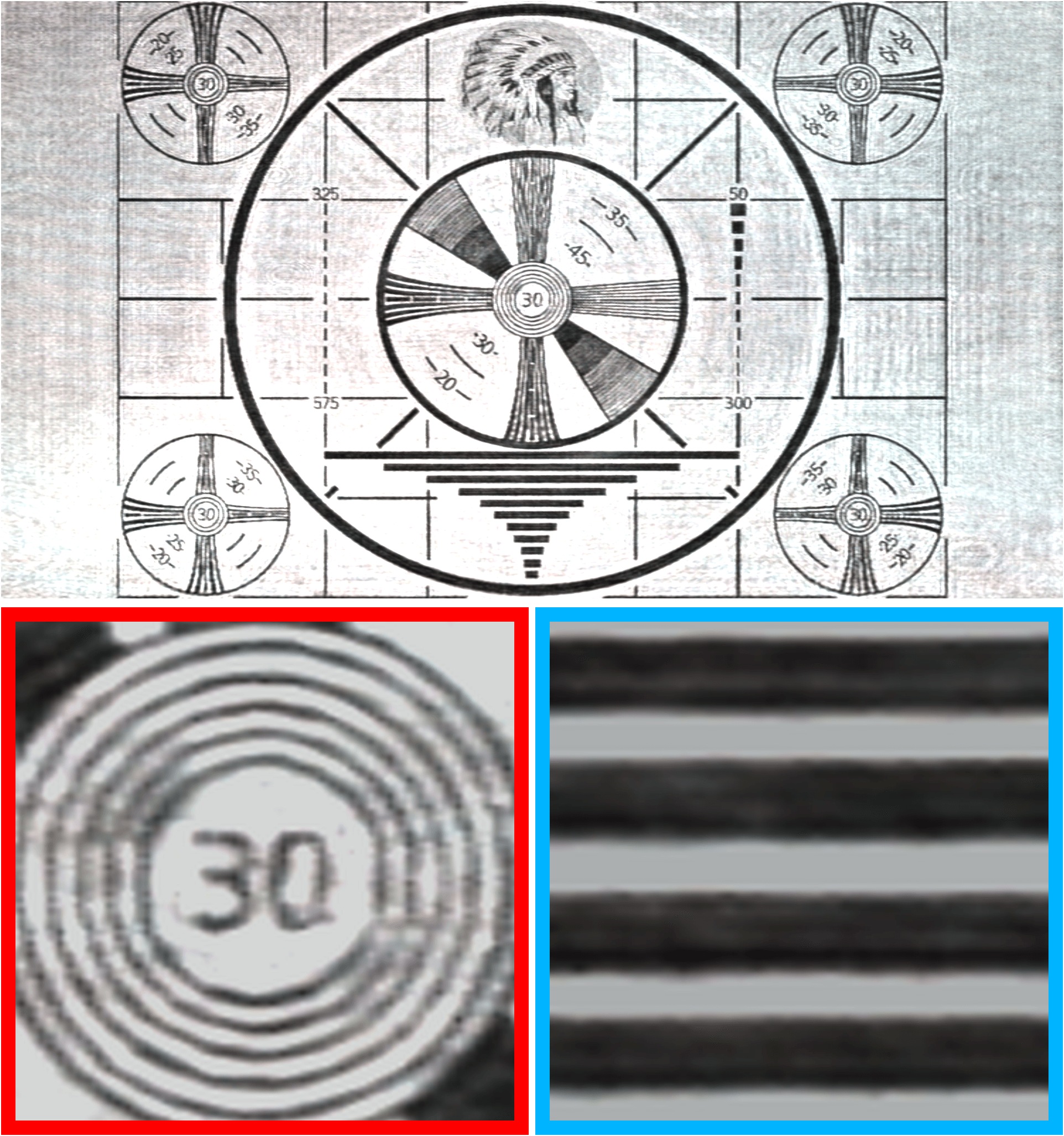} &
\includegraphics[width=3.1cm]{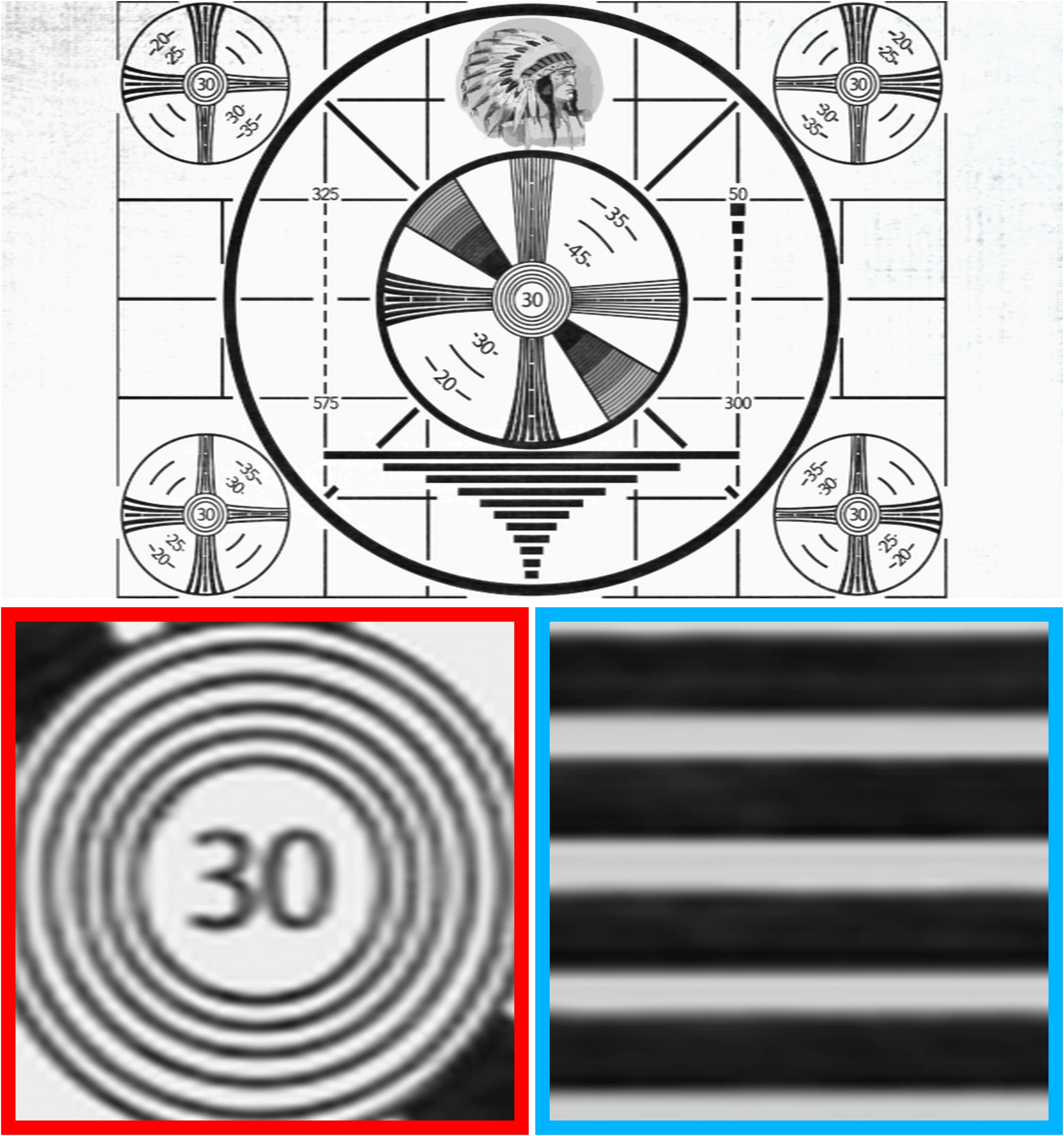} \\
[0.0ex]

\includegraphics[width=3.1cm]{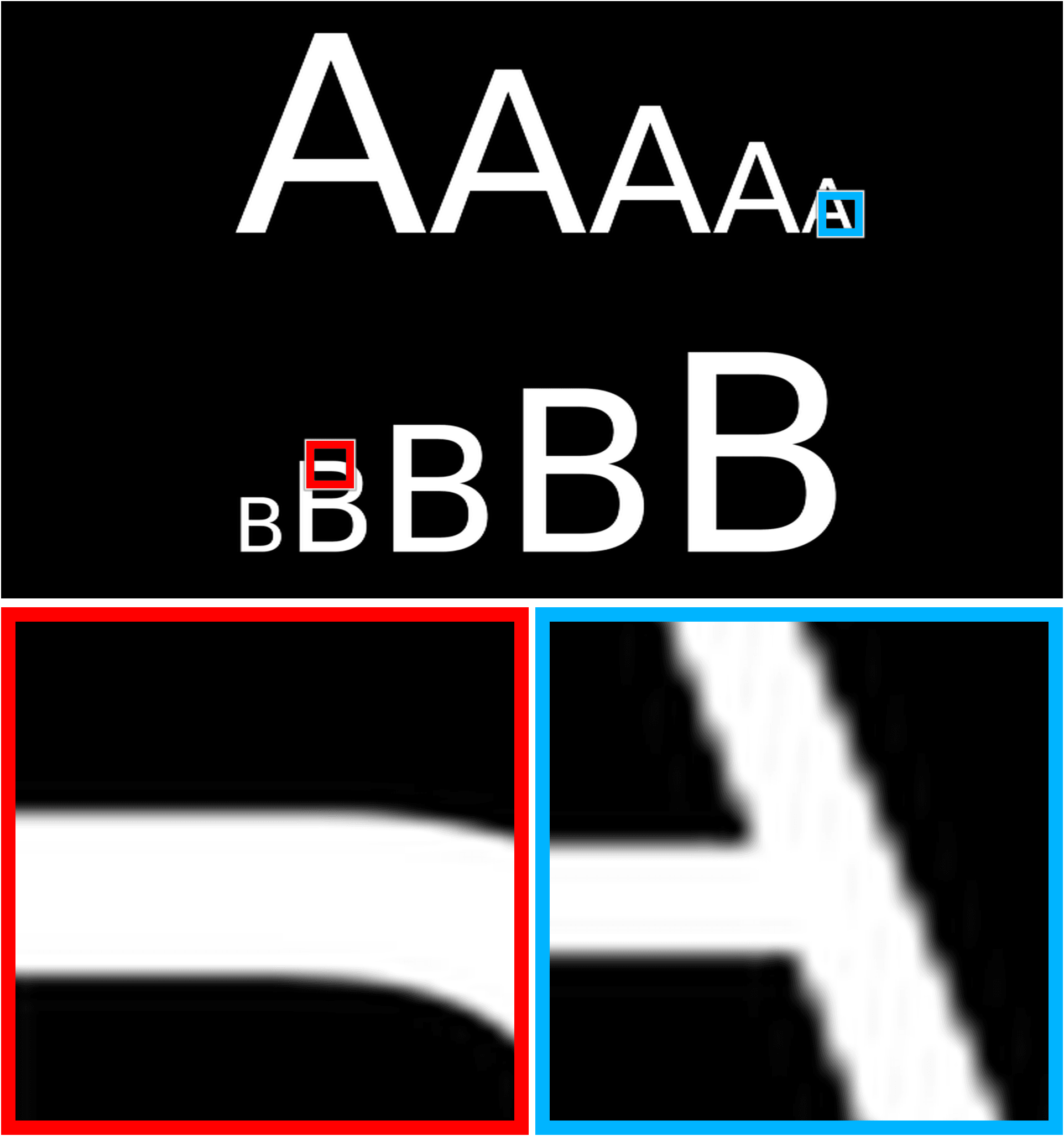} &
\includegraphics[width=3.1cm]{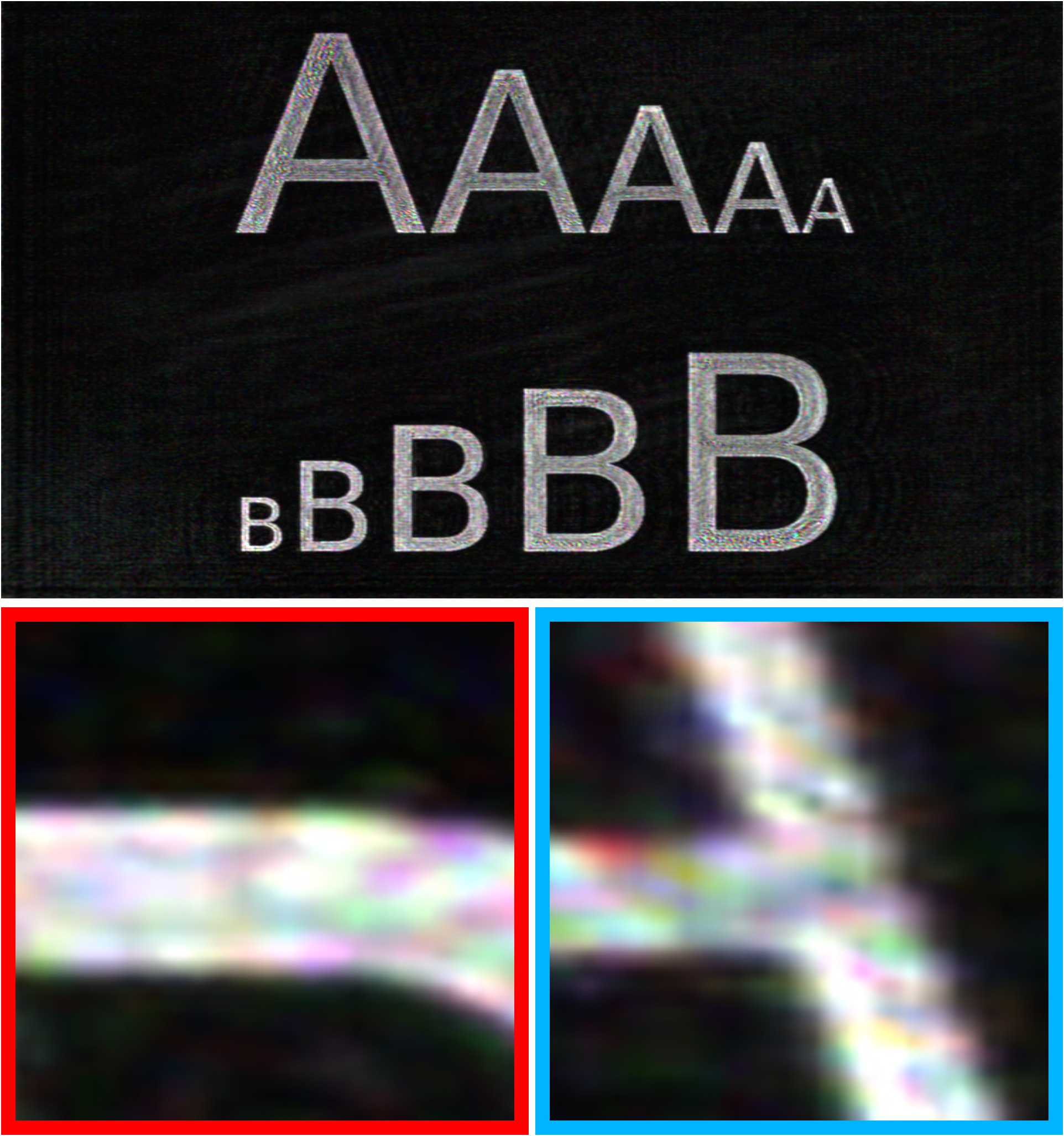} &
\includegraphics[width=3.1cm]{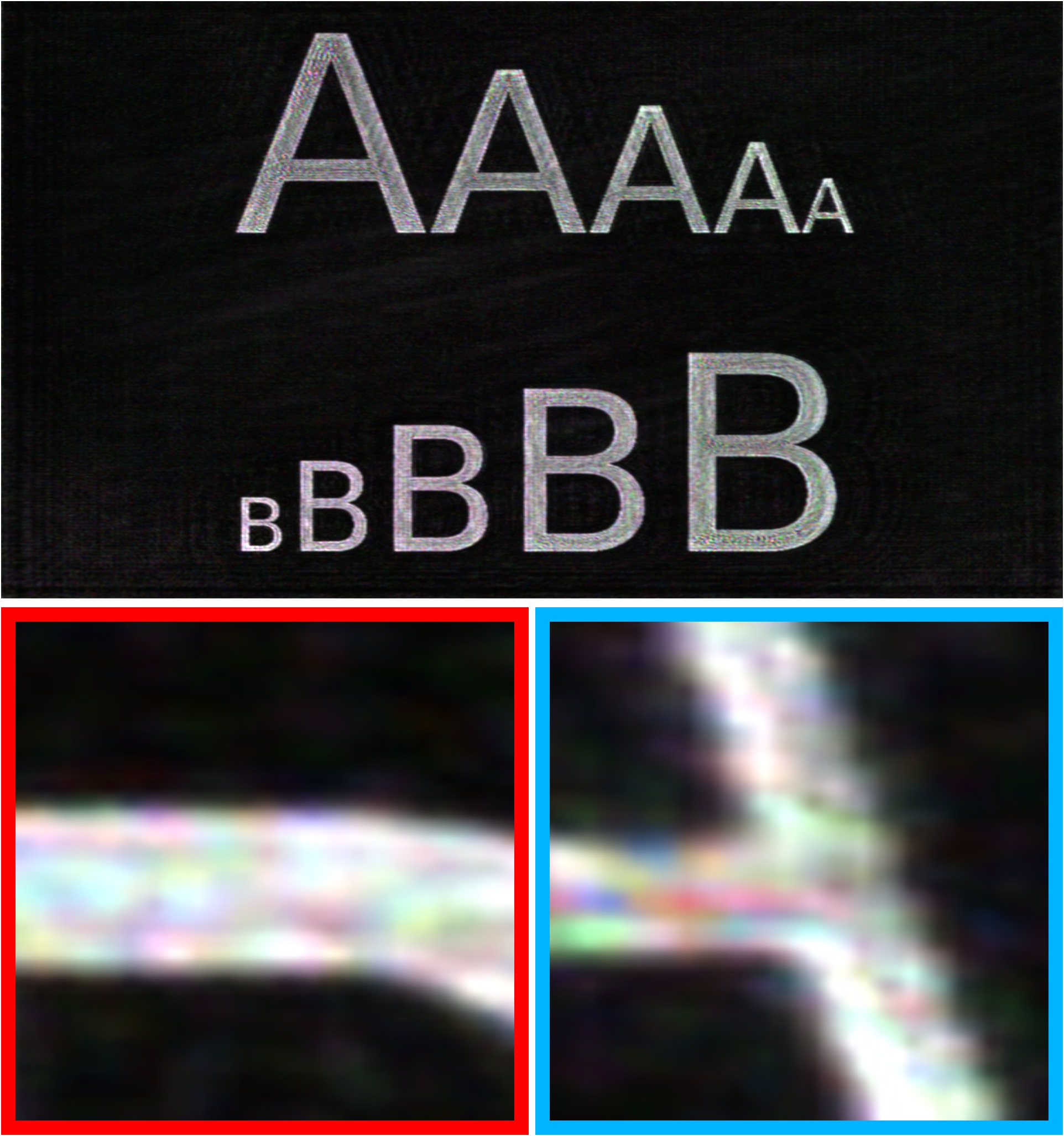} &
\includegraphics[width=3.1cm]{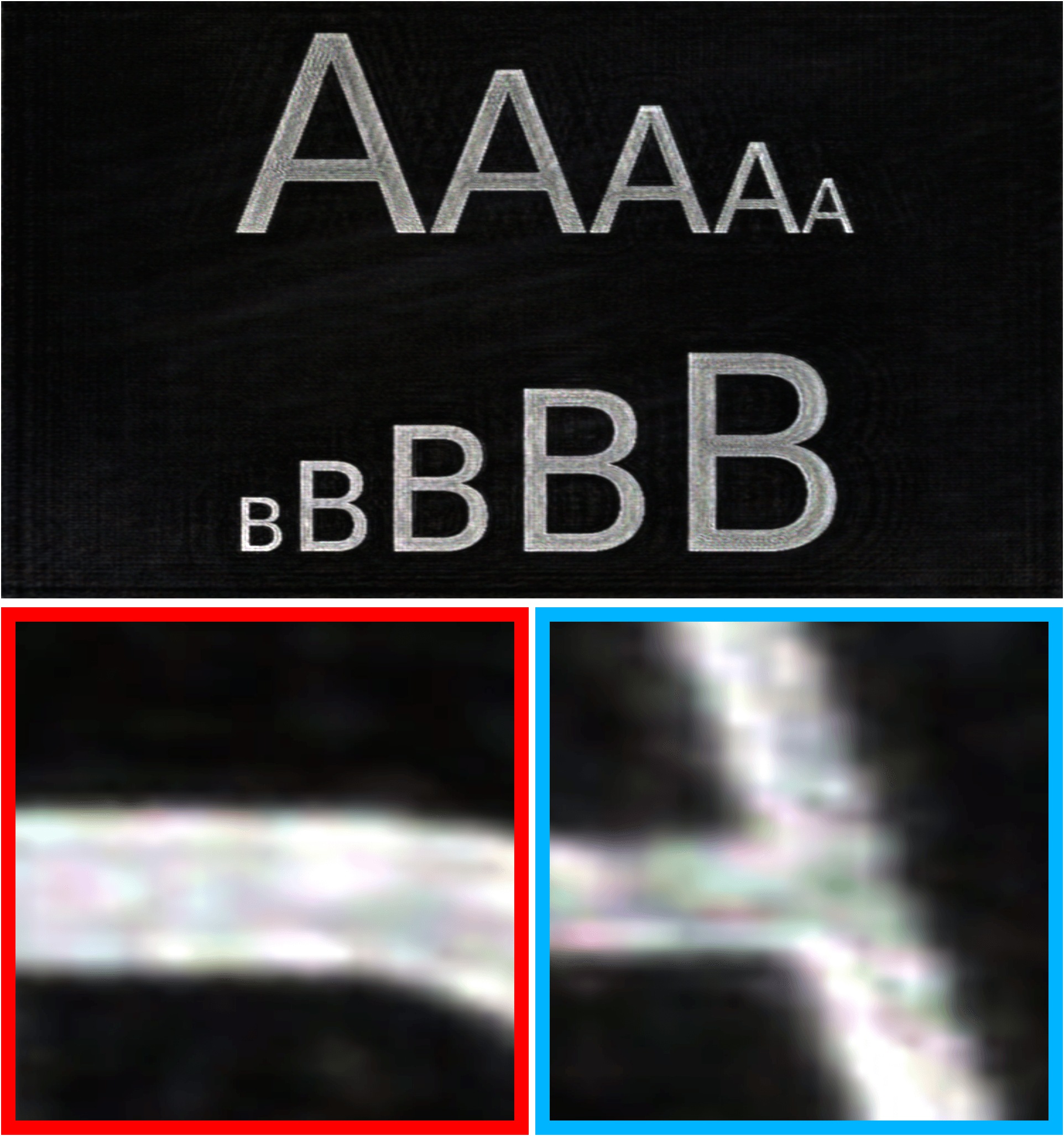} &
\includegraphics[width=3.1cm]{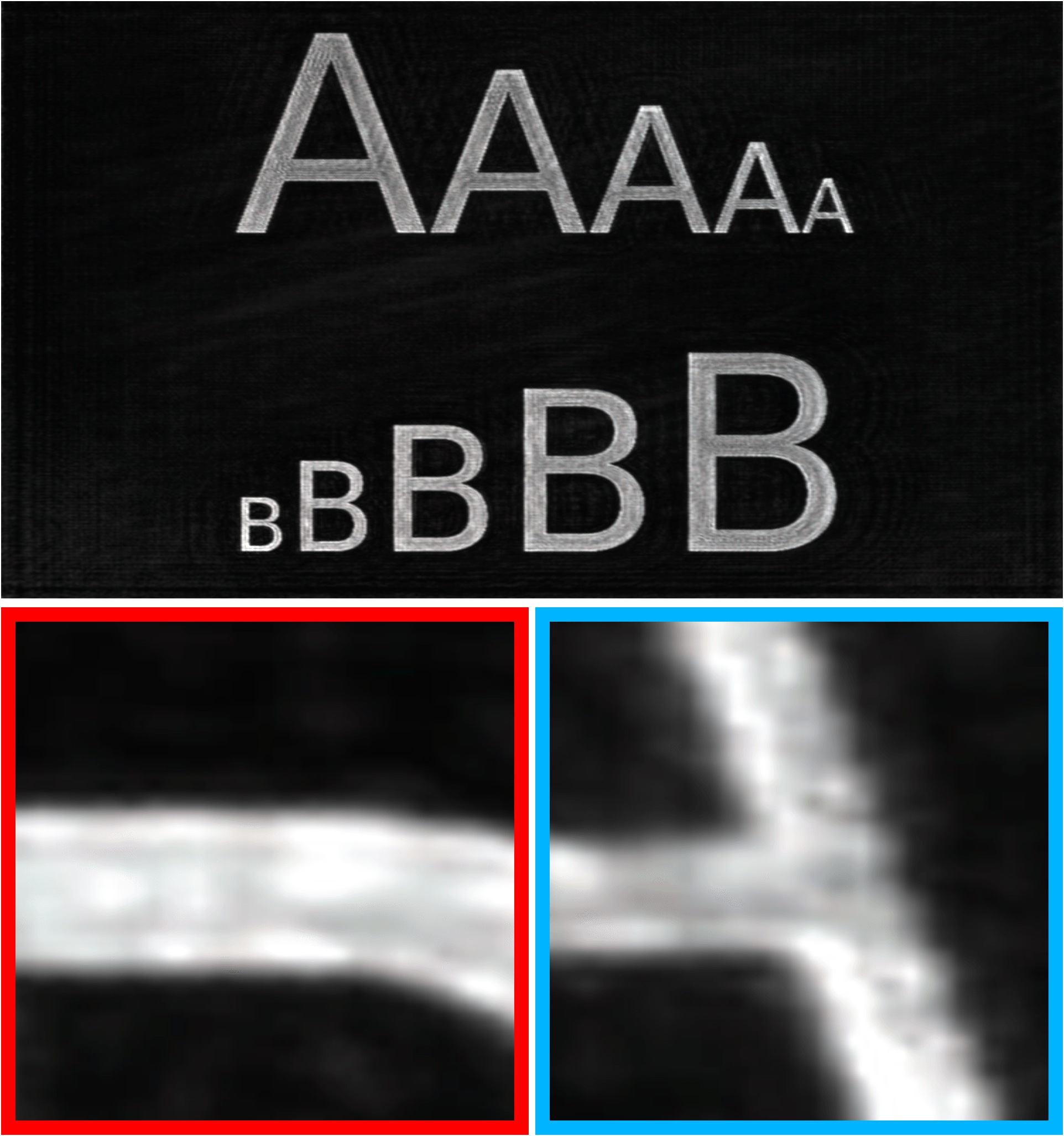} &
\includegraphics[width=3.1cm]{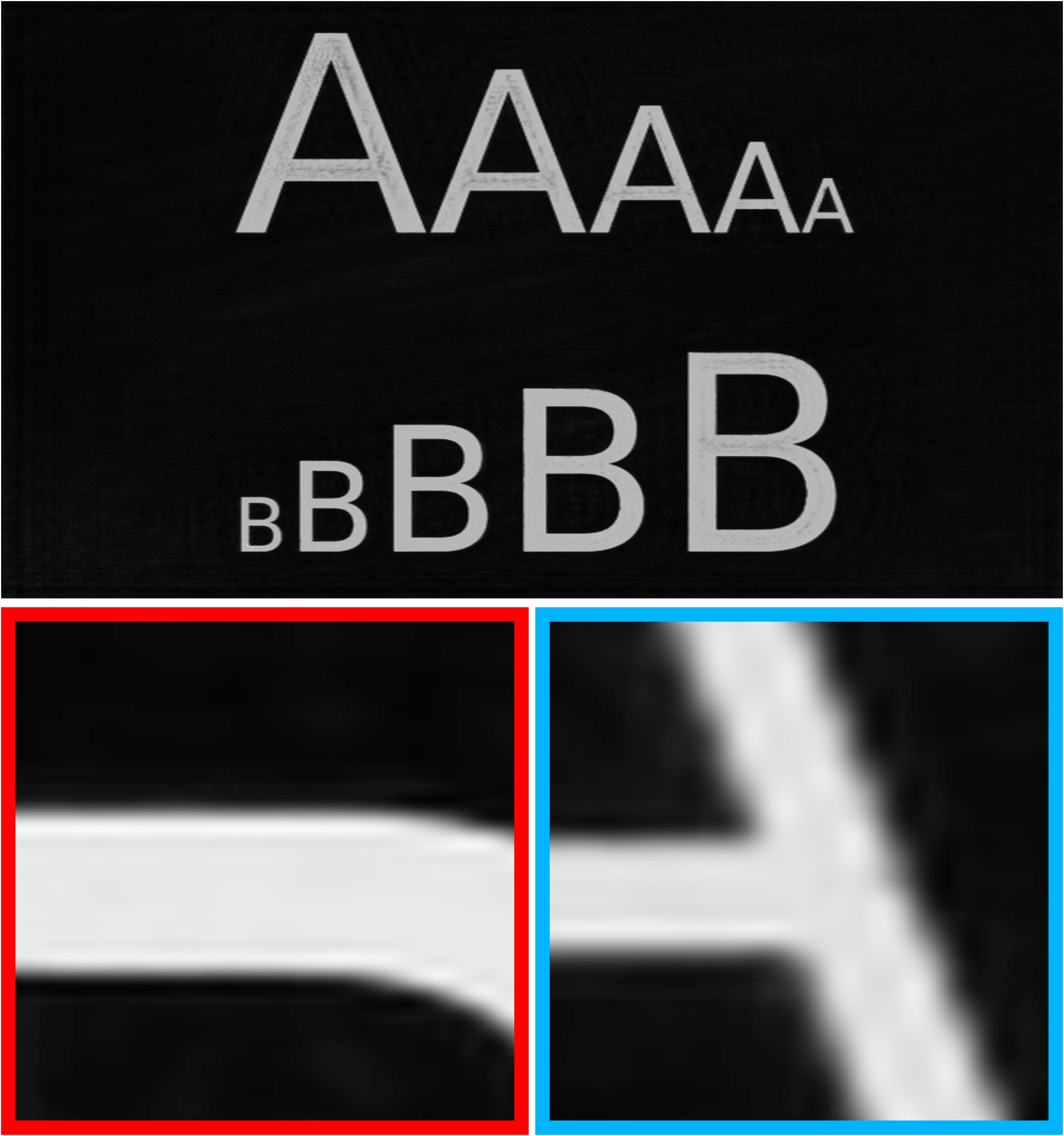} \\
[0.0ex]

\end{tabular}%
}
\vspace{-2mm}
		\caption{ \textit{Temporal and Polarization Multiplexing vs. Single-Shot Ellipsography (Experimental Evaluation).}
We experimentally compare speckle suppression using CITL \cite{peng2020neural} and different averaging methods \cite{choi2022time,nam2023depolarized} with the proposed single-shot ellipsography method. Polarization multiplexing rely on mutual incoherence of two orthogonal polarization components and achieves at most $\sqrt{2}$ speckle reduction. Temporal multiplexing 64 frames reduce the speckle further, at the cost of temporal resolution. In contrast, ellipsography---both with global (GA) and per-pixel (PA) analyzers---achieves equal or better speckle suppression in a single frame, while preserving image sharpness and contrast. }
        
		\label{fig:tm_comparison}
\end{figure*}

\Cref{fig:exp-speckle} presents both qualitative and quantitative evaluations of speckle suppression, 
complementing the results presented in \Cref{fig:2d_captures,figure:3d_figure} and \Cref{tab:exp-metrics}. 
Compared to CITL-optimized reconstructions, ellipsography achieves a \textit{$3\times$ reduction in speckle contrast}, consistent with the trends observed in simulations. 
As shown in the red-bordered inset regions, CITL optimization suffers from loss of fine image details due to residual interference noise, while ellipsography preserves sharp features and produces low-speckle, high-contrast reconstructions from a single exposure. 
We also compare two variants of our method: one using a per-pixel analyzer and one with a global analyzer.
While the global analyzer already outperforms all baselines, it introduces minor artifacts such as brightness non-uniformity in high-contrast regions (\eg the white strips), due to errors from adopting a global analyzer for filtering spatially varying pixel-wise polarization response. 
The per-pixel analyzer, by contrast, shows clean, artifact-free images that faithfully match the target, highlighting the full potential of the vectorial framework. 
These results confirm that ellipsography provides robust speckle suppression under real-world hardware conditions, establishing a new benchmark for single-shot high-fidelity holographic display performance.

\subsection{Ellipsography vs Time and Polarization Multiplexing}

Figure~\ref{fig:tm_comparison} presents an experimental comparison of ellipsography against representative temporal and polarization multiplexing strategies for speckle suppression.
We evaluate (i) traditional scalar holography optimized via CITL optimization~\cite{peng2020neural}, 
(ii) polarization multiplexing using two orthogonal polarization channels~\cite{nam2023depolarized}, 
(iii) temporal multiplexing by averaging multiple independently modulated hologram frames~\cite{choi2022time}, and (iv) the proposed ellipsography method. 
All approaches utilize the same calibrated optical propagation model to ensure fair comparison.

Polarization multiplexing reduces speckle by combining two mutually orthogonal polarization components. However, the number of incoherent polarization channels is strictly limited to two. Because interference persists within each polarization mode, residual speckle remains even after optimization.
Temporal multiplexing achieves further speckle reduction by averaging over multiple independently modulated holograms. Speckle contrast decreases approximately as $1/\sqrt{N}$ with the number of frames $N$. 
This improvement, however, comes at the cost of reduced temporal resolution, increased compute complexity, and sensitivity to motion or synchronization errors \cite{chan2025holospeed}. 
Notably, temporal averaging does not alter the underlying coherent interference pattern of any single frame---it merely averages it out over time.

In contrast, \textit{ellipsography achieves superior speckle suppression within a single exposure}. Instead of averaging multiple realizations of speckle, ellipsography reshapes how coherent interference projects onto the measured scalar channel via polarization-interference engineering. 
By jointly optimizing the vector field and analyzer, residual interference is effectively redirected into per-pixel polarization states rejected by the analyzer. 
This results in clean and high-contrast reconstructions that preserve fine detail, while avoiding the trade-offs inherent to time or channel multiplexing approaches.

\subsection{Frame Multiplexing vs Worst-Case Ellipsography}

\setlength{\tabcolsep}{1.0pt}
\renewcommand{\arraystretch}{0.6}

\begin{figure*}[t]
\centering
\resizebox{\textwidth}{!}{%
\begin{tabular}{cccccc}

\multicolumn{1}{c}{\fontsize{8}{9}\selectfont Target} &
\multicolumn{1}{c}{\fontsize{8}{9}\selectfont CITL Optimization} &
\multicolumn{1}{c}{\fontsize{8}{9}\selectfont 32-Frame Average} &
\multicolumn{1}{c}{\fontsize{8}{9}\selectfont Ellipsography (WC)} &
\multicolumn{1}{c}{\fontsize{8}{9}\selectfont Ellipsography (GA)} &
\multicolumn{1}{c}{\fontsize{8}{9}\selectfont Ellipsography (PA)} \\
[0.1ex]

\includegraphics[width=3.1cm]{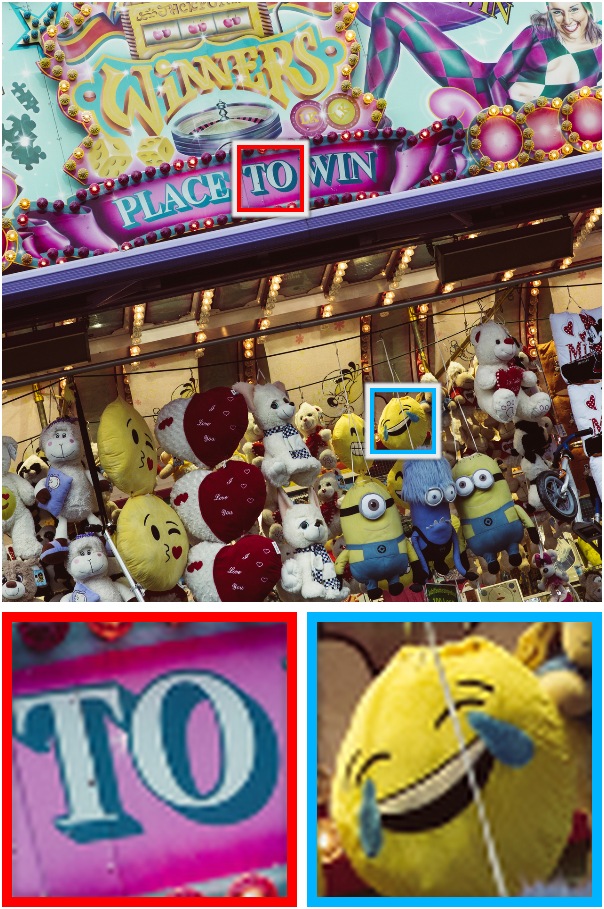} &
\includegraphics[width=3.1cm]{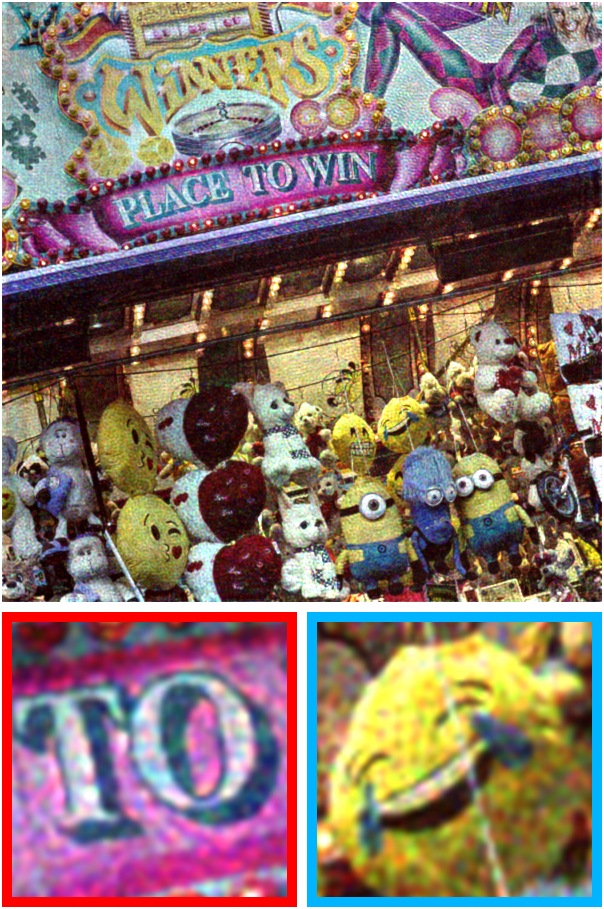} &
\includegraphics[width=3.1cm]{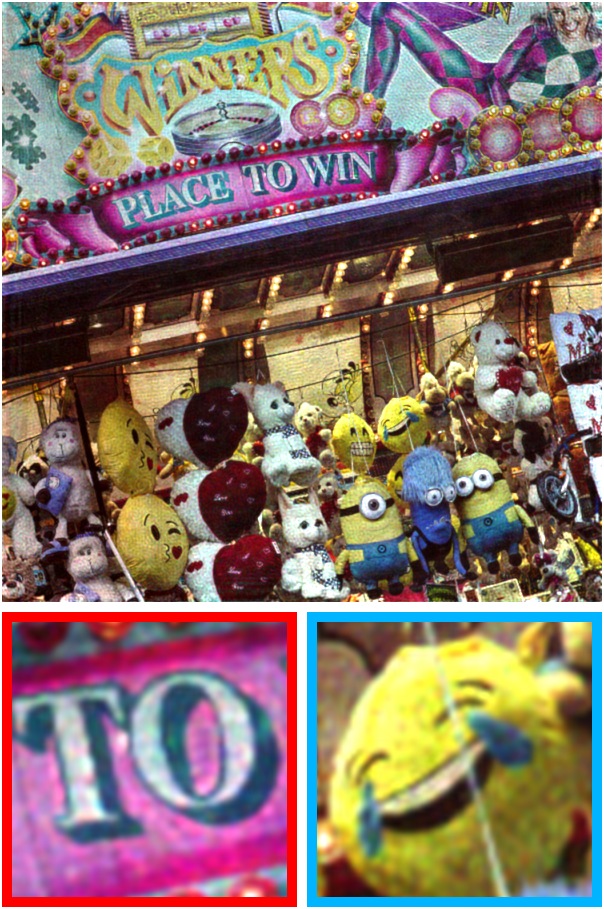} &
\includegraphics[width=3.1cm]{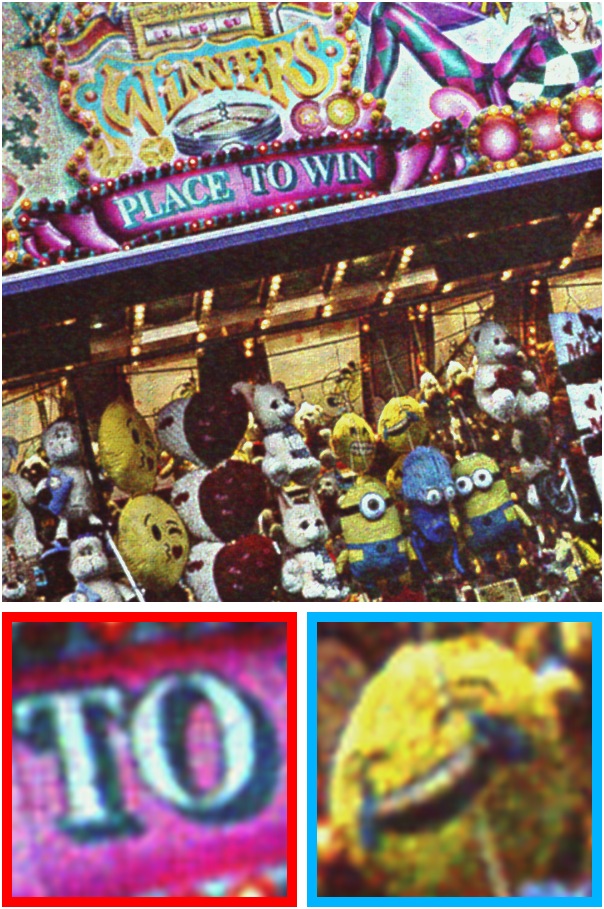} &
\includegraphics[width=3.1cm]{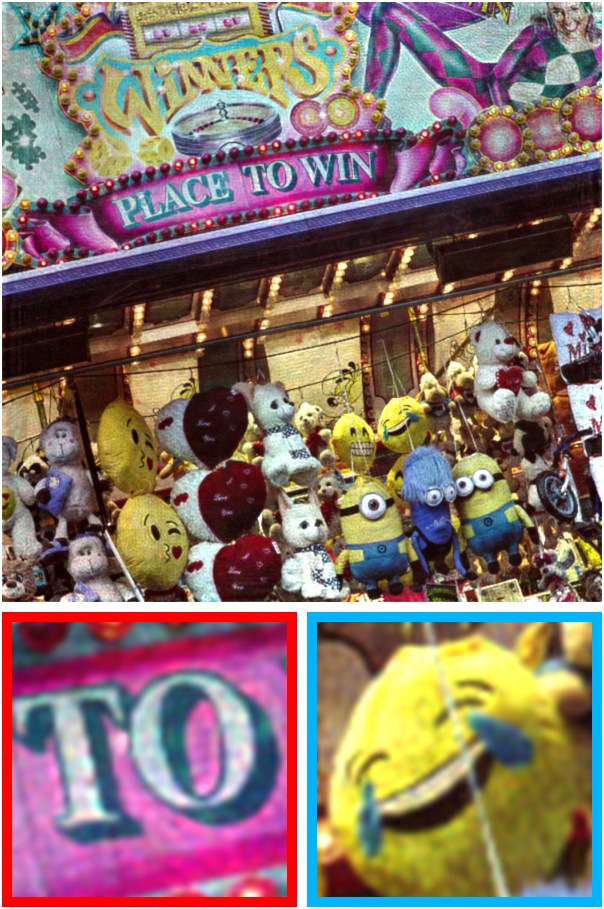} &
\includegraphics[width=3.1cm]{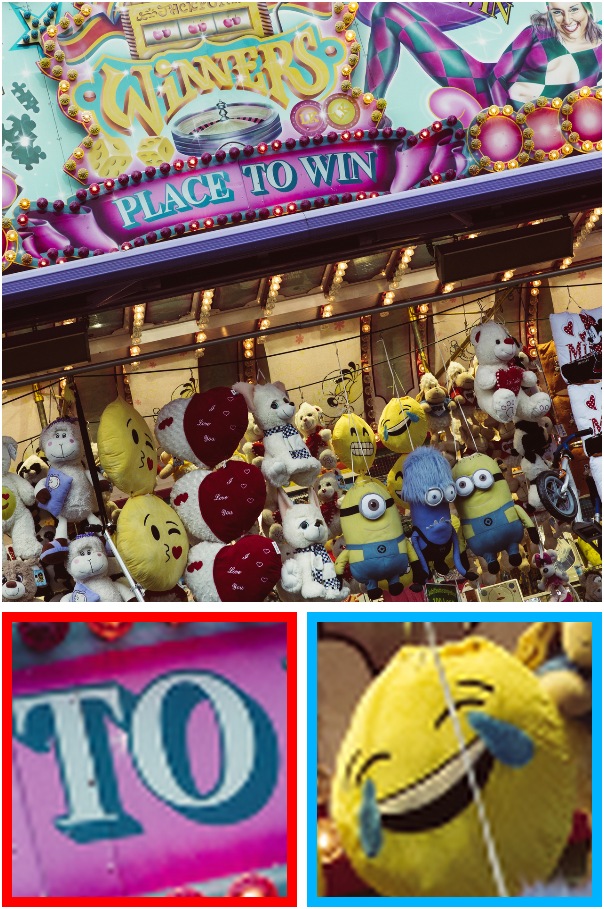} \\
[0.0ex]

\includegraphics[width=3.1cm]{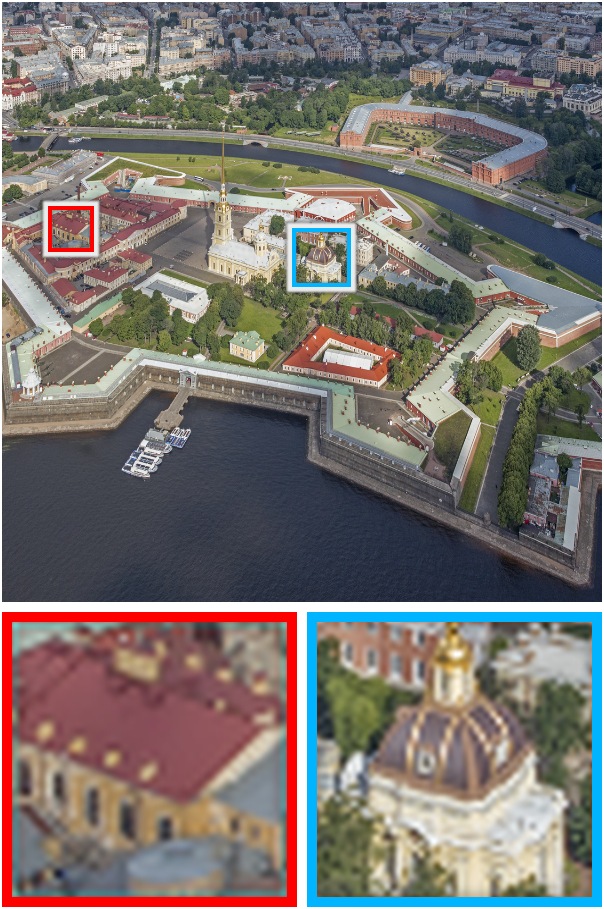} &
\includegraphics[width=3.1cm]{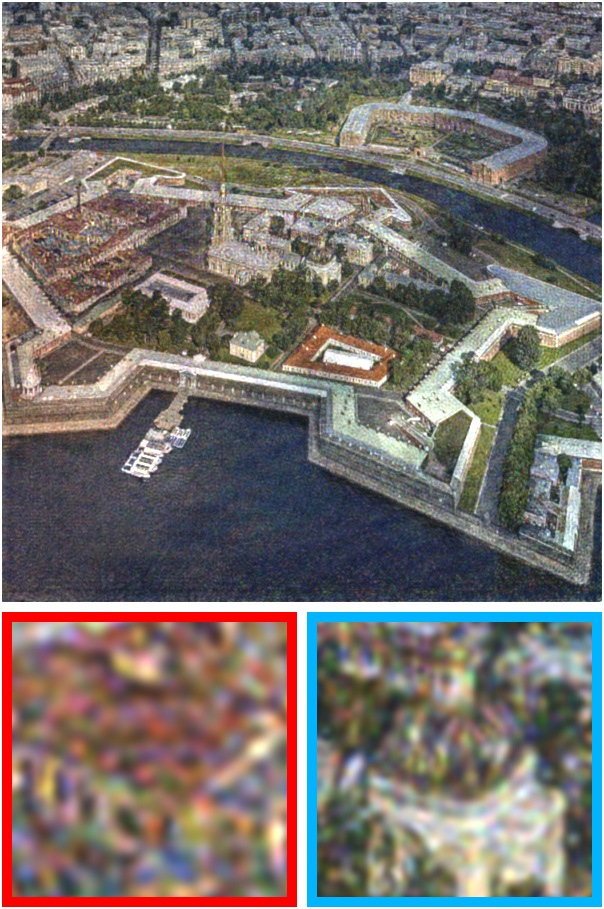} &
\includegraphics[width=3.1cm]{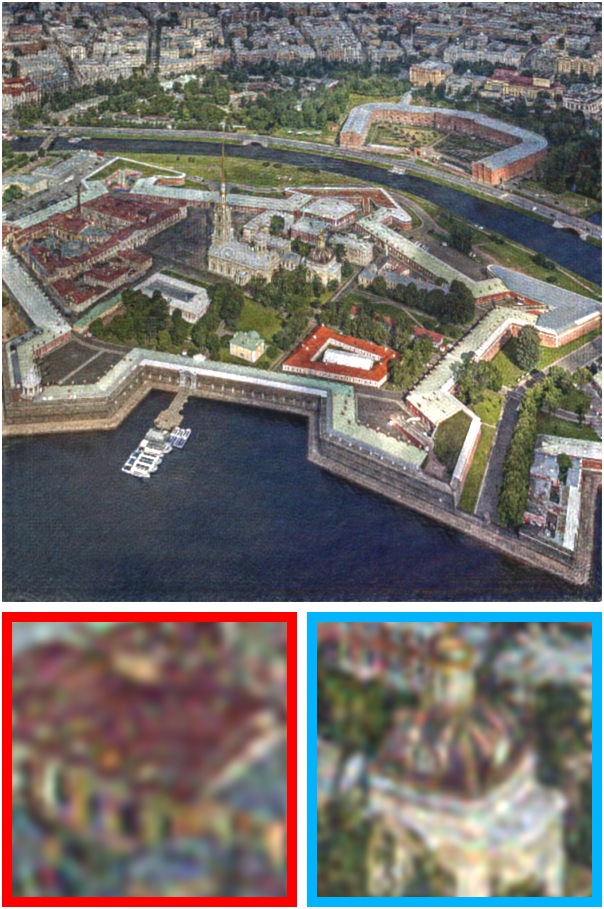} &
\includegraphics[width=3.1cm]{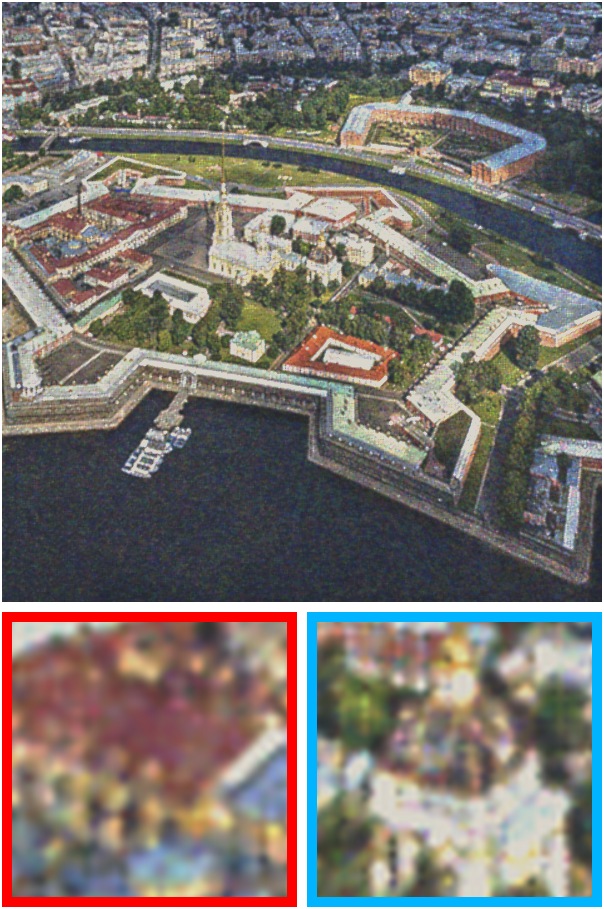} &
\includegraphics[width=3.1cm]{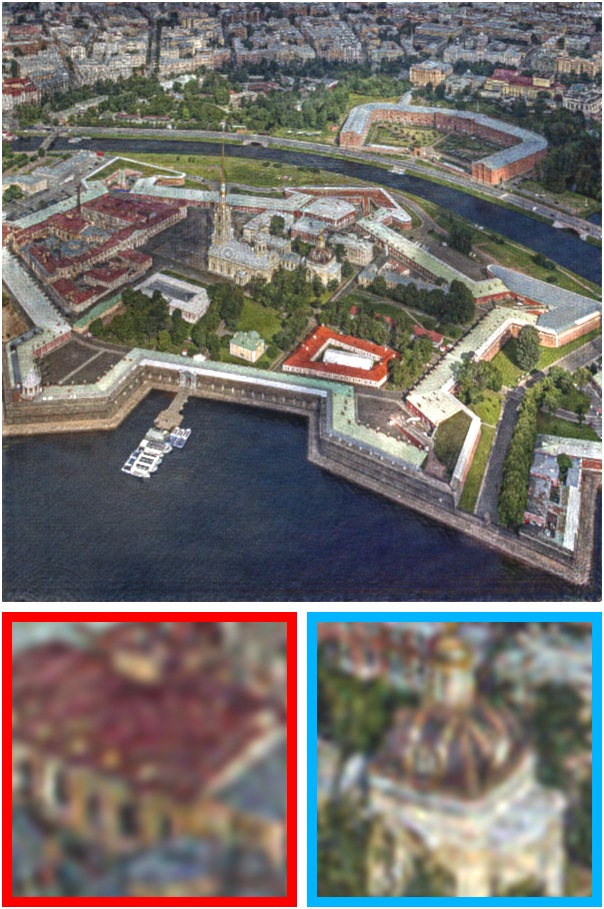} &
\includegraphics[width=3.1cm]{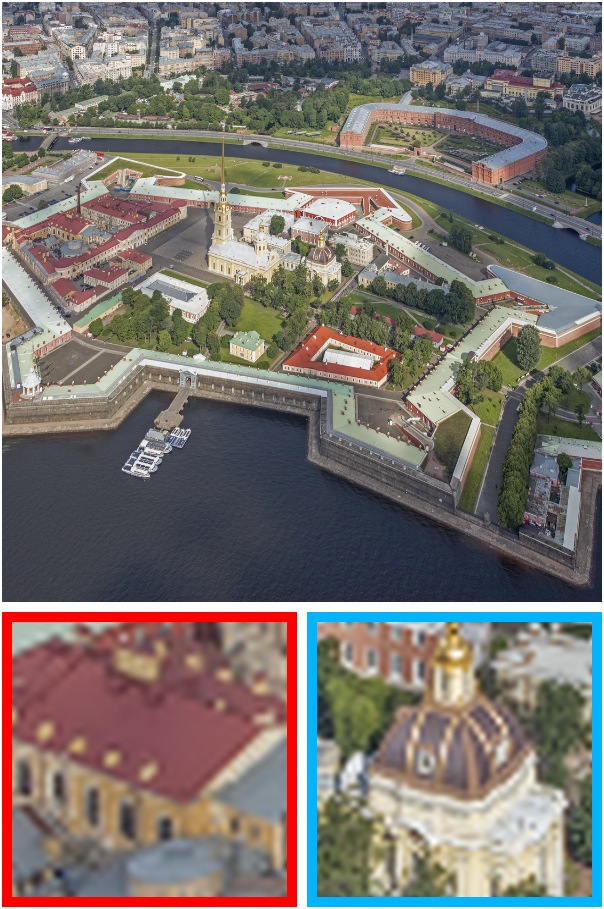} \\
[0.0ex]

\end{tabular}%
}
\vspace{-2mm}
		\caption{\textit{Temporal Multiplexing vs. Worst-Case Ellipsography (Experimental Evaluation).}
We compare speckle suppression using 32-frame temporal multiplexing of CITL--optimized holograms (a strong, conservative baseline) with ellipsography optimized for a per-pixel analyzer but implemented using a global analyzer, representing a worst-case (WC) performance scenario. 
While temporal multiplexing reduces speckle, it introduces averaging artifacts such as blurring and loss of high-frequency detail, evident in fine text and reef surface patterns. 
In contrast, ellipsography---even in the worst-case condition---achieves comparable or superior speckle suppression in a single exposure while preserving image sharpness and contrast.
}
        
		\label{fig:worst}
\end{figure*}

\Cref{fig:worst} presents experimental results comparing 32-frame temporal multiplexing of CITL-optimized holograms---a strong, conservative baseline---with
ellipsography optimized for a per-pixel analyzer but implemented using a global analyzer, representing a worst-case performance scenario.
While temporal multiplexing effectively reduces speckle, it inherently smooths high-frequency fluctions, which leads to blurring of sharp features and loss of spatial detail, particularly evident high-contrast regions such as text and the reef surface in the first-row insets of \Cref{fig:worst}. 
Moreover, temporal multiplexing requires high-speed SLMs and exhibits
sub-linear performance scaling with the number of frames, making further speckle reduction increasingly expensive in both system bandwidth and complexity.

In contrast, ellipsography achieves substantial speckle suppression and high image quality in just a single frame, even under worst-case conditions.
Rather than relying on high-speed temporal averaging, ellipsography leverages spatial resolution and polarization diversity to structure coherent intereference, thereby decoupling speckle suppression from the SLM's temporal bandwidth.
This makes the technique fully compatible with commercially available SLMs and well-suited for practical, real-time, high-fidelity holographic displays without the cost and complexity of high-speed temporal multiplexing.

\section{Discussion and Future Work}
\label{sec:discussion}

Our work takes a step toward practical holographic displays by breaking the longstanding trade-off between phase randomness and image quality, improving key characteristics such as speckle suppression, eyebox energy distribution, and focus cues.
Here, we discuss some of the features of the proposed ellipsography approach, including limitations that arise from various aspects of our design and implementation, and opportunities for future improvements.

\paragraph{Compactness and Miniaturization}
Our current implementation is a benchtop prototype built using two cascaded $4f$ systems. 
However, we envision the core principles of ellipsography built into a compact projector architecture suitable for integration into near-eye display systems \cite{kim2022holographic,gopakumar2024full}.
One pathway to miniaturization is the use of shorter focal length lenses within the $4f$ configuration.
Another promising direction is to replace the $4f$ relay with a single lens of focal length $f'$ placed at a distance of $2f'$, forming a $1:1$ relay between the SLMs. This would enable further compactness while preserving image fidelity \cite{hu2014high}.
Although shorter focal length lenses may introduce spherical and chromatic aberrations, these can be pre-compensated in the hologram phase \cite{maimone2017holographic,shi2021towards}.

\begin{figure}[t]
\centering
\begin{tabular}{cc}
\multicolumn{1}{c}{\fontsize{8}{10}\selectfont Ideal} &
\multicolumn{1}{c}{\fontsize{8}{10}\selectfont Simulated Misalignment} \\
[-0.3ex]
\includegraphics[width=0.49\linewidth]{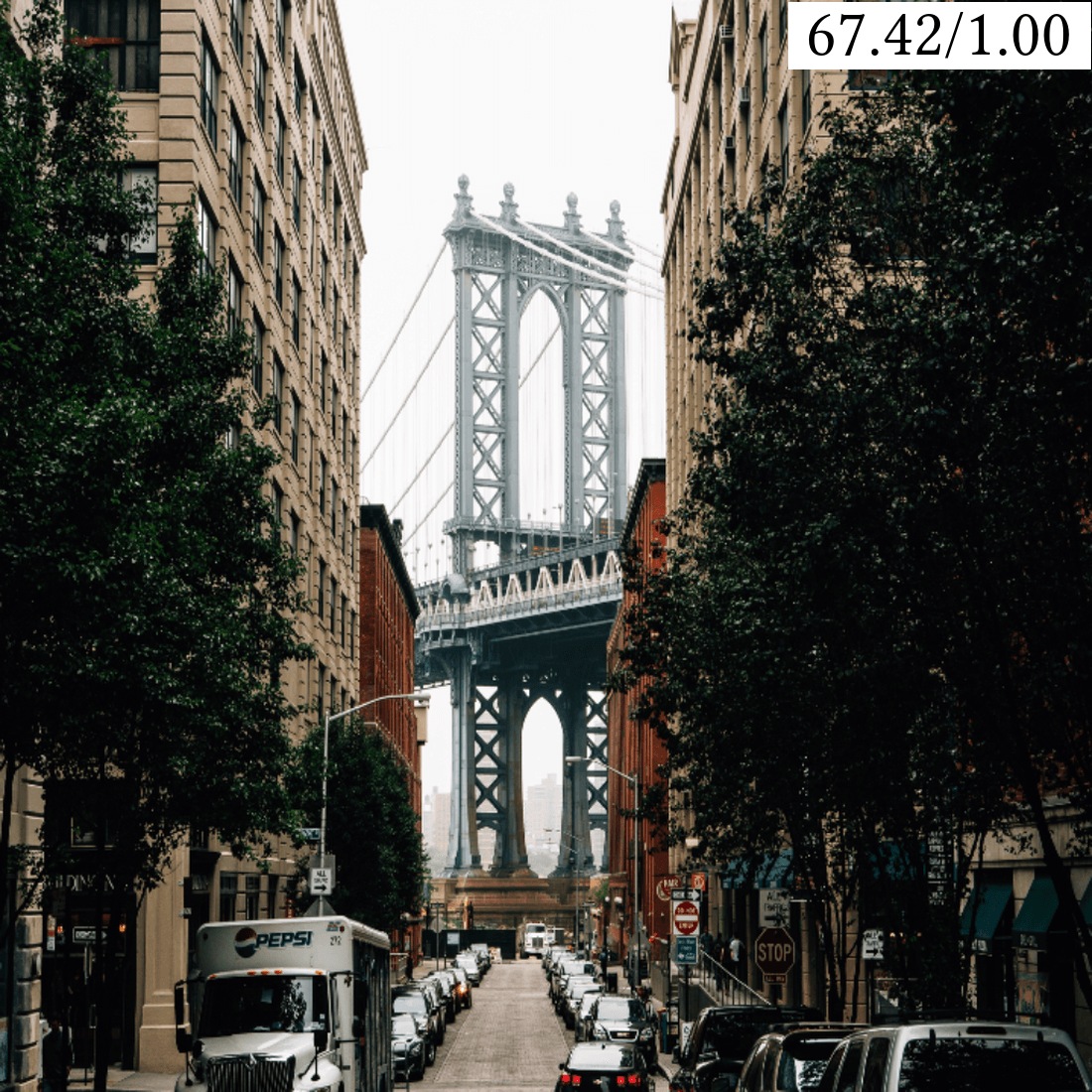} &
\includegraphics[width=0.49\linewidth]{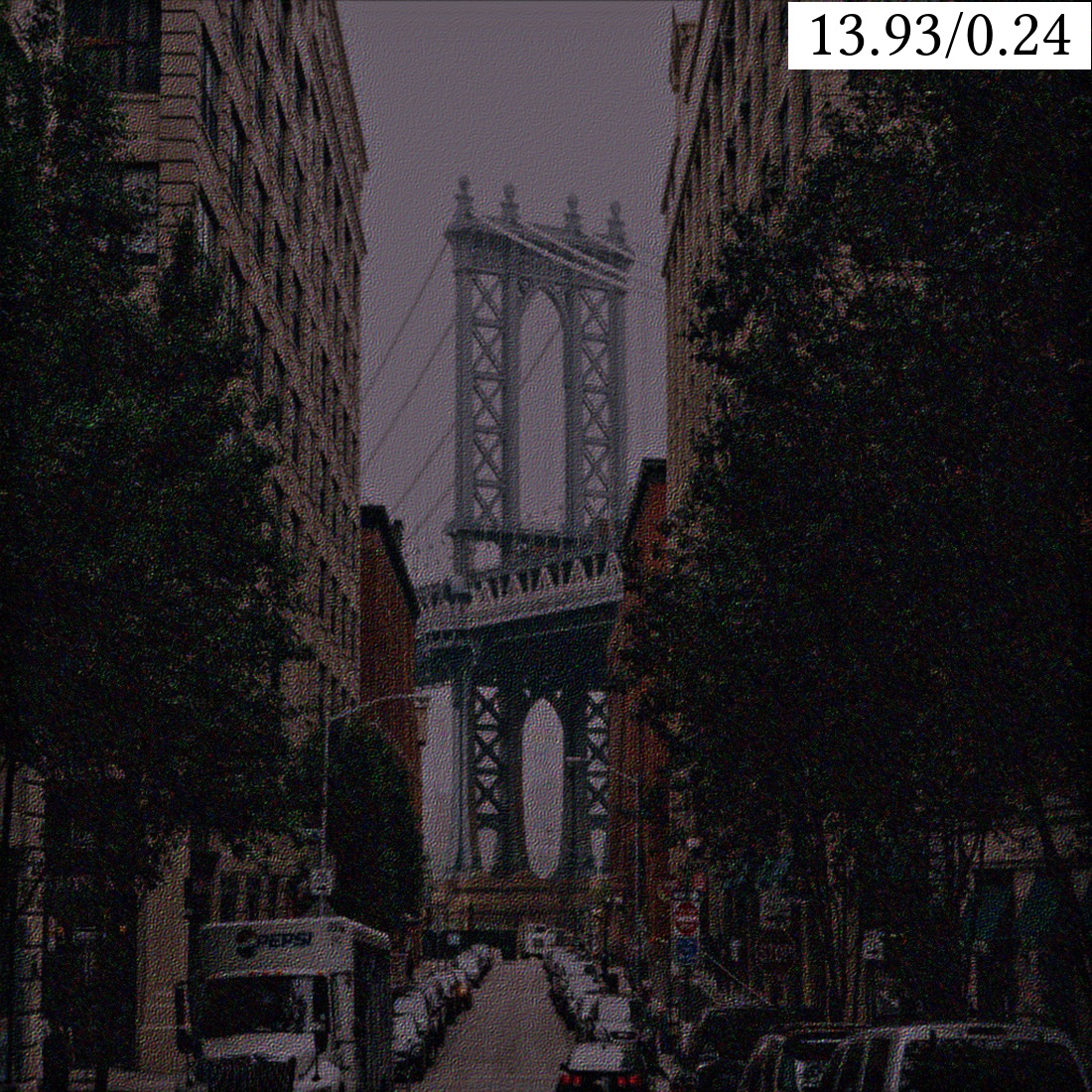}
\end{tabular}

\vspace{-2mm}
  \caption{\textit{Impact of Misalignment on Reconstruction Quality.}
  We visualize the combined effect of 1 pixel misalignment between the SLMs and $1^{\circ}$ misalignment of the analyzer angle to evaluate the alignment sensitivity of our dual-modulation setup. Quantitative evaluations include PSNR/SSIM.
  Please zoom in on the digital version for a clearer view.
  }
  \label{fig:misalignment}
\end{figure}

\paragraph{Optical Alignment and Hardware Considerations} 
While ellipsography demonstrates strong performance in both simulation and physical experiments, optimizing the optical and hardware stack remains an important direction for future work.
In practical setups, imperfections in lenses, polarizing elements, or even minor optical misalignments can introduce light leakage, polarization cross-talk, and interference artifacts that degrade image quality.
\Cref{fig:misalignment} shows the results of a simulation where a deliberate 1-pixel shift and a $1\%$ misalignment between modulated polarization channels were introduced.
The resulting reconstruction exhibits image degradation, highlighting the system's sensitivity to alignment errors.
These real-world deviations account for much of the performance gap observed between the simulation and experiment.
To address these challenges, future hardware implementations could incorporate more robust polarization optics, improved alignment mechanisms, and camera-in-the-loop feedback \cite{peng2020neural,chakravarthula2020learned}.

\setlength{\tabcolsep}{2pt}
\renewcommand{\arraystretch}{0.8}

\paragraph{Computation Speed}

\setlength{\tabcolsep}{2pt}
\renewcommand{\arraystretch}{1.3}
\begin{table}[]
\small
\caption{Runtime performance of different methods. Iterative approaches, including Wirtinger holography, Neural 3D holography and our proposed Ellipsography, generally exhibit longer computational time. In contrast non-iterative methods such as Tensor Holography, HoloNet and DPAC enable near real-time inference. All runtimes are measured on the same hardware as the average per-channel hologram generation time.}\label{tab:running_time}
\begin{tabular}{|m{2.6cm}|l||m{2.5cm}|l|l}
\cline{1-4}
Iterative Methods & Time (s) & Neural Methods & Time (s) &  \\ \cline{1-4} \cline{1-4}
Wirtinger holography   &   $1.3$          & Tensor Holography      &      $0.11$        &  \\
Neural 3D holography   &   $1.87$           &  HoloNet   &     {$0.13$}  &  \\ 
Ours  &   $2.2$           &  DPAC   &     {$0.74$}  &  \\ 
\cline{1-4}
\end{tabular}
\vspace{2mm}
\end{table}

In our current implementation, SLM phase patterns are generated using iterative gradient descent, requiring approximately two seconds to converge. 
As shown in \Cref{tab:running_time},
neural network-based methods offer a promising path toward real-time phase synthesis, having already been successfully applied to traditional scalar holography approaches \cite{peng2020neural,shi2021towards,eybposh2020deepcgh}, although primarily for smooth phase holograms. 
Recent work has also demonstrated the effectiveness of neural networks in compressing holograms to low bit rates, facilitating efficient transmission and display on edge devices \cite{wang2022joint,shi2022neural,qu2025holozip}.
Adapting these learning-based approaches to ellipsography could significantly reduce computation time and enable real-time performance with interactive frame rates.

\paragraph{Perceptual Considerations}
The human visual system is not naturally accustomed to processing images affected by speckle noise \cite{verschaffelt2015speckle}, yet its perceptual impact remains underexplored. 
While prior studies have examined depth cues, parallax, accommodation and artifact resilience in holographic displays \cite{georgiou2023visual,kim2022accommodative,kim2024holographic,Chu2025RealTime}, a comprehensive understanding of how speckle influences image and depth perception is still lacking. 
By suppressing speckle in random phase holograms, our approach improves not only image quality but also eyebox uniformity, focus cues, and depth realism. 
Future work involving rigorous user studies will be essential to quantify these perceptual gains and benchmark holographic displays against traditional and emerging display technologies. Moreover, eye aberrations can reintroduce speckle \cite{chakravarthula2021gaze,kim2024holographic}. Mitigating this `perceived speckle'' will require additional perceptual eye modeling and adaptive optical correction.

\paragraph{{\'E}tendue Expansion}
Our current prototype operates within the native {\'e}tendue of the display optics. 
However, expanding {\'e}tendue is essential for realizing practical holographic displays with a wide field of view and a large eyebox.
A key challenge in {\'e}tendue-expanded systems is that when the eyebox exceeds the pupil size, only a portion of the eyebox energy enters the user's eye, potentially leading to perceived image artifacts \cite{chakravarthula2022pupil,chakravarthula2021gaze}. 
Various optical strategies have been proposed for {\'e}tendue expansion, including binary masks \cite{kuo2020high}, diffractive optical elements \cite{tseng2024neural}, holographic optical elements \cite{xia2020towards}, and lens arrays \cite{chae2023etendue}.
As demonstrated in \Cref{subsec:pupil-invariance}, our method produces pupil-invariant holograms that preserve image quality across the eyebox, independent of pupil size or location. 
Integrating ellipsography with advanced {\'e}tendue expansion techniques presents a promising direction for future work. 

\paragraph{Next-Generation SLM Designs}
Conventional SLMs are typically limited to phase-only or amplitude-only modulation, constraining the degrees of freedom available for wavefront control. 
By incorporating independent polarization modulation, SLMs can leverage the vectorial nature of light, enabling more expressive manipulation of wavefronts and efficient hologram synthesis. 
As demonstrated in this work, joint phase and polarization control enables significantly enhanced speckle suppression, improved depth cues, and more realistic 3D reconstructions from a single frame.
We hope this work motivates the development of next-generation SLM hardware with native support for joint phase and polarization modulation, paving the way for high-fidelity real-time holographic displays.

\section{Conclusion}
\label{sec:conclusion}

We introduced \textit{Ellipsography}, a new class of random phase holography that enables the generation of speckle-suppressed high-fidelity holographic reconstructions. 
To our knowledge, this is the first method to break the longstanding tradeoff between phase randomness and image quality, delivering realistic focal cues, uniform eyebox energy, and full-resolution image fidelity, without requiring spatial or temporal multiplexing.
At the core of ellipsography is the joint modulation of phase and polarization
to structure interference and suppress speckle at its origin. This fundamental capability offers a new degree of control in coherent wavefront synthesis and sets the stage for a rethinking of optical system design.

While our focus in this work is on displays, the broader implications of ellipsography extend to microscopy, biomedical imaging, non-destructive testing, and data storage---any domain where coherent imaging is limited by speckle and interference noise.
We view ellipsography as a foundational step toward practical and high-fidelity holography. 
We believe this opens up a new research direction in computational holography---one that brings holography closer to its long-envisioned potential as the ultimate display technology.


\end{document}